\documentclass[aps,rmp,twocolumn,floatfix,showpacs]{revtex4}
\usepackage{graphics}
\usepackage{color}
\usepackage{amsmath}
\usepackage[hypertex]{hyperref}

\begin{document}
\raggedbottom

\title{Laser-driven nonlinear cluster dynamics}
\author{Th. Fennel}
\author{K.-H. Meiwes-Broer}
\author{J. Tiggesb\"aumker}
\affiliation{Institut f\"ur Physik, Universit\"at Rostock, D-18051 Rostock, Germany}
\author{P.-G. Reinhard}
\affiliation{Institut f\"ur Theoretische Physik, Universit\"at Erlangen, D-91058 Erlangen, Germany}
\author{P.~M. Dinh}
\author{E. Suraud}
\affiliation{Laboratoire de Physique Th\'eorique, Universit\'e Paul
 Sabatier, CNRS, F-31062 Toulouse cedex, France}
\date{\today}

\begin{abstract}
Laser excitation of nanometer-sized atomic and molecular clusters offers various opportunities to explore and control ultrafast many-particle dynamics. Whereas weak laser fields allow the analysis of photoionization, excited-state relaxation, and structural modifications on these finite quantum systems, large-amplitude collective electron motion and Coulomb explosion can be induced with intense laser pulses. This review provides an overview of key phenomena arising from laser-cluster interactions with focus on nonlinear optical excitations and discusses the underlying processes according to the current understanding. A brief general survey covers basic cluster properties and excitation mechanisms relevant for laser-driven cluster dynamics. Then, after an excursion in theoretical and experimental methods, results for single- and multiphoton excitations are reviewed with emphasis on signatures from time- and angular resolved photoemission. A key issue of this review is the broad spectrum of phenomena arising from clusters exposed to strong fields, where the interaction with the laser pulse creates short-lived and dense nanoplasmas. The implications for technical developments include the controlled generation of ion, electron, and radiation pulses, as will be addressed along with corresponding examples. Finally, future prospects of laser-cluster research as well as experimental and theoretical challenges are discussed.
\end{abstract}

\maketitle
\raggedbottom
\tableofcontents
\noindent
\section{Introduction}
\label{sec:100}
Clusters of atoms and molecules frequently appear as a novel state of matter on the one-nanometer scale. For example, different types of bonding or various structural and chemical features can be realized within the same material by just changing the particle size. The opportunity to vary, almost at will, the number of atoms in the clusters thus offers a unique avenue to explore the organization and properties of matter from a fundamental point of view~\cite{Hab94ab,Mar96,Sug98,Alo06}. This also applies to optical phenomena arising from small particles, e.g., due to surface plasmons~\cite{Krei95}, which fascinated scientists since a long time~\cite{RayPM99,Mie08}. Today's lasers open an even more exciting perspective of cluster science, i.e., the opportunity to steer and resolve ultrafast dynamics on the nanoscale.

Due to the progress in laser technology~\cite{KelN03,Rul05}, well-controlled short and intense laser pulses can be routinely delivered these days. This opens the door to explore light-induced dynamical phenomena far beyond the mere analysis of ground state properties. For example, the real-time analysis of nuclear and even electron motion becomes possible, as in the case of molecules or atoms~\cite{Zew94,CorNatP07}. When applied to clusters, short pulses controlled in amplitude and phase allow one to drive and resolve ion and electron dynamics on their natural time scales and under extreme conditions. For instance, electronic relaxation processes or the time-evolution of collective modes can be studied with laser-excited clusters. As a more violent scenario, strong-field exposure transforms clusters into well-isolated nanometer-sized plasmas, with interesting prospects for pulsed particle, radiation, or even neutron sources. With the advent of VUV free electron lasers~\cite{FelJPB05} coherent multiphoton inner-shell excitations are accessible with intense femtosecond pulses. Inspired by such opportunities, the subject of laser-cluster interactions has spawned sustained interdisciplinary activities and experienced enormous developments over the last two decades. It definitely holds the promise to deliver unprecedented insights into the nature of light-matter interactions in complex systems and stimulated challenging efforts in experiment and theory.

In this review we focus on the nonlinear response behavior of clusters subject to laser fields, concentrating on the nonrelativistic intensity regime. It is our aim, in close connection between theory and experiment, to discuss signatures and mechanisms for multiphoton as well as for strong-field excitations. Nevertheless, even single-photon absorption can lead to complex dynamics, e.g., due to electron correlations, structural transitions, or competing electronic decay channels. As a result, the response can clearly go beyond a simple and direct mapping of ground state properties. In any case, pronounced nonlinearities emerge when multiphoton absorption is involved. As a typical example within the still photon-dominated regime, above-threshold ionization can be observed with clusters, showing additional finite-size and many-particle effects when compared to atomic systems. At higher intensities in the so-called field-dominated regime, the immediate excitation of several electrons and laser-driven collisions induce avalanche processes of highly nonperturbative nature. As a surprising feature, clusters very efficiently absorb intense laser radiation~\cite{DitPRL97_3121}, with an energy capture per atom much higher than for atoms or bulk material~\cite{Bat01}. Moreover, strong-field laser-cluster interactions lead to the emission of fast electrons~\cite{SprPRA03}, multiply charged ions~\cite{KoePRL99}, and high-energy photons~\cite{McPN94}, documenting the excitation of core electrons. When compared to atoms, the appearance intensities for these products are strongly reduced with clusters. The discussion of the underlying dynamics and appropriate theoretical treatments is the central topic of this contribution.

Different aspects of laser-excited clusters have previously been reviewed, such as the electronic structure of simple metal clusters~\cite{Hee93,Bra93,Eka99}, low- and moderate-field dynamics~\cite{Rei03a}, ionization mechanisms in strong optical and VUV laser fields~\cite{SaaJPB06}, and excitations with ultraintense pulses~\cite{KraPR02}. The current report aims to deliver a present-day view on cluster dynamics in optical laser fields, with emphasis on the strong-field regime, and incorporates recent findings regarding angular resolved emission, electron acceleration, and processes behind very highly charged ions. Moreover, routes will be reviewed to resolve the cluster response in time by varying the pulse duration or using dual-pulse excitations. Special features of this review are the extensive presentation of experimental and theoretical methods and the attempt of closely combining theory and experiment.

The text is organized in six major parts. Section~\ref{sec:200} offers a quick outlook of the topic and discusses basic physical mechanisms. It thus provides some basic elementary stepping stones on which to build an understanding of the topic. Section~\ref{sec:300} is devoted to a brief survey of available theoretical tools for describing cluster dynamics and tries to show how the various approaches may be linked together in terms of regimes for which they were primarily developed. Section~\ref{sec:400} focuses on experimental techniques, discussing cluster production and laser sources as a starter. Emphasis is essentially put on the detection of emitted particles. In the ensuing presentation of selected results, Sec.~\ref{sec:500} concentrates on the intermediate intensity domain in which photons still count. In this regime experiments have revealed detailed insight into the quantum nature of clusters and allow one to explore the emergence of nonlinear behaviors. Section~\ref{sec:600} finally comes to the main topic of the paper and describes highly nonlinear strong-field induced dynamics where quantum effects are partially wiped out. After a brief survey of initial/original results in the field, a detailed analysis of systematic trends and present day more elaborate approaches are presented. This in particular concerns differential cross-sections and time-resolved analyses. Finally, Section~\ref{sec:700} provides a brief outlook and proposes a few promising future directions of research in the field. We discuss in particular the prospects of laser developments, either in terms of pulse shaping of today's sources or by considering forthcoming devices like the projected XFEL lasers. We also comment on embedded and deposited clusters, high-energy particle acceleration with clusters, and point out some future challenges for theory.

\begin{figure*}[t]
\centering \resizebox{2\columnwidth}{!}{\includegraphics{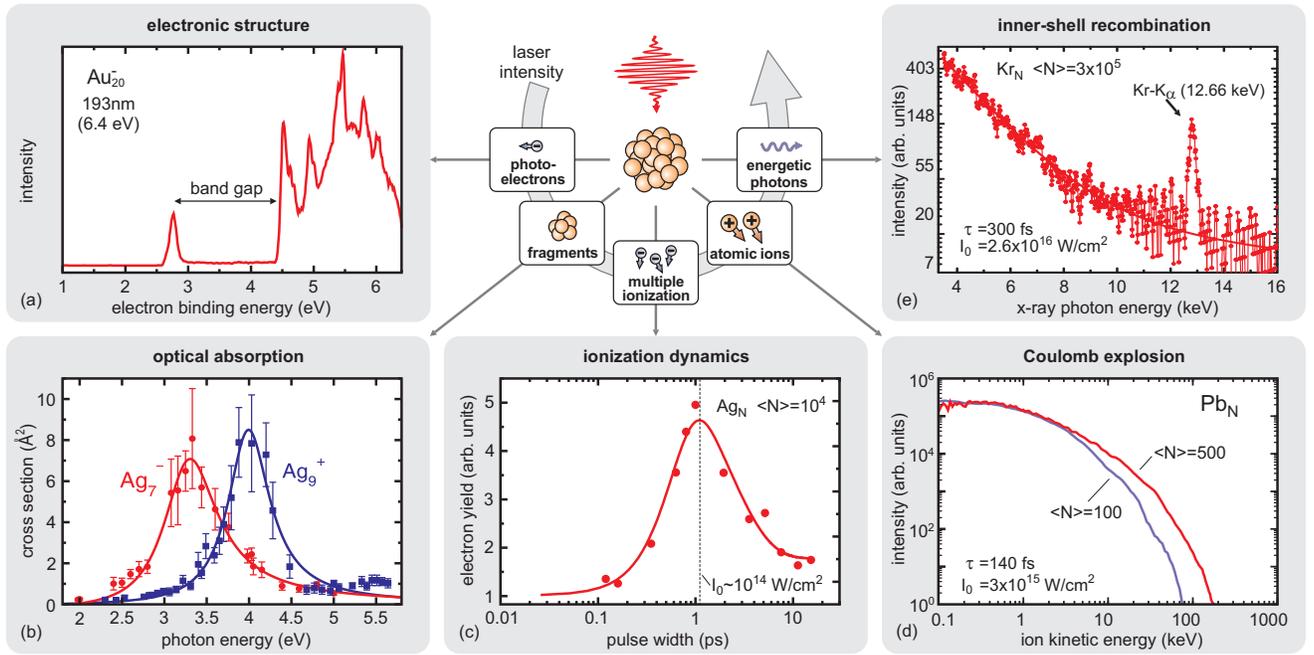}}
         \caption{\label{fig:exper_fiveways} Five decay channels of laser-excited clusters aside with properties/processes that may be resolved from their analysis (see text). (a) electronic structure of negatively charged gold clusters with 20 atoms (Au$_{20}^-$) extracted from the photoelectron spectrum, from~\cite{LiS03}, with permission from AAAS; (b) optical absorption of Ag$_7^-$ and Ag$_9^+$ as determined by photofragmentation, adapted from~\cite{TigPRA93,TigCPL96}, with permission from Elsevier; (c) ionization dynamics of Ag$_N$ in intense laser pulses resolved by measuring the total electron yield as function of pulse width at fixed pulse energy~\cite{Rad04}; (d) Coulomb explosion of Pb$_N$ analyzed by recoil energy spectroscopy of emitted atomic ions, from~\cite{TeuEPJD01}, with kind permission of The European Physical Journal (EPJ); (e) inner-shell recombination in strongly excited krypton clusters measured by x-ray spectroscopy, from ~\cite{IssPP04}, with permission from American Institute of Physics.}
\end{figure*}

\section{General survey of laser-cluster interactions}
\label{sec:200}
Laser irradiation of clusters allows the investigation of a broad spectrum of dynamical processes, ranging from single-photon driven ionization to the strong-field induced explosion of a nanometer-scaled plasma. Irrespective of the regime under consideration, the absence of dissipation into substrate material offers a clean analysis of reaction products, i.e., electrons, ions, cluster fragments, as well as photons. Depending on the cluster material and the chosen laser intensity, quite different properties and response mechanisms can be probed, as will be discussed throughout this review. Exemplarily, Fig.~\ref{fig:exper_fiveways} illustrates a few response channels and properties that may be analyzed and can be viewed as a rough guideline.

As an example for electron emission in the single-photon regime, Fig.~\ref{fig:exper_fiveways}a shows an ultraviolet photoelectron spectroscopy (UPS) result on Au$_{20}^-$ obtained with low intensity laser excitation. The photoelectron energy spectrum images the electronic structure, i.e., binding energies and spectral occupation densities of single electron states, and contains comprehensive information on the system. The large band gap in Fig.~\ref{fig:exper_fiveways}a, for example, reflects the high stability of the tetrahedral Au$_{20}$~\cite{LiS03}. Besides structure analysis, photoelectron spectroscopy (PES) is a powerful tool for monitoring excited states and reactions dynamics, see Sec.~\ref{sec:5A0}.

Laser-induced fragmentation may be analyzed, e.g., to determine optical properties. Fig.~\ref{fig:exper_fiveways}b displays the optical absorption cross-section of size-selected silver clusters measured by photofragmentation~\cite{TigPRA93,TigCPL96}. The spectra exhibit a pronounced resonance, i.e., the Mie surface plasmon, see Secs.~\ref{sec:2A0} and~\ref{sec:2C0}. Collective excitations, as prime examples for multielectron effects, are not only relevant in the single-photon limit, but are important for the cluster response in the multiphoton and strong-field regime as well, see Secs.~\ref{sec:5B1} and \ref{sec:6B0}.

With increasing laser intensity, nonlinear and feedback effects begin to severely influence the cluster response, such as the electron emission. Fig.~\ref{fig:exper_fiveways}c shows an example for larger silver clusters, where the measured total electron yield, i.e., the average cluster ionization, is plotted as a function of the temporal width of the exciting laser pulse~\cite{Rad04}. The strong variation with pulse duration reveals a pronounced ionization dynamics that can be related to the interplay of collective plasma heating and ultrafast relaxation of the ionic structure, see Sec.~\ref{sec:6B1}. In addition, as a result of high charging of cluster constituents, atomic ions are accelerated to high kinetic energies by Coulomb explosion, see Secs.~\ref{sec:6A2} and \ref{sec:6A3}. Examples for ion energy spectra from intense laser excitation of lead clusters are displayed in Fig.~\ref{fig:exper_fiveways}d~\cite{TeuEPJD01} and document kinetic energies of up to hundreds of keV as well as a clear cluster size effect in the recoil energy. Within the strong-field induced excitation process a hot and highly ionized nanoplasma is formed. Clear evidence for the presence of energetic electrons is given by the creation of inner-shell atomic vacancies in the cluster constituents, the recombination of which can be monitored by analyzing the extreme ultraviolet (EUV) and x-ray emission, see Secs.~\ref{sec:6A4} and \ref{sec:6B2}. The example in Fig.~\ref{fig:exper_fiveways}e shows energetic  $K_{\alpha}$-radiation at $12.6\,{\rm keV}$ resulting from irradiation of krypton clusters~\cite{IssPP04}. A detailed analysis of the EUV and x-ray emission can be used for monitoring ion charge state distributions.

The examples highlighted in Fig.~\ref{fig:exper_fiveways} illustrate the wide spectrum of phenomena resulting from laser irradiation of clusters. Before analyzing particular response effects in more detail, a few basic facts about ''protagonists'' of such processes, i.e., clusters and lasers, will be recalled. In the following we furthermore remind basic mechanisms of energy absorption and ionization relying on both individual atomic and cooperative processes and provide a rough classification of different coupling regimes.

\subsection{Basic cluster properties and timescales}
\label{sec:2A0}
Cluster properties are strongly dependent on the type of their constituents. We consider four typical cluster materials: Na as a simple metal, Ag as a noble metal, C as a covalent material, and Ar as a rare-gas systems. Table~\ref{tab:materials} recalls a few basic facts of these elements, e.g., the electronic core and valence levels and corresponding energy gaps. Since cluster properties are by nature also size-dependent (number of constituents between a few and several thousand atoms), atomic, dimer, and bulk values are stated, which fixes typical orders of magnitude.

\begin{table}[b!]
\begin{tabular}{@{}l@{}c@{}|c c c c@{}}
\hline
 & & Na & Ag & C & Ar \\ \hline
{\bf atom} & & & & \\ \hline
ionization potential$^{[1]}$  & [eV]& 5.14 & 7.58 & 11.26  & 15.8 \\
$\varepsilon_\mathrm{val}-\varepsilon_\mathrm{core}$$^{[1]}$ &[eV]& 26.0 & 53.9 &  8.21  & -  \\
valence level && 3s & 5s &  2p & - \\
core level &&  2p & 4p & 2s & 3p \\
lowest dipole exc.$^{[1]}$ &[eV]&  2.1 & 3.66 & 7.48 & 11.62 \\
critical laser intensity
  &\rule[-6pt]{0pt}{16pt}{\big[$\small\frac{\rm W}{\rm cm^2}$\big]}
  &  {3${\times}$10$^{12}$}&  {1$\times$10$^{13}$} & {6$\times$10$^{13}$} & {2$\times$10$^{14}$} \\ \hline
{\bf dimer} & & & & \\ \hline
 bond length$^{[2,3,4,5]}$& [\AA]& 3.08 & 2.53 & 1.20 & 3.83\\
 dissoc. energy$^{[2,5]}$ &[eV]& 0.76 & 1.69 &  6.3 & 0.012\\ \hline
{\bf bulk} & & & & \\ \hline
work function$^{[2]}$&[eV]& 2.75 & 4.26 &  4.8 & 15.8 \\
cohesive energy$^{[2]}$&[eV]& 1.12& 2.95& 7.8 & 0.08 \\
Wigner-Seitz radius$^{[2]}$ &[\AA]& 2.10 & 1.59 & 1.21 &2.21   \\ \hline
\end{tabular}
\caption{\label{tab:materials} Basic atom, dimer, and bulk properties for four typical cluster
materials. Bulk properties for C correspond to graphite which is close to the C$_{60}$ cluster and
carbon nanotubes. The critical laser intensity is estimated with Eq.~(\ref{eq:OFIth}), see
Sec.~\ref{sec:2C0}. The Wigner-Seitz radius $r_s$ characterizes the atomic density. $^{[1]}$NIST;
$^{[2]}$\cite{CRCHBCP};$^{[3]}$\cite{VerJCP83};$^{[4]}$\cite{BeuJCP93};$^{[5]}$\cite{Hirschfelder54}.}
\end{table}

For a given element, the atomic ionization potential (IP) and the bulk work function (WF) indicate the electronic stability of a corresponding atomic cluster with respect to optical excitation. Both IP and WF follow a similar trend over the given materials, i.e., increase from Na to Ar. Typically, metal clusters can be ionized or excited much easier, i.e., with lower photon energies or less intense radiation, than covalent or rare gas systems. This trend is also reflected in the first atomic dipole transition (lowest dipole excitation). The IP further indicates the ionization behavior in strong fields as it determines the critical laser intensity required for atomic barrier suppression, see Sec.~\ref{sec:2C0} for details.

Structural stability is not necessarily linked to that of the electronic system. This becomes evident after comparing dimer dissociation energies or bulk cohesive energies with the IP's, e.g., for C with Ar. Note that the bulk cohesive energies roughly reflect the binding energy per atom of the cluster, while the atomic Wigner-Seitz radius r$_s$ may be used to approximate the cluster radius ($R_{\rm cl}\approx r_{\rm s}N^{1/3}$). The values for the dimer bond length indicate typical interatomic distances.

In the visible and ultraviolet spectral range the optical response is mainly determined by valence electrons. In metal clusters, electron delocalization leads to a strong resonance, the Mie surface-plasmon, as a unique feature of finite objects with sub-wavelength dimension. It corresponds to a collective oscillation of the whole valence electron cloud against the ionic background. When considering schematically a cluster as a metallic drop~\cite{Mie08}, the Mie surface plasmon frequency of a neutral system can be estimated  as~\cite{Hee93,Bra93}
\begin{equation}
  \omega_\mathrm{Mie}=e\,\left(4\pi\epsilon_0m_e r_s^3\right)^{-1/2},
\label{eq:miefreq}
\end{equation}
with $r_s$ the effective Wigner-Seitz radius of conduction electrons, $e$ the elementary charge, $\epsilon_0$ the permittivity of vacuum, and $m_e$ the electron mass. For small Na$_N$, for example, the plasmon energy is around \mbox{$\hbar\omega_\mathrm{Mie}\approx2.8{\rm eV}$}~\cite{SchEPJD99b}, while Eq.~(\ref{eq:miefreq}) predicts a value of \mbox{$3.4{\rm eV}$}. This indicates that the actual Mie response depends on further details (finite size effects, geometrical structure, excitation, net charge, etc.), but Eq.~(\ref{eq:miefreq}) already provides a reasonable order of magnitude sufficient for many forthcoming discussions.

\begin{figure}[t]
\centering\resizebox{0.8\columnwidth}{!}{\includegraphics{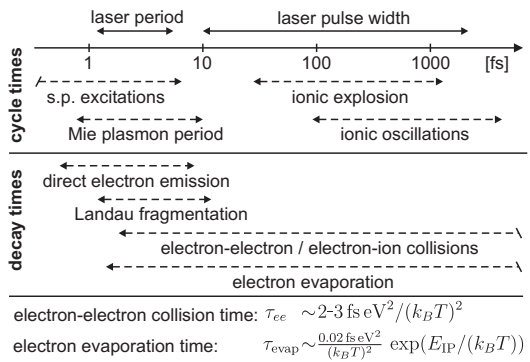}}
\caption{Typical time scales for the dynamics, taking sodium clusters as a prototype. On the top the ranges associated to fs lasers are depicted . Further, processes related to motion (cycle times) and lifetimes due to relaxation (decay times) are indicated. Approximate expressions for electron-electron collisions ($\tau_{\rm ee}$) and electron evaporation ($\tau_{\rm evap}$) are given below. \label{fig:timescales}}
\end{figure}
For considering reaction pathways and energy dissipation it is useful to compare relevant time scales. To that end we consider Na as a typical example for a metal cluster. Fig.~\ref{fig:timescales} provides a schematic overview over times related to laser characteristics, electronic and ionic motion, and lifetimes for relaxation processes. For the moment we ignore the extremely short times associated with core electrons. They certainly play an important role in intense laser fields, but are usually dealt with in terms of simplified rate equations, see, e.g., Sec.~\ref{sec:3C0}. The pulse duration of optical lasers may be varied over a wide range extending from fs to ps or even ns. We focus here on pulse widths of the order a few tens to a few hundred fs.

The shortest time scales in Fig.~\ref{fig:timescales} are related to the electronic motion. The Mie plasmon period as the most basic one is of the order of fs, cf. Eq.~(\ref{eq:miefreq}). In the same range, but with a wider span from sub-fs to several fs, are cycle times for other single-particle excitations and direct electron escape, i.e., single-particle excitation into the continuum. Somewhat slower is the plasmon decay due to Landau fragmentation, in analogy to Landau damping known from plasma physics~\cite{Lif88}. In clusters, Landau fragmentation results from the coupling of plasmons with energetically close single-particle excitations. Viewed in coordinate space, it corresponds to collisions of electrons with the anharmonic potential at the cluster surface. The Landau relaxation time $\tau_{_{\rm L}}$ depends on cluster size and has, e.g., for Na$_N$, its lowest values for \mbox{$N\approx 1000$}~\cite{Bab97a}. For \mbox{$N>1000$} it can be estimated from the time between collisions of an electron with the cluster boundary (``wall friction'') as \mbox{$\tau_{_\mathrm{L}}\approx (r_sN^{1/3})/v_{_\mathrm{F}}$} where \mbox{$v_{_\mathrm{F}}=(\hbar/m)(9\pi/4)^{1/3}/r_s$} is the Fermi velocity~\cite{Yan90}. For \mbox{$N<1000$}, however, $\tau_{_\mathrm{L}}$ increases for smaller $N$ due to the reduced level density. The relaxation time $\tau_{\rm ei}$ describes damping due to electron-ion collisions. It is strongly temperature dependent ($\sim30$\,fs for Na at 273K) and scales as $\tau_{\rm ei}\propto T^{-1}$ at low temperature due to electron-phonon scattering~\cite{Ash76} and follows $\tau_{\rm ei}\propto T^{3/2}$ in a high-temperature plasma~\cite{Spi56}.

The most widely varying times are related to the collisional damping from electron-electron collisions and thermal electron evaporation. Both strongly depend on the internal excitation of the cluster, which may be characterized by an electronic temperature $T$. A simple connection between internal excitation energy per electron \mbox{$\epsilon^*=E^*/N$} and temperature can be established by the Fermi gas model. For \mbox{$k_{\rm B}T\ll\epsilon_{_\mathrm{F}}$}, $T$ can be estimated as \mbox{$k_{\rm B}T=2(\epsilon_{_\mathrm{F}}\epsilon^*)^{1/2}/\pi$}, where \mbox{$\epsilon_{_\mathrm{F}}=\hbar^2(9\pi^2/4)^{2/3}/(2m_er_s^2)$} is the Fermi energy. For the particular case of sodium at bulk density, we have $k_BT=(1.28\,\mathrm{eV}\,\epsilon^*)^{1/2}$. Electron-electron collisions are the key mechanism for electronic thermalization. The $T^{-2}$ law for the corresponding collision time in Fig.~\ref{fig:timescales} is known from Fermi liquid theory~\cite{Pin66,Kad62}. For low $T$, collisions are strongly suppressed due to Pauli blocking of energetically available electronic states. At high $T,$ electron collisions become competitive with Landau damping and sometimes even the dominating damping mechanism. Electron-electron collisions can be described by semiclassical models, see Sec.~\ref{sec:3B3}.

An even more dramatic temperature (or excitation energy) dependence appears for the electron evaporation time, whose trend is dominated by the exponential factor $\exp{({E_{\rm IP}}/(k_BT))}$, where $E_{\rm IP}$ denotes the value of the ionization potential. The more detailed expression for the evaporation time given in Fig.~\ref{fig:timescales} is based on the Weisskopf formulae~\cite{Wei37} \mbox{$\tau_\mathrm{evap}\approx \pi \hbar^3/(8m_er_s^2N^{2/3}) (k_BT)^{-2}\exp{(E_{\rm IP}/(k_BT))}$} and a cluster size of $N$=100. For this size the crossing point \mbox{$\tau_\mathrm{coll}\approx\tau_\mathrm{evap}$} occurs at a temperature of about \mbox{$k_BT=0.8$\,eV}. This corresponds to a hot (nano-)plasma where finite electron clouds are practically an unstable evaporative ensemble. In general, electron evaporation represents a very (sometimes even the most) efficient cooling mechanism for highly excited clusters.

Ionic motion spans a wide range of long time scales. Vibrations, which may be measured by Raman scattering, see, e.g.,~\cite{Por01a}, are typically in the meV regime, i.e., have cycle times of 100\,fs to 1\,ps. In small clusters, ionic vibrations can induce satellites in the optical spectrum~\cite{Ell02a,Feh06b}. Strong laser irradiation usually leads to large amplitude ionic motion and cluster explosion due to Coulomb pressure generated by ionization and thermal excitation. Electron-ion coupling due to Coulomb pressure proceeds at the electronic time scale, i.e., within a few fs. The effect on the ions, however, develops at slower scale, typically beyond 100\,fs, due to the large ionic mass. The time scale of Coulomb explosion can be estimated by considering sudden ionization of cluster constituents to an average atomic charge state $\langle q \rangle$. In this case the cluster expands homogenously and doubles its radius after \mbox{$\tau_{\rm doub}\approx2.3(\sqrt{2\pi\epsilon_0}/e)\,m_{\rm ion}^{1/2}r_s^{3/2}/\langle q \rangle$}, where $m_\mathrm{ion}$ is the ion mass and $r_s$ is the initial atomic Wigner-Seitz radius. For Na$_N$ this yields $\tau_{\rm doub}\approx63\,{\rm fs}/\langle q \rangle$. In consequence, strong ionization drives clusters apart quite rapidly, accompanied with strong changes in the optical properties. Corresponding signatures can be analyzed with pump-probe techniques, see Sec.~\ref{sec:6B0}. For excitations that do not induce explosion, the time scale of electron-ion thermalization reaches up to the ns range~\cite{Feh06b}. Ionic relaxation is even slower, e.g., thermal emission of a monomer can easily last $\mu$s.

As shown above, cluster dynamics comprises a large span of time scales, making their theoretical description to a great challenge. Ionic motion may require a simulation time up to several ps while electronic times scales down to a small fraction of a fs have to be resolved. Theoretical approaches for a corresponding description are subject of Sec.~\ref{sec:300}. Relaxation processes at the ns scale, however, require more phenomenological approaches.

\subsection{Intense laser fields: key parameters}
\label{sec:2B0}
We proceed with a brief summary of basic facts and key parameters of intense laser fields. In the nonrelativistic regime, laser pulses acting on atoms, molecules, or clusters can usually be described as a homogenous time-dependent electric field of the form
\begin{eqnarray}
{\cal E}(t)= {\bf e}_{\rm z} |{\cal E}_0| f(t) \cos(\omega_{\rm las}t+\varphi(t)) , \label{eq:E_las}
\end{eqnarray}
where ${\bf e}_{\rm z}$ denotes linear polarization in z-direction, ${\cal  E}_0$ is the peak field strength, $f(t)$ is the normalized temporal field envelope of the pulse, $\hbar\omega_{\rm las}$ is the photon energy of the carrier, and $\varphi(t)$ is an additional temporal phase. Any other polarization (linear or circular) can be described by superposition. The phase can be written as $\varphi(t)=\varphi_{\rm ce}+\frac{\beta}{2} t^2+\frac{\gamma}{3} t^3+O(t^4)$ where $\varphi_{\rm ce}$ is the carrier-envelope phase, $\beta$ and $\gamma$ denote linear and quadratic chirp, and the last term indicates higher order chirp contributions. Furthermore, the instantaneous frequency reads $\omega_{\rm  inst}(t)=\omega_{\rm las}+\dot \varphi(t)$ and the instantaneous pulse intensity is given by $I(t)=I_0|f(t)|^2$, where $I_0=c\epsilon_0|{\cal E}_0|^2/2$ is the peak intensity and $c$ is the vacuum speed of light. Typically, the pulse duration $\tau$ is given as the full width at half maximum (FWHM) of the temporal intensity profile. A common temporal pulse profile is a Gaussian field envelope, which then reads $f(t)=\exp\left(-2\ln2\,t^2/\tau^2\right)$. In absence of chirp, the bandwidth $\Delta \omega$ (FWHM) of the corresponding spectral intensity profile is related to the temporal pulse width via the time-bandwidth product $\tau_0 \Delta \omega/2\pi=0.441$. Increasing the pulse duration by dispersive pulse stretching to $\tau$ induces a linear chirp of $\beta=\pm 4\ln2\sqrt{s^2-1}/(s^2\tau_0^2)$, where $s=\tau/\tau_0\ge 1$ is the stretching factor with respect to the bandwidth-limited pulse. The chirp direction (up or down) depends on the sign of the group velocity dispersion of the optical element. However, it should be noted that the exact forms of $f(t)$ and $\varphi(t)$ are not always easy to ascertain experimentally. Nonetheless, the pulse duration can nowadays be varied very flexibly over a wide range, e.g., between a few fs up to ns for optical lasers.

In the dipole approximation and using the length gauge, the coupling of the pulse to an electron at position ${\bf  r}$ can be described by an external potential
\begin{equation}
V_{\rm las}({\bf r},t)=e\,{\cal E}(t)\cdot{\bf r}.
\end{equation}
Therefore the system size has to be well below the wavelength \mbox{$\lambda= 2\pi c/\omega_{\rm las}$}, which is well justified for nm clusters and excitation in the optical domain ($\lambda\sim {\rm \mu m}$). The dipole approximation becomes questionable for UV photons and very large clusters, but will be valid in most cases considered below.

To classify coupling regimes it is useful to consider a freely oscillating electron (pure quiver motion, no drift velocity) in the laser field. The cycle averaged kinetic energy defines the ponderomotive potential, which reads
\begin{eqnarray}
U_p&=&\frac{e^2|{\cal E}_0|^2}{4m_{e}\omega_{\rm las}^2} \label{eq_U_pond}
\end{eqnarray}
at the pulse peak. It can be expressed more conveniently by $U_p=9.33\times10^{-14}{\rm eV}\times\,I_0[{\rm W/cm^2}]\,(\lambda[{\rm \mu m}])^2$.
\begin{figure}
\centering\resizebox{0.92\columnwidth}{!}{\includegraphics{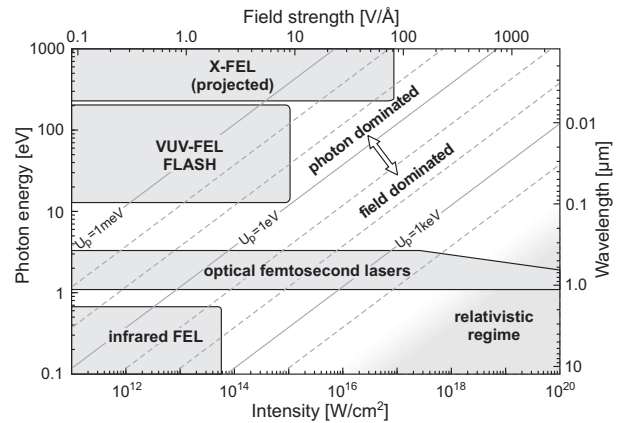}}
\caption{\label{fig:laser-regimes} Intensity-frequency regimes attainable with different high intensity laser systems (shaded blocks). Corresponding wavelengths and electric field strengths are displayed on the additional scales. Lines indicate regions of constant ponderomotive potential $U_p$. The transition from photon- to field-dominated coupling is roughly given by \mbox{$U_p=E_{\rm IP}$}, as schematically depicted for an IP of a few eV. VUV-FEL: vacuum ultraviolet free electron laser; X-FEL: x-ray free electron laser.}
\end{figure}
Fig.~\ref{fig:laser-regimes} displays the dependence of $U_p$ in the frequency-intensity plane aside with the characteristic parameter regions which can be realized with high intensity laser sources. As a rule of thumb, regimes of photon- and field-dominated coupling are separated by a $U_p$ that equals the typical electron binding energy in the considered system, as schematically displayed in Fig.~\ref{fig:laser-regimes}. This condition is related to the Keldysh parameter, as discussed in more detail in Sec.~\ref{sec:2C0}. Figure~\ref{fig:laser-regimes} further illustrates the enormous flexibility of optical lasers to produce high intensities up to the relativistic limit where $U_p$ becomes nonnegligible compared to the electron rest energy. In this review, however, we focus on intensities for which relativistic effects and the magnetic field of the pulses may be neglected. Compared to optical lasers, vacuum ultraviolet and x-ray free electron lasers (VUV-FEL/X-FEL) cover a fundamentally different regime, i.e., photon-driven dynamics at high intensities due to the low ponderomotive potential, see~\cite{SaaJPB06} and Sec.~\ref{sec:7B0}.

\subsection{Ionization and heating mechanisms in clusters}
\label{sec:2C0}
Several basic ionization and energy absorption mechanisms are of relevance for describing laser irradiated particles and will be briefly introduced below. Departing from concepts for atomic and molecular systems we move on to cooperative and collective effects which stem from the many-particle nature of clusters.

\begin{figure}[b]
\centering\resizebox{0.8\columnwidth}{!}{\includegraphics{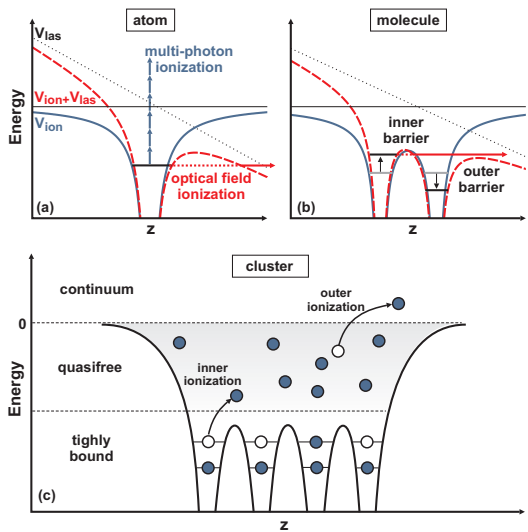}}
\caption{\label{fig:ioni_mechanisms} Schematic view of ionization mechanisms in atoms molecules and clusters. Panels (a) and (b) display potentials of the unperturbed ions $V_{\rm ion}$, the laser $V_{\rm las}$, and their effective sum. In panel (a) the pathways for MPI and OFI of a bound electron are indicated, while panel (b) depicts the charge-resonance-enhanced ionization (CREI). The vertical arrows in (b) indicate the Stark shift. Panel (c) illustrates inner and outer ionization of a  cluster based on an effective potential.}
\end{figure}

On the atomic level, two fundamentally different photoionization processes may be considered. The first is vertical excitation of a bound electron by single- or multiphoton absorption in a rapidly oscillating laser field, see multiphoton ionization (MPI) in Fig.~\ref{fig:ioni_mechanisms}a.
This mechanism proceeds over many laser cycles and prevails for weak and moderate fields in the so-called perturbative domain. A MPI process of order $\nu$ is characterized by the reaction rate \mbox{$\Gamma_\nu=\sigma_\nu I^{\nu}$}, where $\sigma_\nu$ is the corresponding cross-section. MPI, which may be enhanced when intermediate resonant states are available, can promote electrons far beyond the continuum threshold, leading to characteristic peaks separated by units of the photon energy in the electron energy spectrum. This effect, termed as above-threshold ionization (ATI), is well-known from atoms and also appears in clusters, see Sec.~\ref{sec:5B2}. The second mechanism is optical field ionization (OFI). Here the laser acts as a quasistationary electric field. For sufficiently strong fields, bound electrons tunnel through the barrier emerging from the combined potential of the residual $q$-charged ion and the laser field, i.e., \mbox{$V(x)\propto -a/|z|-z$}, with $a=(qe^2)/(4\pi\epsilon_0|{\cal E}_0|)$. This is schematically depicted in Fig.~\ref{fig:ioni_mechanisms}a (dashed curve). The probability for atomic tunneling ionization can be described by the well-known ADK rates found by Ammosov and Delone and Krainov~\cite{AmmJETP86}.

A useful measure for the significance of MPI over OFI is the Keldysh adiabaticity parameter~\cite{Kel65}
\begin{equation}
  \gamma  =  \sqrt{\frac{E_{\rm IP}}{2U_p}} \label{eq:keldysh},
\end{equation}
which compares the IP with the peak kinetic energy of a freely quivering electron (2$U_p$). Single- or multiphoton ionization dominates for \mbox{$\gamma \gg 1$}, where the quiver energy is small compared to the IP. For $\gamma \lesssim 1$, the binding energy can be overcome within a single laser cycle and OFI is promoted. An equivalent expression for the Keldysh parameter is \mbox{$\gamma=\omega_{\rm las}\tau_{\rm tunnel}$}, which gives a ratio of the tunneling time \mbox{$\tau_{\rm tunnel}=\sqrt{2E_{\rm IP}\,m_e/(e^2{\cal E}_0^2)}$} and the optical period. Optical field ionization dominates if the tunneling time is comparable to or smaller than the optical period; MPI is the leading process otherwise.

Within the tunneling regime ($\gamma\lesssim 1$), the ionization probability in one optical cycle approaches unity if the potential barrier can be fully suppressed. For an atomic system, this so-called barrier suppression ionization (BSI) roughly sets in at the threshold intensity
\begin{equation}
  \label{eq:OFIth}
  I_{\rm BSI} = \frac{\pi^2c\epsilon_0^3}{2e^6}\frac{E_{\rm IP}^4}{q^2}\approx4\times 10^9\frac{(E_{\rm IP}[{\rm eV}])^4} {q^2} {\rm
  W/cm}^2,
\end{equation}
which reasonably predicts ion appearance intensities in atomic gases~\cite{AugPRL89}. Note that  Eq.~(\ref{eq:OFIth}) was used to determine the critical intensities in Tab.~\ref{tab:materials}.

The above considerations apply to isolated atoms where the laser parameters govern the dynamics. For extended systems, i.e., from the molecular level on, structural details become increasingly important. Ionization barriers are influenced by the fields from neighboring ions, which, for example, gives rise to {\it charge-resonance-enhanced ionization} (CREI) well-known from strong-field ionization of diatomic molecules~\cite{SeiPRL95,Zuo95}. Within this process, an appropriate internuclear separation results in a simultaneous lowering or suppression of inner and outer potential barriers with respect to the Stark-shifted electronic states (see Fig.~\ref{fig:ioni_mechanisms}b), giving rise to an enhanced ionization rate. For larger or smaller separations either the inner or the outer barriers increase and the ionization probability is reduced. As a truly cooperative effect, CREI has been considered also for very small clusters~\cite{VenPRA01,Sie02}, cf. Sec.~\ref{sec:6B1}.

Very convenient for describing charging dynamics in larger systems is the concept of inner and outer ionization~\cite{LasPRA99}. As indicated in Fig.~\ref{fig:ioni_mechanisms}c, electrons in the cluster may be classified into tightly bound, quasifree, and continuum electrons. Within this picture, \emph{inner ionization} describes the excitation of tightly bound electrons to the conduction band, i.e., electrons are removed from their host ion but reside within the cluster. Correspondingly, the final excitation into the continuum is termed \emph{outer ionization}, which contributes to the net ionization of the system. At moderate laser intensities, systems with initially delocalized electrons, like metallic particles, may undergo outer ionization only. In any case, however, the energy span between the thresholds for inner and outer ionization grows with cluster charge, cf. Fig.~\ref{fig:ioni_mechanisms}c, underlining the growing importance of quasifree electrons for the interaction dynamics. Besides purely laser-induced MPI and OFI, ionization can be driven by cluster polarization (field amplification) or cluster space-charge fields, e.g., subsequent to strong ionization. In addition, quasifree electrons can drive electron impact ionization (EII), as may be described by semiempirical cross-sections~\cite{LotZP67}. The onset and self-amplification of such additional processes is frequently termed {\it ionization ignition}~\cite{Ros97}.

The presence of a \emph{nanoplasma}, i.e., of quasifree electrons and (multi-)charged atomic ions in the cluster, has substantial impact on the energy capture from a laser pulse. If collective effects are negligible, electrons can acquire energy from the laser field via Inverse
Bremsstrahlung (IBS), i.e., by absorbing radiation energy during scattering in the Coulomb field of the ions. IBS relies on the conversion of laser-driven electron motion into thermal energy because of directional momentum redistribution within elastic collisions and is a basic volume-heating effect in underdense plasmas~\cite{KraJPB00}. Considering a fixed collisional dephasing time $\tau_{\rm coll}$ (inverse collision frequency), the IBS heating rate per electron in terms of the ponderomotive potential reads
\begin{equation}
\left \langle \frac{dE}{dt}\right\rangle_{\rm IBS}=2U_p\frac{\tau_{\rm coll}\omega_{\rm las}^2}{\tau_{\rm
coll}^2\omega_{\rm las}^2+1}.
\label{eq:IBS_heating}
\end{equation}
Whereas the heating rate becomes independent of $\omega_{\rm las}$ in the low-frequency case (dc-limit), a \mbox{$U_{\rm p}/ \tau_{\rm coll}$}-dependence is found for \mbox{$t_{\rm coll}\omega_{\rm las}\gg 1$}. It should be noted that the collisional relaxation time, which is a function of electron temperature (cf. Sec.~\ref{sec:2A0}) and becomes frequency dependent \mbox{($\tau_{\rm coll}\propto \omega_{\rm las}^{2/3}$)} for short-wavelength laser excitation, is in general difficult to obtain. For laser-irradiated clusters, pure IBS heating dominates the energy capture of quasifree electrons only at laser frequency far above the Mie plasmon frequency. If the laser frequency becomes comparable to or smaller than $\omega_{\rm Mie}$, the collective response of quasifree electrons in the cluster has to be taken into account. Surface charges from the laser-driven collective electron displacement induce polarization fields, that strongly modify the effective field in the cluster in amplitude and phase. For a spherical plasma and sufficiently small displacements the corresponding restoring force is linear, i.e., the absorption rate per electron for collective IBS heating is described by a Lorentz profile
\begin{equation}
\left \langle \frac{dE}{dt}\right\rangle_{\rm Res}=2U_p\frac{\tau_{\rm coll}\omega_{\rm las}^4}{\tau_{\rm
coll}^2\left(\omega_{\rm Mie}^2-\omega_{\rm las}^2\right)^2+\omega_{\rm las}^2}.
\label{eq:Resonance_heating}
\end{equation}
This expression is equivalent to the heating rate assumed in Ditmire's nanoplasma model, cf. Sec.~\ref{sec:3C0}. Whereas the absorption rates in Eqs.~(\ref{eq:IBS_heating}) and~(\ref{eq:Resonance_heating}) meet in the  high-frequency limit, IBS heating is strongly suppressed for \mbox{$\omega_{\rm las} \ll\omega_{\rm Mie}$} due to efficient screening of the external field by the collective electron displacement. Most importantly, excitation with \mbox{$\omega_{\rm las}\approx\omega_{\rm Mie}$} leads to plasmon-enhanced energy absorption in Eq.~(\ref{eq:Resonance_heating}), cf. the cross-sections in Fig.~\ref{fig:exper_fiveways}b. Resonant collective driving of cluster electrons can produce strong field amplification that supports cluster ionization and direct acceleration of electron~\cite{Rei98b,FenPRL07a}.

In the above discussion the absorption rates have been assumed to scale linearly with intensity ($\propto U_p$), cf. Eqs.~(\ref{eq:IBS_heating}) and~(\ref{eq:Resonance_heating}). This requires
that the dephasing time and the plasmon frequency are constants. In strong fields, however, the large quiver amplitudes actively modify the nanoplasma properties. Hence, both the dephasing time and the plasmon frequency become functions of intensity which introduces additional nonlinear terms.

Another very important aspect for the cluster response to strong optical laser fields is the time dependence of the plasmon energy. It scales as \mbox{$\omega_\mathrm{Mie}\propto \sqrt{\rho_{\rm bg}(t)}$}, where $\rho_{\rm bg}$ is the ion-background charge density. In early stages of the interaction $\rho_{\rm bg}$ is usually too high for being in resonance with the driving IR-field, i.e., the system is \emph{overcritical}. This is the case in metal- and, already after moderate inner ionization, in rare-gas clusters and leads to strongly suppressed IBS heating as explained above. Less efficient surface heating effects like vacuum heating or Brunel-heating~\cite{BruPRL87,TagPRL04} remain active in this overcritical state. Therefore electrons that are pulled away from the surface by the laser field are accelerated outside and contribute their acquired energy upon recollision with the cluster. In any case, as a result of moderate charging and heating, Coulomb forces and thermal electron pressure eventually induce an expansion of the cluster. Corresponding time scales are typically between a few tens of fs to some ps, see Sec.~\ref{sec:2A0} for an estimate of the radius doubling time for pure Coulomb explosion. With cluster expansion the frequency of the collective mode decreases and transiently matches the laser frequency at a certain time, producing a short-lived but strong absorption enhancement, cf. Eq.~(\ref{eq:Resonance_heating}). This idea is a central element of the hydrodynamic approach from Ditmire, see Sec.~\ref{sec:3C0}, however, characterizing the resonance condition in terms of a critical electron density. The latter is justified only for nearly charge neutral systems, such as very large clusters. Since, according to the harmonic potential theorem~\cite{DobPRL94}, the ionic background creates the restoring force for quasifree electrons, the background charge density is the more general parameter applicable also to charged systems. Nonetheless, for sufficiently long pulses the transient resonance induces efficient heating of quasifree electrons and, as a consequence, strongly supports outer ionization and cluster Coulomb explosion. At high laser intensity, this delayed resonant coupling is important irrespective of the cluster material and leaves clear signatures in the absorption as well as in emission spectra, see Sec.~\ref{sec:6B1}.

\subsection{Classification of coupling regimes}
\label{sec:2D0}
While the relative importance of the above mechanisms depends on the specific scenario, regimes can be identified where particular processes prevail. However, such classification cannot be achieved based on a single parameter like laser intensity.  While very low intensities lead to linear and very high ones to nonlinear behavior, other laser characteristics or cluster properties determine the nature of the response for intermediate cases. We briefly discuss a rough sorting of regimes used throughout this article.

The \emph{linear regime} is the domain of weak laser fields associated with single-photon processes and large values of $\gamma$ (cf. Eq.~\ref{eq:keldysh}). Mechanism are sensitive predominantly to the laser frequency. The prevailing examples are optical response spectra. As this is a key tool, there is a huge body of reviews and books see, e.g.,~\cite{Krei95,Bra93,Hee93,Hab94ab}. Early cluster experiments often used ns pulses for studies on structure or low-energy dynamics~\cite{Hab94ab,NaePR97}. Another typical process is single-photon ionization which can be analyzed by photoelectron spectroscopy, see Fig.~\ref{fig:exper_fiveways}a and Sec.~\ref{sec:5A0}.

The \emph{multiphoton regime} is associated with moderate laser intensities where multiphoton processes begin to show up (\mbox{$I\sim10^{8}-10^{13}$ W/cm$^2$} depending on material and frequency). Each laser parameter, i.e., frequency, field strength, and pulse profile, becomes equally important. Typical examples are second harmonic generation~\cite{Goe95,Kle99} and multiphoton ionization. Of particular interest are cases where a multiple of the photon energy can excite an intermediate state of the system. Then, besides direct MPI, a sequential ionization from the (long-living) intermediate state becomes possible~\cite{Poh01}. Another example is above-threshold ionization. Processes emerging in the multiphoton regime are subject of Sec.~\ref{sec:5B0}.

At sufficiently high intensity the laser irradiation produces large ionization and strong heating (\mbox{$I\sim10^{12}-10^{19}$ W/cm$^{2}$}). The excitation of many electrons and strong feedback effects on the response indicate the so-called \emph{strong-field domain} where the dynamics cannot be treated perturbatively. Typically, the excitation leads to cluster Coulomb explosion, accompanied by emission of energetic particles, i.e., electrons and ions, as well as photons. The emitted ions usually carry higher charges than in the case of irradiation of single atoms which underlines the impact of cooperative processes. Moreover, the reactions proceed somehow similar for very different cluster materials (from metals to rare gases) since electrons from atomic shells are activated and the transient nanoplasma determines the dynamics. Such highly nonlinear processes are in the focus of Sec.~\ref{sec:600}.

\begin{figure}[t]
\centering \resizebox{0.75\columnwidth}{!}{\includegraphics{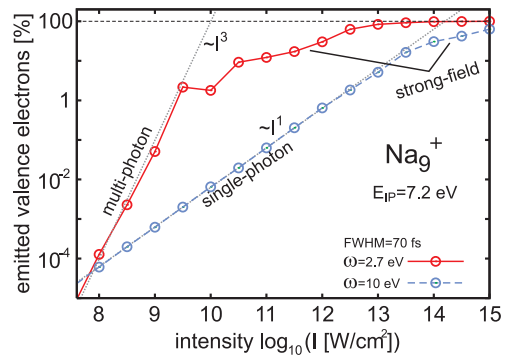}}
\caption{\label{fig:ipowernu} Ionization of Na$_9^+$ as function of laser intensity for excitation by 70\,fs $\cos^2$-shaped laser pulses for two frequencies (as indicated). The ionization potential is $7.2\,$eV. Three photons of $\hbar\omega_{\rm las}=2.7$ eV are required to lift an electron into the continuum (multiphoton ionization) while one photon suffices for $\hbar\omega_{\rm las}=10$ eV (linear behavior). At high intensity both cases become nonperturbative, indicating strong-field conditions. Note, that $\hbar\omega_{\rm las}=2.7$\,eV is close to the Mie plasmon of Na$_9^+$, which leads to the early onset of the strong-field response in this case. Calculations are done in TDLDA. }
\end{figure}

A possible marker for the actual regime is the total ionization yield as function of laser intensity. Lowest order perturbation theory predicts that the yield scales with $\propto I^\nu$, where $\nu$ is the number of photons required to overcome the ionization potential. Figure \ref{fig:ipowernu} gives an example for Na$_9^+$ excited with 70\,fs laser pulses and shows the intensity-dependent electron yield for two different laser frequencies. The slope at low intensities agrees nicely with the $I^\nu$ law, yielding $\nu=3$ (multiphoton) for the lower- and $\nu=1$ (single-photon) for the higher frequency. However, the curves turn over at higher intensities where sorting in orders of photon becomes obsolete (breakdown of perturbation theory). One approaches the \emph{''strong-field domain''}. Note, that the two laser frequencies perform in a very different way. With \mbox{$\hbar\omega_{\rm las}=10$\,eV} excitation, the yield follows the linear behavior and becomes nonperturbative at rather large intensities. With the lower frequency the ionization is a three-photon process and the transition to the nonlinear regime evolves at a much lower intensity. Two effects contribute in the latter case: the near-resonance excitation of the Mie plasmon~\cite{Rei98b} and the stronger impact of optical field effects at lower Keldysh parameters.

\section{Theoretical tools for cluster dynamics}
\label{sec:300}

\subsection{Approaches in general}
\label{sec:3A0}
\begin{table*}[thb]
\scriptsize
\begin{tabular}{|l|l|lll|l|l|}
\multicolumn{7}{c}{\bf \scriptsize Approximations for the electron system}\\
\hline
{\em approach} & {\em scheme} & {\em system}  & $N$ & $E^*/N$ [eV]
 & {\em regime}
 & {\em examples}\\
\hline\hline {\it\bf ab initio}
 & full TDSE
 & He & 2 &  & S D
 & \cite{Par03a}
\\
\cline{2-7}
 & QMC
  & C$_N$& $\lesssim 60,\infty$
  & 0
  & S (D) & \cite{Cep80}
\\
&& pure e$^-$ &&&& \cite{Nee02,Par96}
\\
\cline{2-7}
 & CI
 & any &  &  & S E & \cite{Kra05a,Sch07a}
\\
\cline{2-7}
 & MC-TDHF
 & any &  &  & S E  & \cite{Nes05a,Cai05a}
\\
\hline\hline {\bf quantum}
 & basis expansion,
 & any & $\lesssim 50$  & 0 & S E
 & \cite{Gua95,Mat99}
\\
{\bf DFT}& all electrons &&&&&
\\
\cline{2-7}
 & basis expansion,
 & any & $\lesssim 200$ & $\lesssim 0.1$ & S E D
 & \cite{Saa96,Mat99}
\\
&  pseudopotentials &&&&&
\\
\cline{2-7}
 & coord. space grid,
 &  any & $\lesssim 200$   & $\lesssim 1$ & S E D$^*$
 & \cite{Cal00,Yab96}
\\
& pseudopotentials &&&&&
\\
\hline\hline {\bf semiclassical}
 & Vlasov
 & clusters & $\lesssim 5000$ & $>0.1$ & S D$^*$
   & \cite{FerJPB96,FenEPJD04} \\
   {\bf DFT}
 & VUU  && & & S D$^*$  & \cite{DomPRL98b,KoePRA08} \\
\cline{2-7} & Thomas-Fermi
 & any & $\lesssim 10^6$ & $>0.1$ & S D & \cite{Bla97,DomPRL98} \\
\hline\hline {\bf classical}
 & MD
 & any & $\lesssim  10^6$ & $>0.1$ & D
 & \cite{Hab93,Ros97}
\\
\cline{2-7}
 & rate equations
 & any & $>10^4$ & $>1$  & D
 & \cite{DitPRA96,MilPRE01}
\\
\hline
\multicolumn{7}{c}{\bf \scriptsize Approximations for the ionic system}\\
\hline {\bf quantum}
 & full TDSE
 & $\mathrm{H}_2^+$
 & 1+2
 & any
 & D
 & \cite{Sau07a}\\

\hline\hline {\bf nonadiabatic}
 & MD
 & any
 & $\lesssim  10^6$
 & any
 & S E D
 & \cite{Cal00}
\\
\hline\hline {\bf BO}
 & MD, SE
 & any
 & $\lesssim  200$
 & $E^*_\mathrm{ion}<E^*_\mathrm{el}$
 & D
 & \cite{Bre94b}
\\
\hline
\end{tabular}
\caption{\label{tab:theories}
Hierarchy of approaches for the description of electrons and ions in a cluster. Acronyms are defined in the text. The range of applications is listed in the column {\em regime} where structure is abbreviated as S, excitation spectra (optical response) as E, and dynamics as D. The label D$^*$ indicates the capability to describe electron emission and $E^*$ stands for excitation energy.
}
\end{table*}
Clusters are complex systems and their theoretical description requires approximations to the full quantum-mechanical many-body problem - the more so for truly dynamical situations. As approximations are always a compromise between feasibility and demands, there exists a rich spectrum of methods. Table \ref{tab:theories} tries to provide a brief overview of commonly applied methods - in the upper part for electrons and in the lower part for the ions. Keywords, numbers, and citations are guidelines and by no means exhaustive. They should be understood as examples and estimates of orders of magnitude. For {\it ab-initio} methods some entries for typical sizes and excitation energies $E^*$ are left open as they have, in principle, a huge range of validity, but are, in practice, very limited by quickly growing numerical expense. We add a few remarks while going through the table.

The class of \emph{ab initio} theories covers a huge range of treatments depending on the size of the underlying basis space, in particular for the configuration interaction (CI) and the multiconfiguration time-dependent Hartree-Fock (MC-TDHF) approach. The most general methods, i.e.,  exact time-dependent Schr\"odinger equation (TDSE) and Quantum Monte-Carlo (QMC), are still restricted to very few electrons and presently not applicable to clusters. The vast majority of theoretical investigations of cluster dynamics with quantum aspects relies on density-functional theory (DFT) based methods, with quantum mechanical (QM) or semiclassical propagation, where the latter means Vlasov- or Vlasov-Uehling-Uhlenbeck (VUU) schemes. These will be reviewed in Secs.~\ref{sec:3B1}, \ref{sec:3B2} and \ref{sec:3B3}. Very violent processes exceed the capability of DFT methods and are treated in a purely classical manner, either with molecular dynamics (MD) or, more simple, with rate equations. We will briefly sketch both methods in Secs.~\ref{sec:3B4} and \ref{sec:3C0}.

The large ionic mass usually permits their classical propagation by MD. This may be performed simultaneously with the (nonadiabatic) electron cloud or in Born-Oppenheimer (BO) approximation, if the electrons follow adiabatically the ion field. Light elements (particularly H and He) often call for a quantum mechanical treatment also for the ions. A full quantum treatment for both, all electrons and ions, is extremely demanding and has not yet been applied to clusters. However, a QM treatment of He atoms has been widely used for He clusters~\cite{Ser91,Wei92} and for He material in contact with metal clusters~\cite{Anc95,Nak02a}.

\begin{figure}[h]
\centering\resizebox{0.9\columnwidth}{!}{\includegraphics{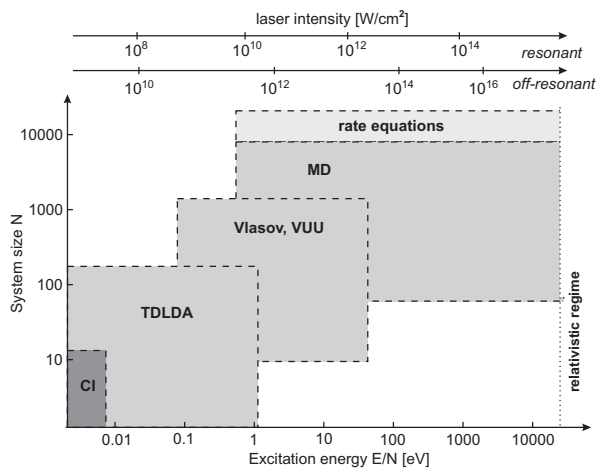}}
\caption{\label{fig:theor_regimes} Schematic view of applicability regimes for different approaches in a landscape of system size vs. excitation energy. The excitation energy can be loosely related to typical laser intensities in the optical range, as indicated by the intensity scales on top for resonant or nonresonant conditions. }
\end{figure}

Figure~\ref{fig:theor_regimes} complements Tab.~\ref{tab:theories} in sketching the regimes of applicability of theoretical models in the plane of excitation and particle number. As the decision for a method depends on several other aspects (e.g., demand on precision, material, time span of simulation), the boundaries of the regimes are to be understood as very soft with large zones of overlap between the models. Note also the two intensity scales on top in Fig.~\ref{fig:theor_regimes}, which indicate that limitations are also sensitive to the nature of the system response, i.e., resonant or nonresonant. The distinction has to be kept in mind when discussing specific systems.

The limitations for CI (and other {\it ab-initio} methods) are purely a matter of practicability. Time-dependent local density approximation (TDLDA) is limited in system size for practical reasons and in excitation energy for physical ones, because of the missing dynamical correlations from electron-electron collisions. The upper limits of VUU are also of purely practical nature while the lower limits are principle ones, e.g., the negligence of shell effects, tunneling, and interference. The same holds for MD and rate equations. The upper limits in energy and/or laser intensity are given by the onset of the relativistic regime, where retardation effects within the coupling begin to severely influence the dynamics. For the particle size, a general upper limit results from the application of the dipole approximation, which typically breaks down beyond some ten thousand atoms. In larger systems the field propagation effects (attenuation, diffraction, reflection) need to be taken into account.

\subsection{Effective microscopic theories}
\label{sec:3B0}
Since a fully {\it ab initio} treatment of cluster dynamics is hardly feasible, simplifications are necessary by eliminating details of many-body correlations. This naturally leads to a description in terms of single-particle states which is well manageable and still maintains crucial quantum features. The eliminated degrees of freedom are moved to an effective interaction to be used in the reduced description.  This leads into the realm of DFT~\cite{Dre90}. TDDFT, i.e., its dynamical extension~\cite{RunPRL84,Gro96}, is widely employed in cluster dynamics~\cite{Rei03a} and still under development, see, e.g.,~\cite{Mar06}.  This section provides a brief overview over the typical approaches used for cluster dynamics these days. We begin with the discussion of the energy functionals, proceed with quantum- and semiclassical DFT methods, and end up with the most simplified treatment, i.e., molecular dynamics.

\subsubsection{The energy functional}
\label{sec:3B1}
\begin{table*}
\scriptsize
$$
\begin{array}{|c|c|ccccccccc|}
\hline \mbox{\bf type} & \mbox{\bf central variables}  & & & \mbox{\bf kinetic} && \mbox{\bf Coulomb} && \mbox{\bf effective}
  && \mbox{\bf external} \\
\hline
\mbox{ions}
&\rule[-0.3em]{0pt}{2.6em}
\;\{\mathbf{R}_I,\mathbf{P}_I\}\;
& E^\mathrm{ion}
 &=
 &\displaystyle
  \sum_I\frac{\mathbf{P}_I^2}{2M_I}
 &+
 &\displaystyle
  \frac{e^2}{8\pi \epsilon_0}\sum_{I,J}\frac{1}{|\mathbf{R}_I-\mathbf{R}_J|}
 &&
 &+
 &\displaystyle
  E_\mathrm{ext}^\mathrm{ion}(\mathbf{R}_I,\mathbf{P}_I)
\\
 \mbox{coupling}
 &&  E^\mathrm{coupl}& = &
 &
 &
 &+
 &
 \rule[-1.8em]{0pt}{2.6em}
 \displaystyle
  \sum_I\!\int\!{\rm d}^3 \mathbf r\,\rho(\mathbf{r})
  {V}^\mathrm{PsP}_I(\mathbf{r})
 &&
\\
\hline
 &   &
\multicolumn{9}{|c|}{\mbox{\rule[-0.6em]{0pt}{1.6em}
      \bf Quantum mechanical }}
\\
\hline
\mbox{electrons}
 & \;\{\varphi_\alpha(\mathbf{r})\}\;
 & E^\mathrm{el}
 & =
 &\displaystyle
  +\sum_\alpha(\varphi_\alpha|\frac{\hat{p}^2}{2m}|\varphi_\alpha)
 & +
 &\displaystyle\rule[-0.9em]{0pt}{2.9em}
   \frac{e^2}{8\pi \epsilon_0}\int {\rm d}^3 \mathbf r \,  {\rm d}^3 \mathbf r'
    \frac{\rho(\mathbf{r})\rho(\mathbf{r}')}
         {|\mathbf{r}-\mathbf{r}'|}
 &+
 &\displaystyle
   E_\mathrm{xc}(\rho)
   -
   E_\mathrm{xc}^\mathrm{(SIC)}(\rho_\alpha)
 &+
 & E_\mathrm{ext}^\mathrm{el}(\rho,\mathbf{j})
\\
 &&&&&&
 \displaystyle\rule[-1.6em]{0pt}{2.2em}
 \rho(\mathbf{r})=\sum_\alpha|\varphi_\alpha(\mathbf{r})|^2
 &&&&
\\
\hline
 &    \Downarrow  &
\multicolumn{9}{|c|}{\mbox{\bf\rule[-0.5em]{0pt}{1.6em}
  Vlasov approximation for electrons}}
\\
\hline
 & \displaystyle
  f(\mathbf{r},\mathbf{p})
 &\displaystyle E^\mathrm{el}
 &=
 &\displaystyle
   \int\!{\rm d}^3 \mathbf r \, {\rm d}^3 \mathbf p \,\, \mathbf{p}^2
  f(\mathbf{r},\mathbf{p})
 & +
 &\displaystyle\rule[-0.6em]{0pt}{2.6em}
   \frac{e^2}{8\pi \epsilon_0}\int {\rm d}^3 \mathbf r \, {\rm d}^3 \mathbf r'
    \frac{\rho(\mathbf{r})\rho(\mathbf{r}')}
         {|\mathbf{r}-\mathbf{r}'|}
 &+
 &\displaystyle
   E_\mathrm{xc}(\rho)
 &+
 & E_\mathrm{ext}^\mathrm{el}(\rho,\mathbf{j})
\\
 &&&&&&
 \displaystyle\rule[-1.6em]{0pt}{3.6em}
 \rho(\mathbf{r})=\int {\rm d}^3 \mathbf p\,f(\mathbf{r},\mathbf{p})
 &&&&
\\
\hline
 &    \Downarrow  &
\multicolumn{9}{|c|}{\mbox{\rule[-0.5em]{0pt}{1.8em}
   \bf Molecular dynamics for electrons}}
\\
\hline
&
\;\{\mathbf{r}_\alpha,\mathbf{p}_\alpha\}\;
 &\displaystyle E^\mathrm{el}
 &=
 &\displaystyle
  \sum_\alpha\frac{\mathbf{p}_\alpha^2}{2m}
 & +
 &\displaystyle
   \rule[-0.9em]{0pt}{2.9em}
   \frac{e^2}{8\pi \epsilon_0}\int {\rm d}^3 \mathbf r \,  {\rm d}^3 \mathbf r'
    \frac{\rho(\mathbf{r})\rho(\mathbf{r}')}
         {|\mathbf{r}-\mathbf{r}'|}
 &+
 &\displaystyle
 \sum_{\alpha\beta}
  V^\mathrm{(eff)}(|\mathrm{r}_\alpha-\mathrm{r}_\beta|)
 &+
 & E_\mathrm{ext}^\mathrm{el}(\mathbf{r}_\alpha,\mathbf{p}_\alpha)
\\
 &&&&&&
 \displaystyle\rule[-1.5em]{0pt}{2.9em}
 \rho(\mathbf{r})
 =
 \sum_\alpha g(|\mathbf{r}-\mathbf{r}_\alpha|)
 &&&&
\\
\hline
\end{array}
$$
\caption{\label{tab:basicE} Composition of the basic energy-density functional for electrons, ions and their coupling $E=E^\mathrm{ion}+E^\mathrm{coup}+E^\mathrm{el}$. The ions are described as classical particles with coordinates $\mathbf{R}_I$ and momenta $\mathbf{P}_I$, $I=1,\ldots,N_\mathrm{ion}$. They correspond to the nuclear centers and the deeper lying, inert core electrons. The coupling to the electrons is mediated by pseudopotentials $V^\mathrm{PsP}_I$ which are designed to incorporate also the impact of the core electrons on the active electrons. The $V^\mathrm{PsP}_I$ counterweight the Coulomb singularity of point charges (see Coulomb coupling term) and install effectively a soft inner charge distribution for the ion.  We show here for simplicity a local pseudopotential which applies throughout all approaches. Nonlocal versions are often used in connection with QM electron wavefunctions. The electrons can be treated at various levels of approximation. The QM stage employs single-electron wavefunctions $\varphi_\alpha$ where $\alpha=1,...,N_\mathrm{el}$.  The semiclassical Vlasov description replaces an orbital based treatment by a phase-space function $f(\mathbf{r},\mathbf{p})$.  In both cases, the Coulomb exchange term and correlations are approximated by effective functionals, usually in local density approximation (LDA) and optionally augmented by a self-interaction correction (SIC).  The fully classical level treats electrons as point particles with specifically tuned effective interaction potentials, e.g., by assuming a charge distribution $g(r)$ having a finite width. The total electronic density $\rho(\mathbf r)$ is computed differently when going from the QM over Vlasov to the MD approaches. Note that the current $\mathbf{j}(\mathbf{r})$ is defined analogously to the density.}
\end{table*}
Since DFT relies on a variational formulation, it aims at well-controlled approximations. The starting point is an expression for the total energy of electrons and ions from which all static and dynamic equations can be derived. Approximations are made only at one place, namely within this energy functional, and everything else follows consistently. Typical energy functionals used in cluster physics (and many other fields) are summarized in Tab.~\ref{tab:basicE}. We comment additional aspects briefly.

Key to success (or failure) is the choice of a reliable functional for exchange and correlations. There are several well-tested functionals within local density approximation (LDA) around, see, e.g.,~\cite{Per92}. These are the workhorses in cluster dynamics. Higher demands, e.g., in describing molecular bonding of covalent materials require more elaborate functionals including gradients of the density, as in the generalized gradient approximation (GGA)~\cite{Per96}.  And even these turn out to be insufficient in some dynamical situations. The spurious self-interaction spoils ionization potentials and related observables. This can be cured to some extent by a self-interaction correction (SIC) or an appropriate approximation to it (for a discussion in the cluster context see~\cite{Leg02}). Recent developments in TDDFT employ the full exchange term and try to simplify that by optimized effective (local) potentials (OEP)~\cite{Sal03a,KueRMP08}. This is still in an exploratory stage and schemes applicable in large-scale dynamical calculations have yet to be developed.

Another source of \emph{effectiveness} are pseudopotentials for ions containing inert core electrons~\cite{Sza85} -- a well-settled topic for static problems. Dynamical applications require to consider the polarizability of core electrons, e.g., in noble metals~\cite{Ser97}. This can be done by augmenting the pseudopotentials with polarization potentials as done in mixed quantum mechanical molecular dynamics approaches~\cite{Gre99a}, for a cluster example see~\cite{Feh06a}.

Table \ref{tab:basicE} finally includes the step down to a fully classical treatment (MD for electrons). This level develops its effective interactions on an independent route, i.e., by explicit adjustment of the effective interactions to basic molecular and/or bulk properties, see Sec.~\ref{sec:3B4}.

\subsubsection{Time-dependent density-functional theory}
\label{sec:3B2}

The time-dependent Kohn-Sham (KS) equations coupled with ionic MD are derived by variation of the given energy (see Tab.~\ref{tab:basicE}) with respect to the single-electron wavefunctions $\varphi_\alpha^\dagger$ and to the ionic variables, for details see, e.g.,~\cite{Rei03a}. They read
\begin{subequations}
\label{eq:TDLDA-MD}
\begin{eqnarray}
  \mathrm{i}\hbar\partial_t\varphi_\alpha
  &=&
  \hat{h}_\mathrm{KS}\varphi_\alpha
  \quad,\quad
  \hat{h}_\mathrm{KS}
  =
  \frac{\delta E}{\delta \varphi_\alpha^\dagger}
  \frac{1}{\varphi_\alpha}
  \quad,
\\[3pt]
  \partial_t\mathbf{R}_I
  &=&
  \nabla_{\mathbf{P}_I}E
  \quad,\quad
  \partial_t\mathbf{P}_I
  =
  -\nabla_{\mathbf{R}_I}E.
\end{eqnarray}
\end{subequations}
Since by far most applications employ the local density approximation (LDA), the electronic part is coined time-dependent LDA (TDLDA).  It is coupled to MD for the ions, yielding together TDLDA-MD.  This treatment where electronic and ionic dynamics is propagated simultaneously is compulsory for strong electronic excitations.

There are many situations where rather slow ionic motion dominates and
the electron cloud acquires only very little excitation energy. For
then, one can switch to the adiabatic Born-Oppenheimer (BO) picture:
\begin{subequations}
\begin{eqnarray}
  \varepsilon_\alpha\varphi_\alpha^{(\mathbf{R}_I)}
  &=&
  \hat{h}_\mathrm{KS}\varphi_\alpha^{(\mathbf{R}_I)}
  \;\Rightarrow\;
  E_\mathrm{BO}(\varphi_\alpha^{(\mathbf{R}_I)},\mathbf{R}_I,\mathbf{P}_I)
  \;,\quad
\label{eq:BOelect}
\\[3pt]
  \partial_t\mathbf{R}_I
  &=&
  \nabla_{\mathbf{P}_I}E_\mathrm{BO}
  \;,\;
  \partial_t\mathbf{P}_I
  =
  -\nabla_{\mathbf{R}_I}E_\mathrm{BO}
  \;.
\label{eq:BOion}
\end{eqnarray}
\end{subequations}
It is assumed that the electronic wavefunctions are always relaxed into the (electronic) ground state for the given ionic configuration and its energy expectation value produces a Born-Oppenheimer energy $E_\mathrm{BO}$ which depends effectively only on ionic variables, see Eq.~(\ref{eq:BOelect}). That ionic energy $E_\mathrm{BO}$ is then used in a standard ionic MD, see Eq. (\ref{eq:BOion}). The method allows one to use larger time steps because only the slow ionic motion is to be propagated. On the other hand, full electronic relaxation takes many static steps. It depends very much on the particular application whether BO-MD is advantageous or not.

The stationary limit of TDLDA (electronic part) is obvious - it is given by Eq.~(\ref{eq:BOelect}). The situation is more involved at the side of the ions. A stationary point is defined by $\partial_t\mathbf{P}_I=0$ and may be reached by simply following the steepest gradient of the potential field. However, the ionic energy landscape is swamped by competing local minima. A straightforward gradient path will end up in some minimum, but not easily in the lowest one, i.e., the ground state. One needs to employ stochastic methods, such as simulated annealing and Monte-Carlo sampling, to explore the high-dimensional landscape of the ionic energy surface, for details see~\cite{Pre92}.

The most time consuming part in TDLDA-MD, i.e., Eqs.~(\ref{eq:TDLDA-MD}), is the electron propagation. There are basically two different approaches: Basis expansion or coordinate-space grid representation, see Tab.~\ref{tab:theories} and references therein. Basis expansions are more efficient in handling different length scales, as typical for covalent systems. Coordinate-space grids, on the other hand, are more adapted for the treatment of highly excited systems where electron emission plays a crucial role. In the latter, absorbing boundary conditions can easily be implemented to avoid unphysical backscattering for the analysis of photoelectron spectra and angular distributions, see, e.g.,~\cite{Cal00,Poh04b}. A very efficient means to find the electronic ground state is the accelerated gradient iteration~\cite{Blu92}. Time stepping is usually based on a Taylor expansion of the time evolution operator. An efficient alternative is the time-splitting method which proceeds by interlaced kinetic and potential evolution~\cite{Fei82,Cal00}. The ionic MD usually employs the Velocity-Verlet-algorithm, see, e.g.,~\cite{Pre92}. Ground state configurations are best searched for by stochastic methods as mentioned above.

\subsubsection{Semiclassical approaches}
\label{sec:3B3}
As particle number and excitation energy grow, an orbital-based treatment of the electronic degrees of freedom becomes practically unfeasible and further approximations have to be made. Less demanding are semiclassical time-dependent density-functional methods, which describe the evolution of the one-body electron phase-space distribution or the electron density and average local currents. The price for such simplification is the loss of the quantized electronic level structure, interference effects, and single electron-hole excitations. However, as these contributions become less important for larger systems with sufficiently narrow energy levels and high excitations, semiclassical methods provide a powerful tool to explore strongly nonlinear laser-cluster dynamics.

A semiclassical equation of motion for the one-particle electron phase-space density  $f({\bf r},{\bf p})$ as an approximation to quantal mean-field dynamics can be found from the well-known $\hbar\rightarrow 0$ expansion, see, e.g.,~\cite{BerPR88,DomAP97a,PlaPRA00,FenEPJD04,FenLNP08}. This, to lowest order, yields the Vlasov equation \begin{eqnarray} \frac{\partial}{\partial t}f+ \frac{{\bf p}}{m} \cdot \nabla_{\bf r}f -\nabla_{\bf p}f \cdot \nabla_{\bf r}V_{\rm eff}({\bf r},t)=0,
\label{eq:3B3_vlasov}
\end{eqnarray}
which is widely used in plasma physics. The effective electron mean-field interaction potential $V_{\rm eff}({\bf r},t)$ in Eq.~(\ref{eq:3B3_vlasov}) follows from the variation of the potential energy \mbox{$E_{\rm pot} = E_{\rm Coul} + E_{\rm xc} + E^{\rm coupl} + E^{\rm el}_{\rm ext}$}, cf. Tab.~\ref{tab:basicE}, with respect to the local electron density $\rho(\mathbf r,t)$, i.e., by \mbox{$V_{\rm eff}= \delta E_{\rm pot}/\delta \rho$}. Ionic motion may be described in the same way as for TDLDA-MD, see Eqs.~(\ref{eq:TDLDA-MD}). Quantum effects, such as exchange and correlation in LDA, are now solely contained in the effective potential and the initial conditions for the distribution function. The latter can be determined from the self-consistent Thomas-Fermi ground state~\cite{Tho27,Fer28} according to $ \displaystyle f^0({\bf r},{\bf p})=2/(2\pi\hbar)^3\Theta(p_{_{\rm F}}({\bf r})-p), $ where $\Theta$ the Heaviside function, $p_{_{\rm F}}({\bf r})=\sqrt{2m[\mu-V_{\rm eff}({\bf r})]}$ is the local Fermi momentum, and $\mu$ the chemical potential. The Thomas-Fermi-Vlasov dynamics resulting from the propagation of the initial distribution $f^0({\bf r},{\bf p})$ according to Eq.~(\ref{eq:3B3_vlasov}) constitutes the semiclassical counterpart of TDLDA.

A generic limitation of mean-field approaches, such as TDLDA and Vlasov, is the negligence of electron-electron collisions. This deficiency may become significant for strong departure from the ground state because of considerably weakened Pauli-blocking. In the semiclassical formulation, binary collisions can be incorporated by a Markovian collision integral of the Uehling-Uhlenbeck (UU) type~\cite{UehPR33}, see~\cite{BerPR88,Cal00,KoePRA08}. This results in the Vlasov-Uehling-Uhlenbeck (VUU) equation
\begin{eqnarray}
\label{eq:vuu}
&\displaystyle
\frac{\partial}{\partial t}f+ \frac{{\bf p}}{m}\cdot\nabla_{\bf r}f
-\nabla_{\bf p}f\cdot\nabla_{\bf r}V_{\rm eff}({\bf r},t)=I_{\rm
UU},\\
&\displaystyle
\mbox{with}\quad I_{\rm UU}({\bf r},{\bf p}) = \int {\rm d}\Omega \:
{\rm d}^3{\bf p}_1 \frac{|{\bf p}-{\bf p}_1|}{m}
\frac{{\rm d}\sigma(\theta,|{\bf p}-{\bf
p}_1|)}{{\rm d}\Omega}&  \nonumber \\ &\times \left[ f_{\bf p'}f_{{\bf
p}_1'}(1-\tilde f_{{\bf p}})(1-\tilde f_{{\bf p}_1}) - f_{\bf
p}f_{{\bf p}_1}(1-\tilde f_{{\bf p}'})(1-\tilde f_{{\bf p}_1'}) \right
].&\nonumber
\end{eqnarray}
The collision term embodies a local gain-loss balance for elastic electron-electron scattering $({\bf p},{\bf p_1})\leftrightarrow({\bf p}',{\bf p_1}')$ determined by the differential cross-section ${\rm d}\sigma(\theta,|{\bf p}_{\rm rel}|)/{\rm d}\Omega$, the local phase-space densities \mbox{$f_{\bf p}=f({\bf r},{\bf p})$}, and the Pauli blocking factors in parenthesis as functions of the relative phase-space occupation for paired spins \mbox{$\tilde f_{\bf p}=(2\pi\hbar)^3f_{\bf p}/2$}. The velocity-dependent scattering cross-section can be calculated for a screened electron-electron potential using standard quantum scattering theory~\cite{DomAP00,KoePRA08}. Since the collision term in the VUU description vanishes in the ground state because of the blocking factors, the Vlasov dynamics is recovered asymptotically in the limit of weak perturbation. Commonly, the Vlasov as well as the VUU equation are solved by the test particle method only for valence electrons, while core electrons are described by ion pseudopotentials, see, e.g.,~\cite{GigAP02,FenEPJD04,KoePRA08}.

Further simplifications can be deduced from hydrodynamic considerations~\cite{BloZP33,BalRMP73}, i.e., by assuming local equilibrium and a slowly varying irrotational velocity field. In this case, the electronic dynamics can be solely described by the time-dependent electron density $\rho({\bf r},t)$ and a velocity field  ${\bf v}({\bf r},t)$. The corresponding equations of motion follow from a variational principle~\cite{DomPRL98}, leading to a standard hydrodynamic problem for an inviscid fluid
\begin{subequations}
\begin{eqnarray}
&\frac{\partial }{\partial t}\rho=&-\nabla \cdot (\rho{\bf v})
\label{eq:2B0_TDTF_continuity},\\
&\frac{\partial}{\partial t}{\bf v}=&-{\bf v}\nabla \cdot {\bf
v}-\frac{1}{m}\nabla\left(V_{\rm kin}[\rho] + V_{\rm
eff}[\rho] \right),\qquad\label{eq:2B0_TDTF_convection}
\end{eqnarray}
\end{subequations}
where $V_{\rm kin}$ and $V_{\rm eff}$ are the potentials of the internal kinetic energy characterizing the local equilibrium and the interaction energy. The continuity equation Eq.\,(\ref{eq:2B0_TDTF_continuity}) and the Euler equation Eq.(\ref{eq:2B0_TDTF_convection}) describe the conservation of mass and momentum explicitly, while the equation of state is implicit in the self-consistent potentials. Analogous to $V_{\rm eff}$, $V_{\rm kin}$ results from variation of the now density-dependent internal kinetic energy. Within the time-dependent Thomas-Fermi (TDFT) approach, the internal kinetic energy is described in Thomas-Fermi approximation by \mbox{$V_{\rm kin}^{\rm TF}({\bf r},t)=\frac{\hbar^2}{2m}(3\pi^2\rho({\bf r},t))^{2/3}$}. TDTF represents the most simple semiclassical time-dependent density-functional approach. The reduction to the propagation of four scalar fields tremendously simplifies the numerical treatment, which is particulary appealing for the study of large systems. For an application to metal clusters see, e.g.,~\cite{DomPRL98}. However, as deformations of the local Fermi sphere are neglected (local equilibrium), TDTF is not capable to describe thermal excitations or highly nonlinear dynamics.

\subsubsection{Classical molecular dynamics}
\label{sec:3B4}
A basic limitation of DFT treatments, quantum or semiclassical, lies in the fact that they are of mean-field nature and thus neglect the effect of fluctuations, even if thermalization due to electron-electron collisions can be accounted for approximately in the semiclassical case.  While mean-field treatments provide a fully acceptable approach for moderately perturbed systems, they cannot account for the large microfield fluctuations arising from strong-field laser excitation. Exploring these fluctuations on a microscopic basis requires the construction of a statistical ensemble of possible trajectories, which exceeds standard mean-field capabilities. However, even if the approximate description of strong-field induced cluster dynamics with the instantaneous ensemble average provided by mean-field DFT methods may be sufficient, technical difficulties hamper their application to realistic systems in this case. The problem arises if energetic quasifree electrons and strongly bound electrons become involved at the same time, which is the typical situation in cluster ionization dynamics in strong fields where highly charged ions are produced. Hence, very different sets of scale in terms of distances and energies need to be resolved numerically, which quickly becomes prohibitive.

Presently, the single practical solution to microscopically resolve ionization dynamics leading to high atomic charge states are classical MD techniques. Numerous groups have developed corresponding methods over the years where quasifree electrons and ions are described purely classically way~\cite{Ros97,Dit98,LasPRA99,Las00,IshPRA00,Tom02,Sie02,SaaPRL03,SiePRL04,Jur04,Bau04a,JunJPB05,Bel06a,Bel06b,FenPRL07b}.

Once innerionized, electrons are explicitly followed according to classical equations of motion under the influence of the laser field and their mutual Coulomb interaction. A striking advantage of the classical treatment is the account of the classical microfield and  many-particle correlations. Nevertheless, there are some difficulties to be circumvented. First, the Coulomb interaction has to be regularized in order to restore the stability of the classical Coulomb system and to avoid classical electron-ion recombination below the atomic energy levels. This is usually done by smoothing the Coulomb interaction, e.g., by inserting a cutoff~\cite{Dit98} or by attributing an effective width to the particle~\cite{FenPRL07b,Bel06a}. The second problem concerns the computational costs. Standard MD simulations scale with the square of the particles number due to the direct treatment of the two-body interactions. For clusters beyond a few thousands of atoms this may easily become prohibitive and more elaborate algorithms such as hierarchical tree codes or electrostatic particle-in-cell (PIC) methods can be used~\cite{Pfa96,Bar86}. Such methods indeed allow the treatment of large clusters on sufficiently long times~\cite{SaaPRL03,JunJPB05,SaaEPJD05,SaaJPB06,Kri06,PetPP08,KunPRL06}. Another option for describing large clusters, even at very high laser intensity including relativistic effects, are electromagnetic PIC codes, see, e.g.,~\cite{JunPRL04,FukPRA06}.

Inner ionization can be treated in various nonexplicit ways. Since deeply bound electrons are associated to large energies and short time scales (typically in the attosecond domain), they are not propagated explicitly in most cases. An exception can be found in~\cite{Bel06a,Bel06b}. In general, however, statistical approaches relying on probabilistic estimates of inner ionization are used. Common strategies for describing atomic field ionization are the consideration of barrier-suppression ionization or the application of tunnel ionization rates, see Sec.~\ref{sec:2B0}. Collisional ionization may be modeled with the semiempirical Lotz cross sections~\cite{LotZP67}. However, this implies that ionization rates, which may be altered by many-particle effects in the systems, become a crucial ingredient of the dynamics.

\subsection{Rate equations and the nanoplasma model}
\label{sec:3C0}
The last step in the hierarchy of approaches from the most microscopic to the most macroscopic ones are the rate equations models, which describe the system in terms of a limited set of averaged global variables. Their time evolution is obtained from a few equations accounting for the major couplings, i.e., the interactions with the laser field and the internal electronic and ionic processes. Such description is based on a continuum picture and thus requires the clusters to be sufficiently large.

The original formulation of a corresponding model for strong-field cluster dynamics was done by~\cite{DitPRA96} and is known as the \emph{nanoplasma model}. This name reflects the assumption that rapid inner ionization of clusters exposed to intense laser fields creates a strongly charged but quasi-homogeneous plasma. The typical cluster size domain for which such picture applies is the nanometer range, whence the denomination. The assumption of a homogeneous plasma requires clusters of sizes larger than the Debye length \mbox{$\lambda_D = \sqrt{\epsilon_0\,k_B T/(e^2 \rho)}$} of the system. Typical density \mbox{$\rho \sim 10^{23}$ cm$^{-3}$} and temperature \mbox{$T \sim 1$\,keV} lead to \mbox{$\lambda_D \sim 5\ \AA$}.

The basic dynamical degrees of freedom in the nanoplasma model are: $N_j$ the number of ions in charge state $j$, $N_e$ the number of ``free" (innerionized) electrons, $E_{\rm int}$ the internal energy of the electron cloud, and $R$  the radius of the cluster. The global character of these variables implies that ions, electrons, and energy are distributed homogeneously in a sphere of radius $R$. The evolution of ion numbers $N_j$ follows the rate equation
\begin{equation}
\label{eq:rate_eq_Nj_1}
  \frac{d N_j}{dt}
  =
  W_j^{\rm tot}N_{j-1} - W_{j+1}^{\rm tot}N_j,
\end{equation}
where $W^{\rm tot}_j$ is the ionization rate for ions in charge state $N_j$ accounting  for tunneling and impact ionization. While tunnel ionization dominates early stages of the evolution, collisional ionization takes the lead at later times. The electron number $N_e$ evolves according to
\begin{equation}
\label{eq:rate_eq_Ne_1}
  \frac{d N_e}{dt}
  =
  \sum_j j \frac{d N_j}{dt} - \frac{d Q}{dt},
\end{equation}
where $Q$ is the total net charge of the cluster whose change is determined by the integrated net flow through the cluster surface. The evolution of the cluster radius $R$ is determined by the total pressure
\begin{equation}
\label{eq:rate_eq_R_1}
  \frac{\partial^2 R}{\partial t^2}
  =
  \frac{p_{\rm \, C} + p_{\rm \, H}}{n_i m_i} \frac{5}{R},
\end{equation}
which is composed of Coulomb pressure $p_{\rm \, C}$ due to net charge and thermal pressure  $p_{\rm \,H}$ of  the hot electron gas (treated as an ideal gas of temperature $T_{\rm e}$ and internal energy $E_{\rm int}=3/2 N_e k T_e$). Here, $n_i$ and $m_i$ denote the number density and the mass of the ions.

The internal energy $E_{\rm \, int}$ of the electron cloud follows
\begin{subequations}
\begin{eqnarray}
\label{eq:rate_eq_E_1}
  \frac{d E_{\rm int}}{d t}
  &=&
  P_{\rm \,abs}
  - \frac{2E_{\rm int}}{R} \frac{\partial R}{\partial t}
    - \sum_j I_p^{(j)} \frac{\partial N_j}{\partial t} - P_{\rm \, loss},\qquad\\
  P_{\rm \, abs}
  &=&
  - \frac{V\epsilon_0}{2} \,{\rm {\bf\cal E}_{\rm int}^2\,Im}[\epsilon(\omega_{\rm las})],
\label{eq:Eelmagn}
 \end{eqnarray}
\end{subequations}
due to absorption of electromagnetic energy ($P_{\rm \, abs}$), to cooling through global expansion ($\partial R/\partial t$ term), to ionization processes ($\partial N_j/\partial t $ term), and to energy loss by electron flow through the cluster surface ($P_{\rm \, loss}$). Here $I_p^{(j)}$ are ionization potentials of ions with charge state $j$. The cycle-averaged heating rate ${P}_{\rm \, abs}$ involves the volume $V$ and the internal electric field amplitude in a dielectric sphere \mbox{${\bf\cal E}_{\rm int} = 3 {\bf\cal E}_{\rm 0}f(t)/|2 + \epsilon(\omega)|$}, where ${\bf \cal E}_{\rm 0}f(t)$ is the vacuum laser field envelope. The dielectric constant $\epsilon(\omega)$ is usually taken from the Drude model \mbox{$\epsilon(\omega) = 1 - \omega_p^2/\omega(\omega + i \nu)$} with  \mbox{$\omega_p^2 = {n_e}e^2/(m_e\,\epsilon_0)$} the plasma (or volume plasmon) frequency and the collision frequency $\nu$ for electron-ion scattering. With these assumptions, the cycle-averaged heating rate is equivalent to the expression given in Eq.~(\ref{eq:Resonance_heating}) in Sec.~\ref{sec:2C0} and exhibits a resonance when the electronic density fulfills $n_e=3\,n_e^{\rm crit}$, where \mbox{$n_e^{\rm crit} = m_e \epsilon_0\omega^2/ e^2$} is termed critical density. This condition reflects the Mie plasmon resonance of a neutral spherical particle~\cite{Krei95}, see the discussion in Secs.~\ref{sec:2A0} and \ref{sec:2C0}.

The Eqs.~(\ref{eq:rate_eq_Nj_1}) through (\ref{eq:Eelmagn}) constitute the dynamics of the nanoplasma model. In spite of its simplicity, the model contains the basic competing processes in the dynamics of the irradiated cluster in a nanoplasma state. It is technically simple, but requires several empirical ingredients, such as, e.g., the various ionization rates. It also involves strong simplifications such as a thermal electron distribution, an intensity independent heating rate and very crude treatment of space-charge effects and electron emission. Nevertheless, it was applied to many experimental results with some successes and its original formulation was extended in several respects.

The original model may be questioned at various places, e.g., regarding to the assumption of homogeneous distributions of all species in the cluster. This constraint was relaxed in~\cite{MilPRE01} by considering a radius-dependent distribution. Further, the damping effect of the cluster surface is neglected. It can be introduced by using a modified collision frequency $\nu = \nu_{ei} + A {v}/{R}$, which then contains electron-ion collisions through ($\nu_{ei}$) and an additional term for surface-induced Landau damping $A {v}/{R}$ ($v$ - average electron velocity). The surface contribution has been shown to play an important role for energy absorption~\cite{MegJPB03}. More recently, detailed cross-sections were computed to include high-order ionization transitions involving intermediate excited states for describing the x-ray emission from Ar clusters~\cite{Mic07}. Another important contribution is the lowering of ionization thresholds in the cluster due to plasma screening effects~\cite{GetJPB06}, which was shown to significantly alter the ion charge distribution as well as the heating dynamics~\cite{HilLP09}.

One should further remind that the nanoplasma model, as a statistical continuum picture, may only describe the gross features of the interaction of intense lasers with clusters. In particular, it cannot access experimental results beyond average values. The model may thus fail in describing the profiles or far tails of, e.g., ion charge state or energy distributions. More detailed insight can, for example, be gained from MD simulations. Nonetheless, even in its crudest version the nanoplasma model may serve as an acceptable starting point to get first insights into the time evolution of charging or the explosion dynamics for large (nanometer) clusters.

\section{Experimental methods}
\label{sec:400}
With the modern molecular beam machines, the variety of radiation sources from the infrared to the x-ray regime, and the multiply parallel detection and data processing possibilities, challenging and highly sophisticated experiments on clusters can be performed. It is possible to prepare targets with narrow size distribution or even completely size-selected, partially at low or ultralow temperature. Vast literature exists on cluster production, e.g.,~\cite{EchIsspic5,Hab94ab,Mil99,Pau00,HeJCP01}. Optical single or many-electron excitation, in some cases also being followed by a probing ultrashort light pulse, has led to far-reaching insight into fundamental processes of the light-matter interaction in clusters. In this chapter, rather than covering the vast multitude of experimental methods, we review selected current techniques used for probing dynamics on free clusters.

\subsection{Generation of cluster beams}
\label{sec:4A0}
\emph{Rare-gas} or \emph{molecular clusters} are produced from an adiabatic expansion through a continuously working or pulsed nozzle with nozzle diameters ranging from a few to $500\,\mu$m, usually restricted by the pumping speed of the apparatus. Mixed clusters are generated by a co-expansion of a gas mixture or by using a pick-up technique with a cross-jet. The cluster size may be varied by changing the nozzle temperature or the stagnation pressure. Typically, the width $\Delta N$ (FWHM) of the size distribution roughly equates the average number $\langle N\rangle$ of atoms per cluster. According to semiempirical scaling laws~\cite{Hag74,HagZPD87,HagSS81}  derived from general considerations about condensation kinetics, $\langle N\rangle$ scales with the "condensation parameter"
\begin{equation}
{\Gamma}^{\star} = k\frac{\left(p_0/{\rm mbar}\,d/{\rm \mu m}\right)^{0.85}}{{(T_{0}/{\rm
K})}^{2.2875}}, \label{eq:Hagena}
\end{equation}
where $p_0$ is the stagnation pressure, $T_0$ is the nozzle temperature, $d$ is the effective nozzle diameter. The gas constants $k$ (in units of $[{\rm K}^{2.2875}{\rm mbar}^{-1}]$) can be calculated from the molar enthalpy at zero temperature and the density of the solid according to~\cite{HagZPD87}, ranging from \mbox{$k=185$} for Ne, over \mbox{$k=1646$} for Ar and \mbox{$k=2980$} for Kr, to \mbox{$k=5554$} for Xe. Equation (\ref{eq:Hagena}) holds for monoatomic gases; otherwise the exponents of $d$ and $T_{0}$ are different. For conical nozzles, $d$ has to be replaced by an equivalent diameter that depends on the half opening cone angle. The scaling laws developed for rare gases have been modified afterwards for metal vapors.

\begin{figure}[t]
\centering\resizebox{0.9\columnwidth}{!}{\includegraphics{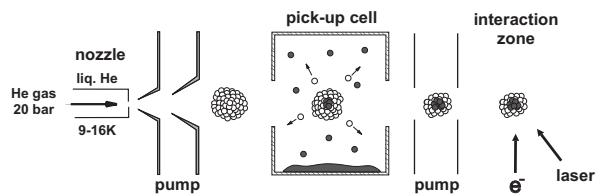}}
\caption{Schematics of a He droplet pick-up cluster beam machine. Atoms from the vapor in the pick-up cell can be loaded
into the droplets at 0.4\,K. After~\cite{DiePRA05}}
\label{fig:helium_machine}
\end{figure}

For experiments at ultra-low temperatures, \emph{helium droplet pick-up sources} prove to be very versatile~\cite{BarPRL96,GoyPRL92,TigPCC07}. A sketch of a typical setup is shown in Fig.~\ref{fig:helium_machine}. He droplets are produced by the supersonic expansion of precooled helium gas with a stagnation pressure of 20\,bar through a 5\,\mbox{$\mu$}m diameter nozzle. By choosing the temperature at the orifice (9-16\,K), the log-normal droplet size distributions can be adjusted in the range of $\langle N\rangle$ = 10$^{3}$-10$^{7}$ atoms. After passing differential pumping stages the beam enters the pickup chamber containing a gas target or a heated oven, where atoms are collected and aggregate to clusters inside the He droplets. With this setup it is possible to record clusters with up to 150 silver atoms~\cite{RadPRL04} or 2500 magnesium atoms~\cite{DiePRA05}, respectively. Downstream another differential pumping stage, laser light or an electron beam ionizes the doped droplets. The benefits of pick-up sources rely on the feasibility to embed clusters into a well-controlled environment. In the case of He, the embedding medium is superfluid, weakly interacting, and ultracold with a temperature of about 0.4\,K~\cite{HarPRL95}, being an ideal nanomatrix for spectroscopic studies~\cite{HeJCP01}. Similarly, droplets or particles of other elements might serve as pick-up medium, e.g., Ar, Kr, or Xe. Subsequent atom agglomeration also can lead to the formation of electronically excited species~\cite{IevCPL00}. While in particular in the case of helium the nanomatrix is mostly transparent under low laser intensity conditions, it may become an active part in the interaction process under strong laser fields that substantially alters the clusters dynamics. Subsequent to plasma formation in the embedded cluster, the nanodroplet may be ionized as well, giving rise to a core-shell type nanoplasma.

In these days \emph{pure metal clusters} are mainly produced with laser vaporization or plasma-based methods. In both cases the material is vaporized, being partially ionized, and then undergoes cooling and expansion in a rare gas. This can be pulsed, allowing for a hard expansion of the seeded clusters into vacuum, or continuously streaming at lower pressure. In a \emph{laser vaporization cluster source} a rotating target rod or plate of the desired material is mounted close to a piezo or magnetically driven pulsed gas valve. Usually He pulses with an admixture of Ne or Ar at backing pressures of 2-20\,bar serve as seeding gas. Intense ns laser light pulses with about 50 to 100\,mJ/pulse erode target material by producing a plasma plume, which is flushed by the seeding gas through an about 1\,mm diameter channel and a nozzle into high vacuum. The close contact with the cold gas leads to supersaturation and efficient aggregation already in the source channel. The nozzle - often elongated by an extender - can be cone-shaped or merely be a cylinder. In some cases an additional small mixing chamber between source body and extender might increase the intensity within a desired mass range. Depending on material and operation conditions, different types of nozzles are in use, partially with very long extenders of 10\,cm or more. There is no optimal photon energy, but the intensity must be sufficient to induce vaporization or create a plasma. However, frequency doubled Nd:YAG laser are often used, as its green color facilitates the beam adjustment. With laser vaporization sources practically all solid materials can be vaporized. As a significant fraction ($\sim10\%$) of the clusters is charged, no additional ionization is necessary for studies on mass selected species.

\begin{figure}[t]
\centering\resizebox{0.85\columnwidth}{!}{\includegraphics{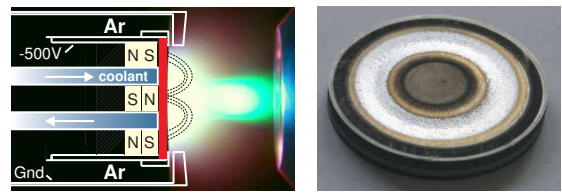}}
\caption{Plasma plume of an uncovered Haberland-type magnetron sputtering cluster source during operation. The ion and electron motion is guided by permanent magnets behind the target. Right: picture of a used silver target.} \label{fig:magnetron}
\end{figure}

Several types of plasma-based sources are commonly used, the most prominent being the \emph{magnetron sputtering cluster source}, going back to developments in the group of Haberland~\cite{HabJVSCA92}. The basic erosion process is high pressure (1\,mbar) magnetron sputtering. This versatile tool operates with a few cm in diameter plane solid target mounted close to an axial permanent magnet, see Fig.~\ref{fig:magnetron}. In the presence of the seeding gas, a high voltage between a ring-shaped electrode and the target initiates and drives a discharge, efficiently eroding the material and producing a circular well after several hours of operation. The mainly charged vapor is cooled by the seeding gas and transported through a nozzle. Conducting materials can be sputtered by this source, whereas ferromagnets may cause difficulties.

In contrast to the magnetron sputtering source which operates with a high voltage discharge, \emph{arc cluster ion sources} make use of high current arcs. Such are known as vacuum arcs, self-stabilizing at about 40\,Volts and 40\,Amps. The discharge can be sustained in vacuum once a spark has initially brought some metal into the vapor phase. It is important that the discharge is carried by the metal vapor rather than by the seeding gas. In order to accomplish this, the temporal development of the high voltage-driven sparc needs special care. Once the metallic component in the source rules the conductivity, the discharge voltage switches to a low level so that the seeding gas will not directly be ionized. Two variants of the arc sources are in use, pulsed ones and continuously working ones. The concept of the pulsed arc cluster ion source PACIS~\cite{SieZPD91,ChaRSI92} is very similar to the laser vaporization cluster source, only that the laser is replaced by a pulsed high-current arc between two electrode rods at about 1\,mm separation. An offspring of the PACIS uses one rotating electrode, then being called ''Pulsed Microplasma Source''~\cite{BarJPD99}. When operated continuously we obtain the Arc Cluster Ion Source ACIS~\cite{MetEPJD01,KleJAP07}. Here the target is a water-cooled hollow cathode, a water-cooled counter electrode serves as anode. Magnet coils around the hollow cathode help to control the arc. Again, the plasma is flushed by an inert seeding gas into vacuum, producing a cluster beam with a high amount of charged species (about 80\%, depending on the material). The beams from the ACIS can be focussed by aerodynamical lens systems. These are sets of orifices and/or confining tubes connected to the nozzle. By choosing appropriate dimensions the on-axis intensities increase, which goes along with a narrowing of the particle size distribution~\cite{PasRSI06}. This type of source can generate large metal particles from 2 to 15\,nm in diameter, an interesting size range for future studies of the intense laser-cluster interactions.

All cluster sources described above are housed inside well-pumped vacuum chambers in order to reduce the gas load at the point of investigation. Ideally, only the central filament of the jet passes a narrow skimmer and enters as collimated cluster beam the photoexcitation chamber. Further differential pumping can lead to sufficiently low pressure for the spectroscopy on isolated species. However, many strong-field experiments do not make use of single cluster excitation. In particular for rare gas clusters, the laser is often focussed onto the beam in the high pressure zone close to the nozzle. In such cases many interacting clusters are simultaneously excited, thus the observed signal might originate from a dense cluster ensemble rather than from isolated systems.

\subsection{Sources for intense radiation}
\label{sec:4B0}
Within the last 20 years ultrashort-pulse lasers have undergone dramatic improvements with respect to pulse width, power, and repetition rate. This was first enabled by the technique of colliding pulse modelocking (CPM) within a ring dye laser~\cite{ForAPL81} and later by the invention of the chirped pulse amplification (CPA) scheme by Maine et al.~\cite{MaiJQE88}. Nowadays, the broadband fluorescent (690-1050\,nm) laser crystal Ti:sapphire operating at a central wavelength of 800\,nm is the working horse in delivering ultrashort and intense optical radiation. Laser pulse durations as short as some femtoseconds~\cite{BraRMP00,KelN03} or attoseconds~\cite{CorNatP07} as well as pulse powers in the Petawatt regime~\cite{LedS03} are available.  To avoid damage of the optical components, the pulses from a modelocked femtosecond laser oscillator is first stretched to some ps before amplification and then re-compressed in the final step~\cite{MaiJQE88}. For energy enhancement regenerative amplifiers or bow-tie shaped multi-pass configurations are typically used. Stretching as well as compression of the pulse is achieved by introducing diffractive elements, e.g., reflection gratings~\cite{StrOC85} in the optical path. High energy pulses in other wavelength regions can be realized, e.g., by amplification of the third harmonic in a KrF amplifier operating at 248\,nm~\cite{BouJOSAB93}. Due to the limited bandwidth of the transition the pulse duration in this type of laser is limited to some hundreds of femtoseconds. With high harmonics (HH) generated by focussing intense pulses into atomic gases the short wavelength regime becomes accessible opening up the route towards attosecond pulses~\cite{PapPRL99}. Pulse intensities as high as $1.3\times10^{13}$\,W/\,cm$^2$ have been reported for the 27th harmonic~\cite{NabPRL05}. Only recently the vacuum ultraviolet free electron laser (VUV-FEL) FLASH at DESY has been setup, currently delivering pulses with wavelengths down to 6.5\,nm at peak energies up to 100\,$\mu$J~\cite{AyvEPJD06}.

In the optical domain single-shot autocorrelators or more sophisticated setups~\cite{Tre02} are applied for pulse characterization. In many experiments only the pulse width is varied by detuning the compressor length. This introduces a linear chirp (Sec.~\ref{sec:2B0}) and allows continuous variation of the pulse duration between sub-100\,fs to many ps. To generate dual-pulses with variable optical delay (pump-probe) the initial pulse may be split into two replica, e.g., by a Mach-Zehnder setup. Moreover, liquid crystal spatial light modulators, acousto optical modulators, and deformable mirrors allow one to modify the pulse structure at will~\cite{WeiRSI00}. Besides amplitude and phase, also the polarization can be altered, e.g., to drive reactions selectively into a desired channel in coherent control experiments~\cite{BruSA95,TanJCP86}. This scheme connected to a feedback algorithm~\cite{JudPRL92} is capable of optimizing the laser-matter coupling, see e.g.~\cite{AssS98} and Sec.~\ref{sec:7A0}.

For pulse focussing, lenses or parabolic mirrors can be used. The latter avoids pulse modification due to the propagation through optical elements, i.e., pulse broadening, self-focussing, or phase modulation. The waist radius of a Gaussian beam at the focus is $w_0=2\lambda f/\pi$, where the $f$-number relates the size of the unfocussed beam diameter $D$ to the focal length of the lens $d_{\rm f}$ by $f=D/d_{\rm f}$, and $\lambda$ is the wavelength. Typical spot sizes are a few tens of $\mu$m. For a qualitative description of nonlinear laser-matter interactions the intensity profile in the focal region has to be taken into account. For a given peak intensity $I_0$, the intensity profile $I(r,z)$ is given by~\cite{Mil88aB}
\begin{equation}
        I(r,z)=\frac{I_0}{1+z^2/z_0^2}\exp\left[-\frac{2r^2}{w_0^2(1+z^2/z_0^2)}\right],
\label{eq:gauss-focus}
\end{equation}
where $r$ and $z$ are the axial and transverse distances to the focus and \mbox{$z_0=\pi\,w_0^2/\lambda$} specifies the Rayleigh length, where the beam radius has increased to $\sqrt{2}w_0$. The focal intensity profile leads to volumetric weighting, which has been used to determine intensity thresholds in the strong field ionization of atoms~\cite{HanPRA96,GooJPB05,BryNP06} and molecules~\cite{BenPRA04}. Applied to clusters, this intensity-selective scanning method has revealed a dramatic lowering of the threshold intensities for producing highly charged ions when compared to atoms~\cite{DoeEPJD07,DoePRL09}.

\subsection{Particle detection techniques}
\label{sec:4C0}
Optical excitation of clusters can lead to extensive fragmentation. Usually fragment mass spectra are analyzed in terms of stabilities, similar to nuclear fission processes~\cite{SchBook92}. In strong fields, however, dedicated techniques are needed to resolve the emission spectra of ions and electrons in detail.

\subsubsection{Determination of charge state distributions}
\begin{figure}[b]
\centering\resizebox{0.7\columnwidth}{!}{\includegraphics{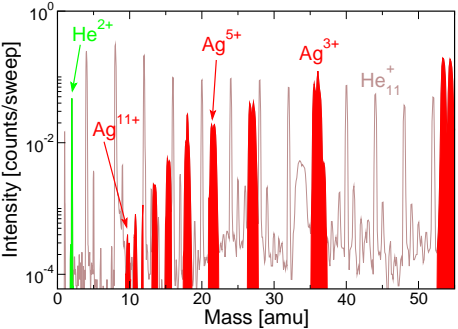}}
\caption{Charge state spectrum from a time-of-flight analysis of Ag$_N$ in He
droplets with $\langle N \rangle$\,=\,40, exposed to 400\,fs laser pulses at
$4\times10^{13}{\rm W/cm^2}$ and 800\,nm. The resulting Ag$^{q+}$
signals from the Coulomb explosion are highlighted. Ions with up to $q=11$ are been detected. The occurrence of He$^{2+}$ stems from charge transfer with the Ag ions at the chosen laser intensity. From~\cite{DoePRL05}. \label{fig:mass_spectrum}}
\end{figure}

The most straightforward method to determine charge state distributions of clusters and their fragments is ion mass spectrometry. Irrespective of the particular method, the mass separation will always be connected to the charge-to-mass ratio.
In particular time-of-flight (TOF) methods with accelerating electrical fields are widely used for analyzing charged products after photoionization. Fig.~\ref{fig:mass_spectrum} shows an example of highly charged atomic ions emerging from silver clusters embedded in He droplets after irradiation with intense fs laser light.
The TOF spectrum exhibits contributions of He and Ag clusters with high masses (not shown here). At short flight times a situation appears like in Fig.~\ref{fig:mass_spectrum}. Whereas the background peaks are signatures of the He droplet fragments, the highlighted series can uniquely be assigned to atomic ions in high charge states from the Coulomb explosion of Ag$_N$. As a matter of fact, the Ag ions carry high recoil energies due to the violent expansion. Therefore TOF methods that use an acceleration of the ionic ensemble by electric fields in the few kV range loose part of their resolution and transmission. Consequently, the TOF spectra only prove the occurrence of the ions but do usually not image the real charge state distribution.

\subsubsection{Acquisition of ion recoil energy spectra}
A simple and versatile tool to investigate ion recoil energies is the acceleration-free TOF spectroscopy. Two preconditions have to be met in order to allow a unique interpretation of the results: First, there has to be a defined source point for the ion emission. Second, the nature (mass) of the ions must be known, which often is a point difficult to achieve. However, the excitation of single-element clusters with sufficiently strong laser fields leads to complete fragmentation into atomic ions with known mass. In this case, the kinetic energy is determined by TOF measurements through a field free drift tube of about 0.5\,m, without initial electric field. For reducing noise caused by secondary electrons and, moreover, to restrict the ion detection to the Rayleigh region of the laser focus, an adjustable narrow slit confines the ion trajectories. Resulting TOF spectra can then directly be converted into kinetic energy spectra, see, e.g., Fig.~\ref{fig:exper_fiveways}d.

\begin{figure}[b]
\centering\resizebox{0.75\columnwidth}{!}{\includegraphics{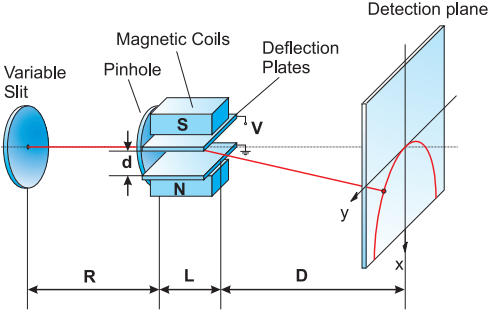}}
\caption{Sketch of the Thomson analyzer. Ions enter a region of parallel electric and magnetic fields trough a tiny hole. The resulting deflection gives characteristic parabolas from which the charge state selective recoil energy can be deduced. A multi channel plate detector with an imaging system serves to record the data. From~\cite{DoeEPJD03}, with kind permission of The European Physical Journal (EPJ).} \label{fig:thomson_spectrometer}
\end{figure}
The field-free ion TOF yields recoil energies irrespective of the ion charge states. For a detailed analysis it is necessary to resolve charge state dependent recoil energies. To this end two methods have successfully been applied, both of which simultaneously measure the ion charge state and energy. The first one uses magnetic deflection time-of-flight (MD-TOF) mass spectrometry~\cite{LezPRL98}. This technique bases on TOF measurements at different positions behind a magnetic field. With the MD-TOF, highly energetic (up to 1\,MeV), multiply charged ions could be recorded.

The second method is of static nature and bases on a principle first applied by Thomson~\cite{ThoPM07}. Fig.~\ref{fig:thomson_spectrometer} sketches the Thomson analyzer for the simultaneous measurement of energy and charge of ions expelled from an exploding cluster. It consists of parallel electric and magnetic fields, followed by a field free drift zone in connection with a position-sensitive detector. The experimentally obtained raw data reflect momentum and energy per charge and have to be transformed to energy \emph{vs.} charge  spectra. For Ag$_N$ the charge state resolved ion energy distribution is rather narrow and the maximum energy grows almost linearly with ionization stage~\cite{DoeEPJD03}.

\subsubsection{Energy and angular resolved electron detection}
\label{sec:4C3}
The experimental challenge in photoelectron spectroscopy results from the notoriously low densities in mass-selected charged cluster beams. To cope with this, time-of-flight electron spectroscopy has been developed with a magnetic field gradient. When the clusters are ionized at a certain spot within an electron {\it magnetic bottle spectrometer} the complete photoelectron spectrum can be recorded by time-of-flight measurements with up to 100\% detection efficiency~\cite{KruJPE83,TayJCP92,GanPRA88,ArnJCP91}. Whereas this method turned out to be extremely fruitful to reveal the electronic level structure of many mass selected cluster anions, the magnetic fields involved hamper the retrieval of satisfying angular information. In the case of a neutral cluster beam, the target density can be sufficiently high in order to get a spectrum even without the magnetic field. Electron emission and drift occur within a field-free tube, equipped with a time-resolving detector. By rotating the polarization direction of the laser, angular resolved photoelectron spectra are obtained. An increasing length of the drift tube increases the energy resolution on the expense of signal intensity. Acceptable results can be achieved with magnetically well-shielded tube of about 0.5\,m length.

In contrast to the electron TOF method, where kinetic energy release information is contained in the electron drift times, imaging techniques extract energy and angular distributions from spatially resolving detection. The striking advantage of this method is that the full emission characteristics can be reconstructed from the 2-D image by means of an Abel inversion. The energy resolution is limited by the quality of the 2-D detector~\cite{HecARPC95}. An improvement of the 2-D imaging technique has been obtained by introducing a lens optics which maps all particles with the same initial velocity vector onto the same point on the detector~\cite{EppRSI97}. So far, this technique has mainly been used to record low-energy electron spectra. With modified electrode configuration energetic electrons from clusters driven to Coulomb explosion are accessible as well~\cite{Skr09}.

\section{Single- and multiphoton processes in clusters}
\label{sec:500}
The previous sections have provided basic tools for the description and analysis of laser-induced cluster dynamics. In the following presentation of specific examples we begin with single-photon processes in Sec.~\ref{sec:5A0} and move on to multiphoton effects in Sec.~\ref{sec:5B0}. In both cases clear signatures of the photon energy persist. Single-photon excitations are typically investigated by photoelectron spectroscopy (PES), which is usually interpreted as a static image of the density of states and so indirectly of the underlying geometry. When carried out with angular resolution, PES reveals structural details of the electronic orbitals being excited. However, even single-photon photoemission goes beyond a mapping of system properties in a static and direct way, as it reflects a dynamical process. Pump-probe studies, as a time-resolved version of PES, give access to ultrafast structural dynamics and energy redistribution pathways. Additional reaction channels emerge with the absorption of multiple photons, as will be the subject of Sec.~\ref{sec:5B0}. Besides above-threshold ionization, as a prime example for multiphoton signatures, thermalization and its effect on electron spectra will be discussed. Another issue are plasmons, which often govern the response of metal clusters and become broadened by nonlinear contributions at higher intensity. However, they remain a dominant doorway process up to the strong-field domain, which is subject of Sec.~\ref{sec:600}.

\subsection{Single-photon electron emission}
\label{sec:5A0}

\subsubsection{Probing the density of states}
\label{sec:5A1}
\begin{figure}[t]
\centering\resizebox{\columnwidth}{!}{\includegraphics{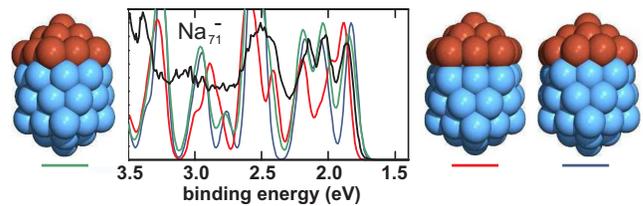}} \caption{PES spectrum of Na$_{71}^-$: experimental result (black curve) from nanosecond laser excitation with $\hbar\omega=4.02$\,eV at $T\simeq$~100~K and theoretical DOS calculated by DFT using different ground state structures (as shown). From the matching of the spectra the left structure is favored while the right ones show less agreement. After~\cite{Kos07b}} \label{fig:PES_simple} \end{figure}

For studying single-electron excitations by photoemission it is often useful to assume, motivated by Koopmann's theorem~\cite{Wei78}, that the essential structures of the electron and ionic systems do not change significantly upon electron emission. The photoelectron energy spectrum thus basically images the density of states (DOS). Based on this assumption, PES has become a powerful tool to explore the electronic structure of mass-selected clusters. Figure~\ref{fig:PES_simple} displays an example from Na$_{71}^-$. The measured data (black curve) exhibits pronounced peaks at binding energies between 1.8 and 3.5\,eV. Such \emph{electronic fingerprints} reveal details of the quantum confinement and change dramatically with cluster size or structure. With DFT calculations it has become possible to obtain theoretical DOS for comparison with experimental PES spectra. Fig.~\ref{fig:PES_simple} displays an attempt to identify the cluster ground state geometry out of theoretically suggested candidates by matching the DOS.
A vast amount of photoelectron spectra on different systems has been accumulated since first successful experiments~\cite{Leo87,Ho90,Pet88,Che90,McH89,GanPRA88,Gan96}. During the course of time, developments in cluster production and electron detection have made it possible to cover large size ranges at high energy resolution. For instance, in~\cite{Wri02} PES spectra of Na$_N^-$ for \mbox{N=31-500} show peaks that can be assigned to the electronic shell structure. For small systems a higher level of theoretical understanding can be obtained from ab-initio quantum chemical methods~\cite{Bon91}.

To date, most PES studies rely on low-energy photon excitations, i.e., \emph{valence-band} PES. Inner shell photoionization, i.e., \emph{core-level} PES, has been demonstrated as well~\cite{WertZPD89,SieZPB93,EbePRL90}. These studies, however, dealt with deposited clusters excited with high photon energy lamps or synchrotron radiation. Common results are shifts of core levels with cluster size. Due to the surface contact a thorough understanding remains difficult since core-hole screening, chemical shifts, electronic relaxation or charge transfer dynamics contribute to the spectra.

With third-generation synchrotron sources, experiments on free neutral (not mass-selected) clusters became possible. One issue of such studies is the absorption site as a probe of the local environment~\cite{PieEPJD06,Hat05}. In rare gas clusters the measured line profiles~\cite{Tch04} show well-separated features that can be attributed to the ionization of surface and volume atoms, respectively~\cite{Ama05,Ber06}. Such analyses can also provide an indirect size measurement, as has recently been shown for neutral nanometer clusters of various metals, i.e., Na~\cite{Per07a}, Pb~\cite{Per07b}, Cu, and Ag~\cite{Tch07}.

Latest progress in core-level PES has been achieved at the free electron laser FLASH which delivers intense pulses with up to 200\,eV photon energy. The energy range and high brilliance open new possibilities to interrogate both the complete valence regions as well as shallow core levels of numerous systems. For example, PES on free mass separated Pb$_N^-$ revealed a pronounced $N$-dependent shift of the 5$d$ core level~\cite{Sen09} which is in accordance with the metallic droplet picture for large $N$. However, strong deviations starting below $N\leq20$ indicate a transition from metallic to nonmetallic bonding due to less efficient core-hole screening.

A solid theoretical understanding of the photoionization process requires the complete toolbox of computational many-particle physics. One example where DFT calculations for Na$_N^-$ are compared to experimental PES was shown above in Fig~\ref{fig:PES_simple}. In the same spirit, Si$^-_N$ for $N=20-26$ have been investigated theoretically in~\cite{Gul05} and compared to data from~\cite{Hof01}. In both cases, not all peaks could be fully reproduced by theory, especially for deeply bound electronic states. Nevertheless, from comparison of the calculated DOS with the experiment the ground state geometry can be identified and discriminated against competing isomers in many cases. Remaining discrepancies reflect that static DFT calculations based on the Kohn-Sham eigenvalues are insufficient to fully describe the photoemission. It is well-known that the interpretation of eigenvalues as single-particle energies requires attention~\cite{KueRMP08,Mun06}. This concerns the meaning of single-particle eigenvalues itself as well as dynamical aspects, as Koopmann's theorem does not hold in a strict way. In other words, photoemission is a highly correlated process. The photoelectron interacts with the residual system during its removal and may substantially modify the level structure. The effect becomes important with low energy electrons and dramatic in the zero electron kinetic energy measurements.

\begin{figure}[t]
  \centering \resizebox{0.7\columnwidth}{!}{\includegraphics{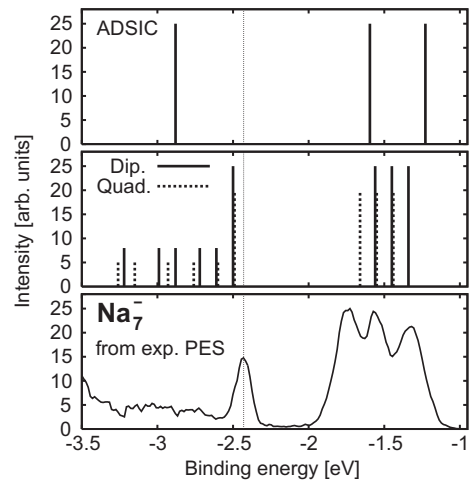}}
\caption{Comparison of measured PES spectra for Na$_7^-$ (lower panel)~\cite{Mos03} and two different theoretical predictions. The upper panel shows the single-electron levels from a (static) Kohn-Sham calculation applying ADSIC. The middle panel presents the theoretical result deduced from the excitation spectrum of neutral Na$_7$, the final product after photoemission. The excitations were computed with TDLDA~\cite{Mun07}. \label{fig:pes_na7m}}
\end{figure}
The question whether PES reflects parent or daughter cluster DOS or a dynamical mixture of both
has been tackled in the case of sodium cluster anions, see Fig.~\ref{fig:pes_na7m}. The comparison between the experimental spectrum~\cite{Mos03} and the Kohn-Sham eigenvalues of the (\emph{parent}) cluster anion calculated with average-density self-interaction correction (ADSIC) is clearly not satisfying~\cite{Leg02}. A way to circumvent the use of the Kohn-Sham eigenenergies is to perform a time-dependent DFT calculation of the response to a small pertubation. In~\cite{Mun07}, the energies of excited states of the neutralized \emph{daughter} cluster are extracted from the time evolution of the dipole and quadrupole moments and are related to the photoelectron kinetic energies by energy conservation (middle panel). While some discrepancies still remain there is a clear improvement over mere static considerations which points out the key role of final state interactions.

\subsubsection{Angular distributions}
\label{sec:5A2}
Besides pure energy spectra, which reflect the electronic level structure, photoemission may also reveal details of the involved orbitals and thermalization phenomena. To this end the emission has to be analyzed with angular resolution, a subject that still is in its early stage. The directionality of the photoelectron angular distribution (PAD) can be quantified by a Legendre expansion: \begin{equation} \label{eq:beta} \frac{\textrm d \sigma}{\textrm d \theta} = \frac{\sigma_{\rm tot}}{4 \pi} \left[ 1 + \beta_2 P_2(\cos \theta) + \beta_4 P_4(\cos \theta) + \ldots \right] \ , \end{equation} where $\theta$ denotes the emission angle with respect to the laser polarization axis. The anisotropy parameter $\beta_2$ ranges from $-1$ (emission perpendicular to polarization), over 0 (isotropic emission), to 2 (emission parallel to polarization) and depends on the orbital symmetry of the initial and final states and the electron kinetic energy.

\begin{figure}[b]
\centering \resizebox{0.7\columnwidth}{!}{\includegraphics{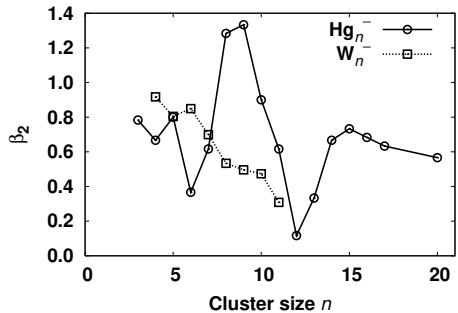}} \caption{Anisotropy
parameter $\beta_2$ extracted from photoelectron angular distributions as a function of cluster
size $N$~: W$_N^-$ (squares) exposed to 4.025 eV laser light, from~\cite{Pin99}, with kind permission from Springer Science+Business Media;
\label{fig:beta2}Hg$_N^-$ (circles) irradiated at 3.15 eV, from~\cite{Ver04}, with permission from American Institute of Physics.}
\end{figure}

Only a few PAD experiments have been performed on clusters so far. Among them are results on W$_N^-$, $N=4-11$~\cite{Pin99,Bag01}, and Hg$_N^-$, $N=3-20$~\cite{Ver04}. The corresponding $\beta_2$ as a function of $N$ are presented in Fig.~\ref{fig:beta2}.
For W$_N^-$, $\beta_2$ is changing from a more directed behavior with small clusters ($\beta_2\sim1$) to nearly isotropic emission ($\beta_2\rightarrow 0$) when $N\rightarrow 11$. From the rapidly reached isotropic behavior it was concluded that larger W$_N^-$ show an indirect emission process, where electron-electron collisions lead to a loss of coherence. This is in line with the tendency of W$_N^-$ to undergo thermionic emission~\cite{LeiEPJD91}. Figure~\ref{fig:beta2} further shows results on Hg$_N^-$ with strongly size-dependent asymmetries. Although the physical origin of these $\beta_2$-fluctuations could not be clarified yet, the data illustrate the high system-sensitivity of angular resolved photoemission.

\begin{figure}[t]
\centering \resizebox{0.95\columnwidth}{!}{\includegraphics{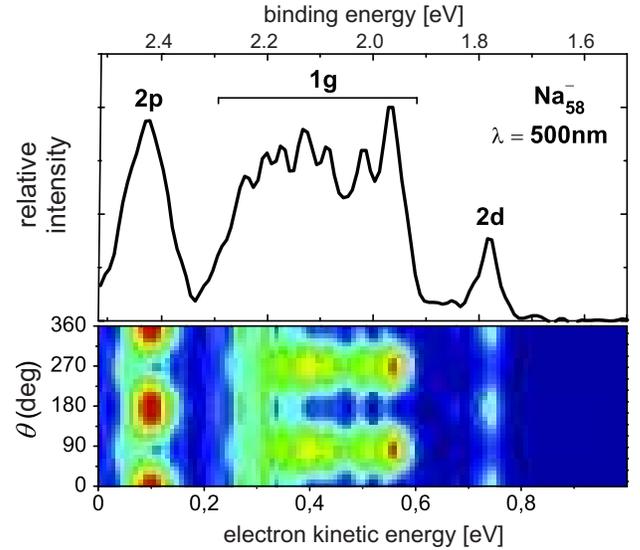}}
\caption{Photoemission from Na$_{58}^-$ obtained with \mbox{$2.48$\,eV} photon energy. The top panel displays angle-integrated spectra while the bottom panel shows the corresponding angular resolved results. The emission angle $\theta$ is defined with respect to the laser polarization. From~\cite{BarS09}, reprinted with permission from AAAS. \label{fig:na58m_pes_pad}}
\end{figure}

A clear dependence of the PAD on the electronic level being excited was demonstrated with medium-sized Na$_{N}^{-}$~\cite{BarS09}. Exemplarily, Fig.~\ref{fig:na58m_pes_pad} compares a standard PES spectrum of Na$_{58}^-$ (top) with the corresponding angular resolved result (bottom panel).

The peaks in the top panel can easily be attributed to emission from the 2$p$, 1$g$ and 2$d$ shells, see~\cite{Hee93,Bra93} for details on the shell nomenclature. For the given photon energy, the comparison with the PAD shows that the $2p$ and $2d$ electrons are emitted parallel to the laser polarization, while the $1g$ emission is aligned perpendicularly. These results demonstrate that the cluster valence electrons preserve their angular momenta after excitation. Further, similar angular distributions for electronic sublevels of a particular shell (e.g. 1$g$) show that the ionic background does not destroy the free angular momentum eigenstate character within an electronic shell. Thus, the results justify a single-particle picture of almost free delocalized electrons for describing the PAD from simple metal clusters.
\begin{figure}[htb]
\centering \resizebox{0.8\columnwidth}{!}{\includegraphics{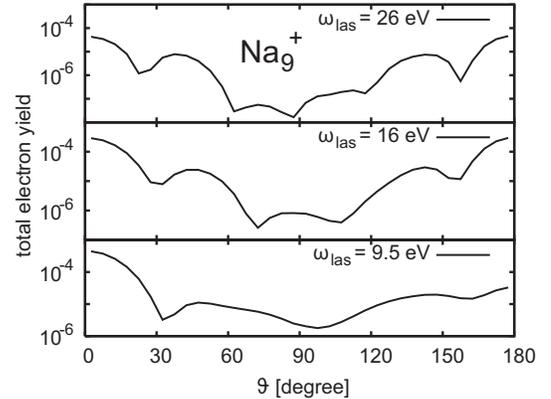}}
\caption{Photoelectron angular distribution of Na$_9^+$ irradiated with three different laser
energies as indicated, calculated in TDDFT. After~\cite{Poh04b}. \label{fig:pad_na9p}}
\end{figure}

Besides a state-sensitivity, PAD spectra are also dependent on the photon energy. A theoretical study of this effect was reported in~\cite{Poh04b} and is illustrated in Fig.~\ref{fig:pad_na9p}. The figure shows angular distributions for three excitations close to and far above the ionization threshold by considering an aligned cluster. The patterns depend on $\hbar \omega_{\rm las}$ and reflect that the nodal structure of the outgoing wave $\varphi_{\bf k}({\bf r})$ changes with momentum ${\bf k}$.  Note that the latter is asymptotically related to the excitation energy by $|{\bf k}|=\sqrt{2m(\hbar\omega_{\rm las}-E_{\rm IP})}$. Systematic scanning of $\hbar\omega_{\rm las}$ modulates the zeroes and maxima of $\varphi_{\bf k}$ and thus in principle allows one to systematically probe the orbitals of cluster electrons and, in turn, the background ionic field. Therefore, PAD is a promising method to gain insight into structural cluster properties. Further, as we shall see below, photoelectron spectroscopy is an excellent tool for analyzing dynamical processes.

\subsubsection{Time-resolved analysis}
\label{sec:5A3}
Excited states populated by cluster photoactivation can decay in different manners, i.e., by emission of radiation, by internal conversion with energy transfer to the ionic degrees of freedom, or by Auger processes. The real time dynamics of such processes can in principle be explored by tracing the occupation and spectral positions of electronic states within time-resolved PES (TRPES). For example, structural changes of the ionic frame might open fast radiationless decay channels due to transient crossings of the potential energy curves of the excited and the ground state (conical intersection). The depletion of the excited level, when explored in TRPES experiments with fs pump-probe techniques, thus offer insight into internal energy conversion processes and ionic relaxation time scales. Below we discuss two recent examples.

\begin{figure}[t]
\centering \resizebox{0.65\columnwidth}{!}{\includegraphics{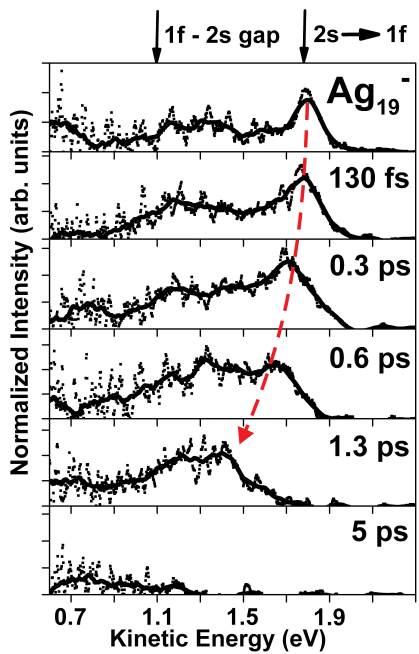}} \caption{Time-resolved photoelectron spectra of Ag$_{19}^-$ obtained with laser pulses of about $3\times 10^9$\,W/cm$^2$, $\hbar\omega_{\rm pump}=1.55$\,eV, and $\hbar\omega_{\rm probe}=3.1$ eV. The indicated times are the delays between the pump and probe pulses. The dashed arrow emphasizes the temporal development of the initially populated 1f level. After~\cite{Nie07}. \label{fig:pes_Ag19m}}
\end{figure}

TRPES on mass-selected Ag$_{N}^-$ with N=3-21 has been investigated in a two-color pump-probe experiment~\cite{Nie07}, using a 1.55\,eV pump photon safely below the vertical detachment energy (VDE). Fig.~\ref{fig:pes_Ag19m} depicts a series of spectra from Ag$_{19}^-$ obtained for time delays up to 5\,ps. The peak at 1.82\,eV observed in the top panel is consistent with the value of the VDE from the 1$f$ level previously excited by the pump pulse. The feature then shifts with time to lower kinetic energies (higher binding energy) and stabilizes between 1.1 and 1.5 eV after about 1.3\,ps (see arrow). Subsequently this peak loses intensity and vanishes after about 5\,ps. This evolution can be explained by a continuous Jahn-Teller deformation of the excited cluster which eventually opens a nonradiative transition channel to the ground state potential. For Ag$_{19}^-$, the decay time is estimated to be 630\,fs, i.e., much shorter than for radiating transitions.

\begin{figure}[htbp]
\centering \resizebox{\columnwidth}{!}{\includegraphics{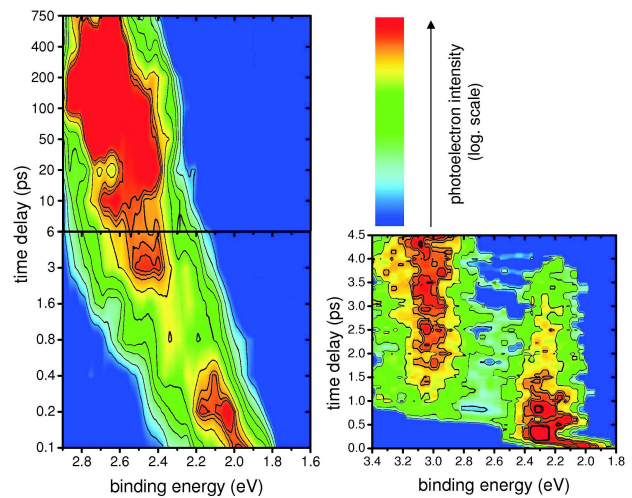}} \caption{Time-resolved photoelectron spectra of Au$_7^-$: (a) Experimental data obtained with 40\,fs pulses at intensities below $10^{11}$ W/cm$^2$ with $\hbar\omega=1.56/3.12$\,eV for pump/probe; (b) Simulated PES obtained from TDDFT coupled to ensemble-MD ``on the fly''. Note that the experimental data is plotted with a logarithmic delay axis while it is linear for the calculation results. From~\cite{Sta07}, with permission from American Institute of Physics. \label{fig:pes_au7m}}
\end{figure}

Another example on gold cluster anions revealed an extremely strong size dependence of the excited state lifetime. More specifically, Au$_6^-$ shows an exceptional long lifetime of more than 90\,ns~\cite{Wal07}. Corresponding DFT and linear-response TDDFT calculations predict decay times of 730\,ns. In contrast to that, Au$_7^-$ and Au$_8^-$ show clear indications for fast internal conversion on the ps time scale driven by ionic motion. As displayed in Fig.~\ref{fig:pes_au7m}a, the excited state peak in the experimental data of Au$_7^-$ is initially observed around a binding energy of 2\,eV.
This value just reflects the energy difference between the excited anion and the neutral ground state. As time evolves, the peak first slightly shifts to higher binding energies, then loses intensity, and finally becomes washed out after about 1\,ps. At the same time a new feature appears between 2.4 and 2.8\,eV which has reached high signal intensity around 3\,ps. The time evolution of the experimental peak maximum yields an exponential decay with a time constant of 1.8\,ps. Results from a corresponding linear-response TDDFT calculation based on a propagation of an ensemble of classical trajectories ''on the fly''~\cite{Sta07} are depicted in Fig.~\ref{fig:pes_au7m}b. Note the linear scale used here, whereas the experimental data is plotted on a logarithmic scale. The calculated population dynamics show a decay time of 1.9 ps, which is in good agreement with the experimental result. A closer analysis of the calculation results reveals the following process: The excited anion relaxes towards a crossing with the anionic ground state within about 340\,fs, see the rapid lowering of the initial peak at 2\,eV in Fig.~\ref{fig:pes_au7m}b. At the crossing the excited state begins to populate the vibrationally excited ground state. This mixing is expressed in the new feature emerging around 3.0\,eV. The rapid bleaching of the 2.3\,eV feature beyond 3\,ps and the stabilization of the 3.0\,eV peak reflect internal energy conversion and further indicate a melting of the cluster~\cite{Sta07}.

An even more elaborate way for investigating cluster dynamics via photoemission is offered by recording time-resolved photoelectron angular distributions (TRPAD). The applicability has been demonstrated with molecules, see~\cite{Suz06}. Femtosecond TRPAD studies on clusters have been performed by the Neumark group using imaging techniques to investigate the relaxation dynamics of small Hg$_{N}^-$, see~\cite{Bra05}.

\subsection{Multiphoton signatures}
\label{sec:5B0}
At intermediate laser intensity, single-photon processes as discussed above begin to be accompanied by multiphoton effects. The current section is devoted to this transition region where linear and nonlinear excitations simultaneously occur. We will focus on processes like above-threshold ionization, resonance broadening, and the onset of electronic thermalization. In all cases the laser intensity remains sufficiently low to resolve the influence of the chosen photon energy.

\subsubsection{Competition of linear and nonlinear excitation}
\label{sec:5B1}
\begin{figure}[b]
 \centering \resizebox{0.9\columnwidth}{!}{\includegraphics{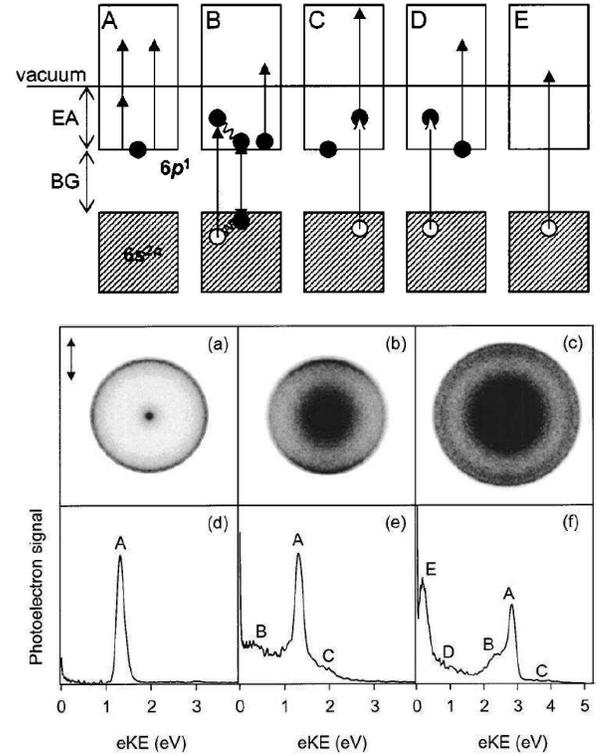}}
\caption{Top: Sketch of possible electron emission pathways from negatively charged mercury
clusters. Note that before irradiation the additional electron is located above the band gap (BG)
of the corresponding neutral system. Bottom: Angular resolved and integrated photoelectron spectra
of Hg$_{14}^-$ at different photon energies, (a) and (d) with $\hbar \omega_{\rm las}=1.57$ eV, (b)
and (e) with $\hbar \omega_{\rm las}=3.15$ eV, and (c) and (f) with $\hbar \omega_{\rm las}=4.58$
eV. The labels correspond to the processes sketched in the top panel. From~\cite{Ver04}, with permission from American Institute of Physics.
\label{fig:pes_principles}}
\end{figure}

Both single- and multiphoton processes can be demonstrated with the example of photoelectron spectroscopy on Hg$_{14}^-$, see Fig.~\ref{fig:pes_principles}~\cite{Ver04}. These spectra are extracted from PAD measurements, see the polar plots in panels $a$, $b$ and $c$ and the corresponding angle-integrated results at the bottom.
Mechanisms leading to specific peaks are schematically indicated in the top panels. Process A depicts the direct emission of the extra electron in the 6$p$ level with one high-energy photon or with two low-energy photons. No excitation energy is transferred to other decay channels and the peak labeled A reflects the energy of the photons and the VDE of the 6$p$ electron. Other pathways leading to electron emission are feasible if the photon energy exceeds the $s-p$ band gap: single-photon detachment of an electron in the 6$s$ band (case E); two-photon interband excitation of an electron from the $s$ band {\it via} the $p$ band (processes C and D); or single-photon Auger processes, labeled B. The latter excitation scheme is observed in the spectra as broad features with an onset consistent with the VDE of the $s$ and the $p$ band in Hg$_{14}^-$. This example thus demonstrates the complex pathways in a seemingly simple case of laser-exposed clusters.

The scenarios exemplified in Fig.~\ref{fig:pes_principles} call for a more detailed theoretical analysis. Some studies in the simpler case of Na clusters were performed in a series of papers in the framework of TDLDA-MD~\cite{Poh03,Poh04a}. For our purpose, we concentrate on the impact of intermediate states in multiphoton induced photoemission~\cite{Poh01}. In this laser intensity regime, the electron single particle energy $\varepsilon_0$ is {\it a priori} deduced from the recorded electron kinetic energy by writing
\begin{equation}
E_{\rm kin} = \varepsilon_0 + \nu \hbar \omega_{\rm las}, \label{eq:pesmulti}
\end{equation}
where $\nu$ is the number of photons involved. Chasing such a multiphoton process is in principle straightforward as the signal shifts according to the change in the photon energy. The picture becomes more complicated if an intermediate state (say, of energy $\varepsilon_1$) can be populated by a one or multiphoton process from an initial state $\varepsilon_0$. Then the original direct $\nu$-order process competes with a sequential $\nu-1$ (or $\nu-2,\ldots,$) process from state $\varepsilon_1$. Since sequential processes are less shifted than direct ones, discrimination is again possible by slightly changing the photon energy. A somewhat similar situation occurs when the photon energy is close to the plasmon excitation~\cite{Poh01}. In this case the spectra exhibit contributions pinned to the resonance energy.

\begin{figure}[htb]
\centering \resizebox{0.8\columnwidth}{!}{\includegraphics{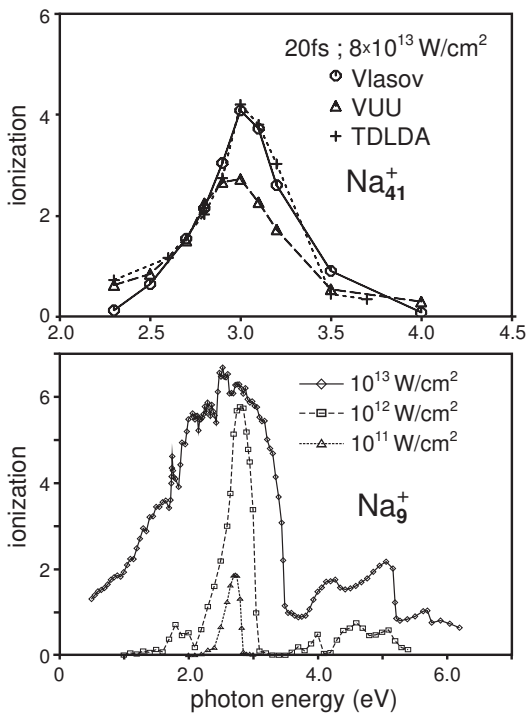}} \caption{Top: Comparison of the total valence electron emission from Na$_{41}^+$ calculated with TDLDA, Vlasov, and VUU calculations as function of photon energy for excitation with a 20\,fs Gaussian laser pulses of intensity \mbox{$I=6\times 10^{11}\,{\rm W/cm}^2$}~\cite{Gig03}. Bottom: TDLDA results of the valence electron emission from Na$_9^+$ over photon energy for excitation with 100\,fs pulses for three different intensities (as indicated). From~\cite{Ull97}. \label{fig:nesc_theo}}
\end{figure}

The plasmon plays a major role in the optical excitation of simple metal clusters, see, e.g., Fig.~\ref{fig:exper_fiveways}b for experimental examples. Even at higher laser intensities where details of level spectroscopy go lost the plasmon remains quite robust. This is exemplarily illustrated in Fig.~\ref{fig:nesc_theo} for Na$_{41}^+$ and irradiation with about 10$^{12}$\,W/cm$^{2}$ (top panel).
The plasmon peak centers around 3\,eV and a comparison between various calculations is performed. The high number of emitted electrons calls for calculations going beyond the mere mean-field and thus requires the inclusion of dynamical correlations and electron-electron collisions, as is done, e.g.,  with VUU. However, the strong impact of the plasmon is obvious in all models. As it provides a resonant coupling channel, enhanced energy absorption and increased electron emission are observed~\cite{Cal00}. It should be noted that TDLDA-MD and Vlasov-LDA-MD match almost perfectly, but for details in the tail of the distributions (e.g. at 3.5\,eV). The peak height for VUU is lower than for the pure mean-field approaches due to damping of the resonance by electron-electron collisions. This damping (or resonance broadening) leads to a higher absorption~\cite{KoePRA08} and slightly enhances ionization for off-resonant excitation.

The nonlinear nature of the response can be probed directly by varying the laser intensity, see the bottom panel of Fig.~\ref{fig:nesc_theo}. Although the calculations on Na$_9^+$ have been restricted to the TDLDA level (no dynamical correlations), they qualitatively reflect major trends. In general, the yield is maximal for photon energies in the vicinity of the plasmon. The peak height strongly increases with laser intensity, but does not show a linear scaling - a clear sign of nonlinearity. In addition, the shape of the spectrum is substantially altered at higher laser intensity. The peak width increases dramatically from about 0.3\,eV at $\rm 10^{11}\:W/cm^2$, 0.6\,eV at $\rm 10^{12}\:W/cm^2$ to almost 2\,eV at $\rm 10^{13}\:W/cm^2$. This indicates clearly a transition from the frequency dominated domain to a dynamical regime where spectra become less sensitive to the frequency of the laser to the benefit of the field intensity.

\subsubsection{Above-threshold ionization and thermalization}
\label{sec:5B2}
\begin{figure}[b]
\centering \resizebox{0.7\columnwidth}{!}{\includegraphics{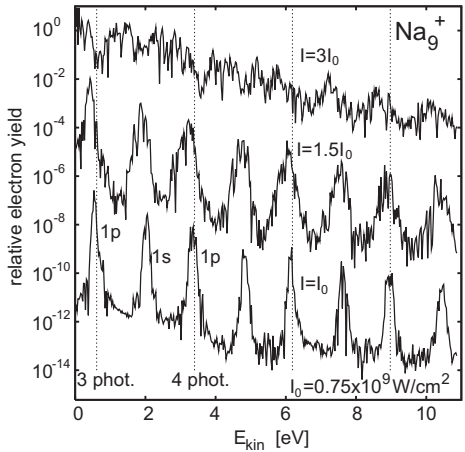}} \caption{\label{fig:pes_schem} Photoelectron spectra from Na$_9^+$ exposed to 100\,fs laser pulses with $\hbar\omega=2.7$\,eV at different intensities (indicated) as calculated with TDLDA~\cite{Poh00}. Vertical dotted lines show energies corresponding to $E_{\rm IP}$ plus a certain number of photons as indicated. Bound states from which electrons originate are indicated in a few cases. }
\end{figure}

At sufficiently high photon density, two and more photons can cooperate almost simultaneously in the excitation of a single electron. One of the consequences is direct electron emission, even far beyond the ionization threshold, although the photon energy stays below the IP. Fig.~\ref{fig:pes_schem} shows computed above-threshold ionization (ATI) spectra for Na$_9^+$, where the photon energy of 2.7\,eV is to be compared with $E_{\rm IP}$ of 7.5\,eV (1$p$ state) and the binding energy of the 1$s$ state of 8.8\,eV. The spectrum at the lowest intensity $I_0$ shows distinct peaks which can be associated with emission from these states and a well-defined number of photons $\nu$, as defined in Eq.~(\ref{eq:pesmulti}).
The intensity $I_0$ is already intermediate, as a sufficient photon density is required to drive multiphoton excitation. On the other hand, the emission from a $\nu$-photon process evolves as $I^\nu$. This produces a steep increase with intensity such that there remains only a small intensity window before the signal becomes blurred. This is demonstrated in Fig.~\ref{fig:pes_schem} with the two higher intensities increased by factors of 1.5 and 3.0, respectively. Note that the spectrum is almost structureless for the highest intensity, which can be explained in the following way: The ongoing ionization increases the bonding and downshifts the single-particle levels. Moreover, the spectrum collects electrons from all stages such that the level motion first induces a broadening of the peaks ($I$=1.5$\,I_0$) and finally a complete blurring. It should be noted that the fully smoothed distribution shows an exponential decrease (after removing the $\sqrt{E_\mathrm{kin}}$ phase-space element). In fact, as a result of the $I^\nu$-law, the exponential trend already appears for lower intensity when connecting individual peaks, for details see~\cite{Poh04a}.

\begin{figure}[b]
\centering \resizebox{0.95\columnwidth}{!}{\includegraphics{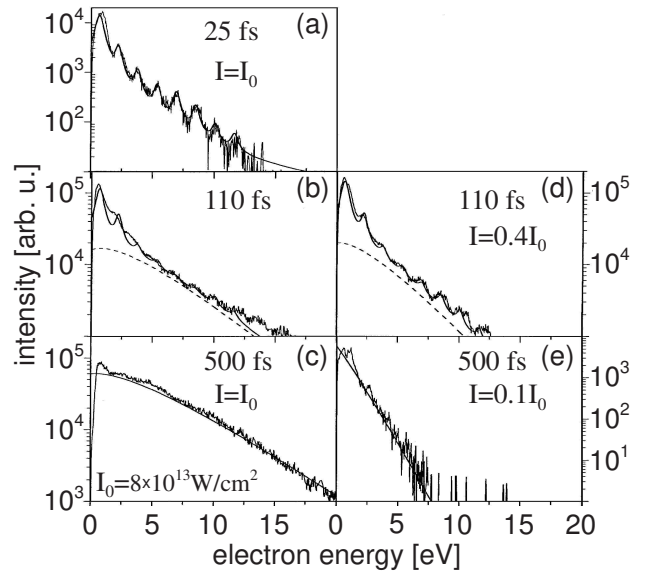}} \caption{\label{fig:pes_C60} Measured photoelectron spectra from C$_{60}$ exposed to laser pulses (790\,nm) for various pulse widths and intensities (as indicated). Dashed lines show estimated thermal contributions. After~\cite{Cam00}. }
\end{figure}

The example in Fig.~\ref{fig:pes_schem} with fixed pulse length deals with the intensity effect on ATI from clusters. However, the pulse duration plays a central role as well because competing perturbation by electron-electron collisions and ionic motion come into play with increasing interaction time. This aspect is worked out in Fig.~\ref{fig:pes_C60} with an experimental ATI result on C$_{60}$.
The two middle panels corroborate the previous observation that increasing the laser intensity (step from right middle to left middle panel) smears out the multiphoton peak structure. Going through the figure from top to bottom, i.e., along increasing pulse duration, one also obtains a disappearance of the detailed pattern, but this time due to an increase in the pulse width. A quick glance back on Fig.~\ref{fig:timescales} helps to an interpretation. After a time span of the order of the electron-electron collision time ($\tau_\mathrm{ee}$), the photon energy is distributed over the whole electronic system. The thus thermalized cloud evaporates one or more electrons at later times. The corresponding thermal emission spectrum is a smooth exponential (times phase-space factor), as indicated by the dashed curves in Fig.~\ref{fig:pes_C60}. The intermediate pulse width of 110\,fs obviously excites a transitional stage where direct and thermal emission compete. The longer time of 500\,fs is safely in the thermal regime and even lowering of the intensity (lower right panel) does not revive any detailed ATI structures.

\begin{figure}[t]
\centering \resizebox{0.7\columnwidth}{!}{\includegraphics{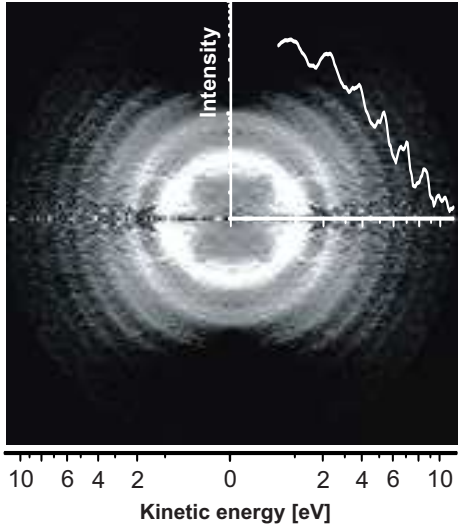}}
\caption{\label{fig:C60_ready} Photoelectron angular distribution of C$_{60}$ for irradiation with 800\,nm laser pulses of intensity \mbox{$I=10^{13}$\,W/cm$^2$} and pulse width of 60\,fs. The laser polarization is oriented along the horizontal axis. The rings correspond to above-threshold ionization, with increasing alignment for larger numbers $\nu$ of absorbed photons. The inset gives the angle-integrated intensity, in qualitative agreement with the result in Fig.~\ref{fig:pes_C60}. After~\cite{Skr09}}
\end{figure}
Since pronounced ATI signatures can be observed with complex systems like C$_{60}$, it is a logical next step to ask for the angular electron distributions as in the case of single-photon excitations discussed before in Fig.~\ref{fig:na58m_pes_pad}. In a corresponding experiment on C$_{60}$ the energy and angular resolved spectra where measured for excitation with 60\,fs laser pulses, see Fig.~\ref{fig:C60_ready}. Integration of the signal over the emission angle basically reproduces the sequence of ATI peaks, as seen for the short pulses in Fig.~\ref{fig:pes_C60}. The PAD exhibits as an additional information that the ATI peaks are well collimated in the direction of the laser polarization, an effect which increases with the number of photons involved. This indicates a direct emission process immediately induced by the pulse for high-order ATI. On the other hand, the more filled inner rings correspond to isotropic emission, which is most probably related to collisional thermalization. A solid understanding of the PAD structures, however, remains a challenging task for future theoretical investigations.

\begin{figure}[t]
\centering \resizebox{0.8\columnwidth}{!}{\includegraphics{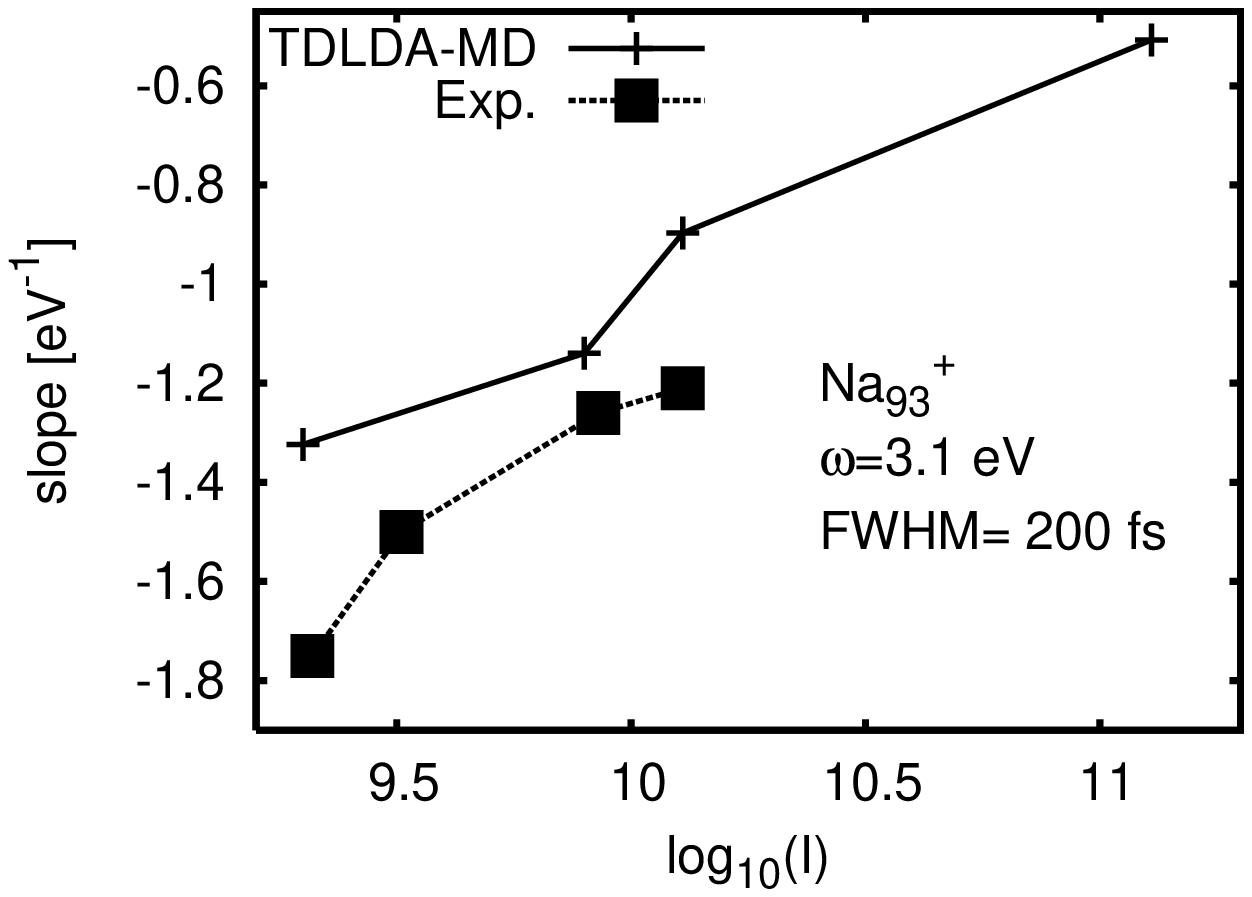}} \caption{\label{fig:slope_i_exp} Slope of the PES from ${\rm Na}_{93}^+$ irradiated by a 200\,fs laser pulse of 3.1\,eV. Results are drawn versus intensity (as log$_{10}(I)$ with $I$ in units of W/cm$^2$). Results from TDLDA-MD~\cite{Poh04a} are compared with the experimental results of~\cite{Sch01} using comparable experimental conditions.  }
\end{figure}

The above results indicate that electronic thermalization becomes increasingly important with increasing reaction time, i.e., if the laser pulse length exceeds the time required for collisional relaxation. Indeed, the underlying energy distributions in Figs.~\ref{fig:pes_C60} and~\ref{fig:C60_ready} show a roughly exponential behavior. For ${\rm Na}_{93}^+$ perfect exponentials have been measured~\cite{Sch01}. The slopes $s$ in $\exp{(-sE_\mathrm{kin})}$ of these distributions depend on the laser intensity, see the filled boxes in Fig.~\ref{fig:slope_i_exp}.
Clearly the slopes decrease with rising intensity, which can be interpreted as an increased heating of the system. As a matter of fact, the situation might be even more complex: TDLDA calculations (without electron-electron collisions) on the same ${\rm Na}_{93}^+$ also yield spectra with smooth exponentials while it is for certain that the signal stems from direct emission only. Trend and magnitude agree fairly well with the experimental data, cf. Fig.~\ref{fig:slope_i_exp}. This indicates that an interpretation as thermal emission is not compulsory from these data alone. A combined analysis including angular distributions and ideally also time resolved measurements would be required in order to unambiguously distinguish between direct and thermal processes. These effects are taken care of in the models particularly suited for highly excited dynamics, i.e., VUU simulations (Sec.~\ref{sec:3B3}), classical molecular dynamics (Sec.~\ref{sec:3B4}), and rate-equations (Sec.~\ref{sec:3C0}). In the experiments, however, a clear identification of thermalization effects remains difficult. Here imaging techniques as demonstrated above provide a promising tool for further investigations.

\section{Cluster dynamics in strong fields}
\label{sec:600}
For the previously considered low and intermediate laser intensities, where nonlinearities like plasmon broadening, the onset of saturation effects, and ATI have been discussed, the electronic configuration and ionic structure of the cluster remains in many respects at least similar to that of the initial state. This situation changes for laser-cluster interactions in the strong-field domain, see examples \mbox{Fig.~\ref{fig:exper_fiveways}c-d}. In this regime the interaction leads to radical changes of the structure and properties of the clusters, such as, e.g., light-induced metallization of rare-gas systems through strong inner ionization, the creation of multiple core-vacancies, or full cluster destruction via Coulomb explosion as a result of extreme charging. The cluster response is mostly (at least in the IR) field strength dominated, as may be deduced from the Keldysh parameter or comparison with the BSI-threshold, cf. Sec.~\ref{sec:2C0}. However, such an assignment based on atomic adiabaticity parameters may not be useful in all cases. The above phenomenological classification seems to better express what is widely understood under ,,strong-field cluster dynamics''.

Several illuminating early experiments in the nineties have sparked the rapid development of this topic. For an overview, we first concentrate on key phenomena and their possible relevance for technical applications. Subsequently we review aspects of the underlying microscopic mechanisms according to the current status of knowledge. Along this line, the following discussion is divided into two parts. Section~\ref{sec:6A0} highlights early surprises and experimental key results, ranging from measurements of energy absorption and emission of energetic particles to short wavelength radiation, and is restricted to a generic discussion of mechanisms and typical trends. Section~\ref{sec:6B0} reviews routes towards a more detailed microscopic understanding by closer relating theory and experiment and by pursuing more elaborate schemes like time- or angular resolved analysis. For a space-saving description of laser parameters, i.e., peak intensity $I_0$, pulses duration $\tau$ (FWHM ), and wavelength $\lambda$, we use the compact notation ($I_0$\,;\,$\tau$\,;\,$\lambda$).

\subsection{Early surprises and basic trends}
\label{sec:6A0}

\subsubsection{Laser energy absorption}
\label{sec:6A1}
A remarkable property of clusters in intense laser fields is very efficient energy absorption. At intensities of the order of $10^{15}\,{\rm W/cm^2}$ the average energy capture per atom can attain values of tens to hundreds of keV and by far exceeds that of atomic and molecular targets. Basically all of the violent processes discussed below have their starting point in this enhanced absorption, in conjunction with the absence of dissipation into surrounding material.

\begin{figure}[htbp]
  \centering
  \resizebox{0.9\columnwidth}{!}{\includegraphics{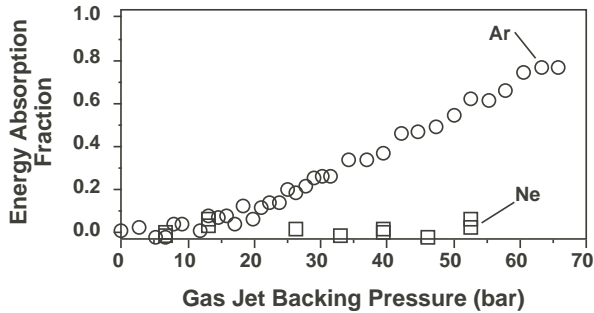}}
  \caption{Energy absorption of Ar$_N$ and Ne exposed to intense laser pulses \mbox{($7\times10^{16}$\,W/cm$^{2}$\,;\,2\,ps\,;\,527\,nm)} as a function of backing pressure. The estimated Ar$_N$ cluster diameter is 80\,\AA~at 40\,bar, and 100\,\AA~at 55\,bar. A major fraction of the laser pulse (up to 80\%) is absorbed by the large Ar clusters while atomic Ne gas remains nearly transparent. Adapted from~\cite{DitPRL97_3121}.} \label{fig:6A1-DitPRL97_3121}
\end{figure}

The direct measurement of the absorption from the relative loss of laser pulse energy in the interaction region requires a high target density \mbox{($10^{13}-10^{15}$\,clusters/${\rm cm}^3$)}. This situation can be realized close to the nozzle ($\sim$1\,mm) of supersonic gas expansion sources and corresponds to an effective particle spacing of $\gtrsim$100\,nm. As a typical example for the different behavior of clusters and gases, Fig.~\ref{fig:6A1-DitPRL97_3121} displays the relative energy absorption ${\cal A}_{\rm L}$ of high intensity laser pulses in a dense jet of Ar$_N$ compared to a Ne gas as function of backing pressure.
Since Ne does not condense at room temperature, the overall low absorption in Ne provides a reference for a gas of similar atomic density. In contrast to that, Ar clusters become increasingly opaque with cluster size beyond the onset of cluster formation (at about 5\,bar in this particular example). It should be noted, that laser pulse depletion from light scattering was found to be insignificant~\cite{DitPRL97_3121}. Further, the increase of gas flux with pressure can be ruled out as a major origin for the higher pulse depletion~\cite{ZwePP02}. Substantial absorption of up to \mbox{${\cal A}_{\rm L}^{\rm max}=0.8$} for the highest pressure in Fig.~\ref{fig:6A1-DitPRL97_3121} is typical for dense atomic or molecular cluster beams and has also been observed with (H$_2$)$_N$, (D$_2$)$_N$, Kr$_N$, Xe$_N$~\cite{DitPRL97_3121,MiuJJAP01,DitN99, LinChPL01}. In most cases, similar to Fig.~\ref{fig:6A1-DitPRL97_3121}, laser attenuation scales roughly linearly with stagnation pressure and then saturates, see also~\cite{JhaAPL06}.

When analyzed as a function of pulse intensity, substantial absorption sets in at relatively sharp thresholds. Beyond the onset intensities, e.g., \mbox{$I_{\rm th}\approx3 \times 10^{13}$ W/cm$^2$} for Ar$_N$ and \mbox{$I_{\rm th}\approx 4 \times 10^{12}$ W/cm$^2$} for Xe$_N$ at 527\,nm, the pulse depletion increases rapidly and attains \mbox{${\cal A}_{\rm L}\approx0.5\,{\cal A}_{\rm L}^{\rm max}$} at one order of magnitude higher intensity ~\cite{DitPRL97_3121}. As the thresholds roughly follow the trend of the corresponding atomic BSI intensities, the behavior indicates an avalanche breakdown process triggered by atomic optical field ionization to establish efficient absorption. A clear signature of the dynamical nature of the energy capture, which turns out to be largely driven by resonant collective electron excitation, is the pulse length dependence of ${\cal A}_L$. This will be further worked out in Sec.\ref{sec:6B0}.

\subsubsection{Highly charged atomic ions}
\label{sec:6A2}
\begin{figure}[b]
  \centering \resizebox{0.9\columnwidth}{!}{\includegraphics{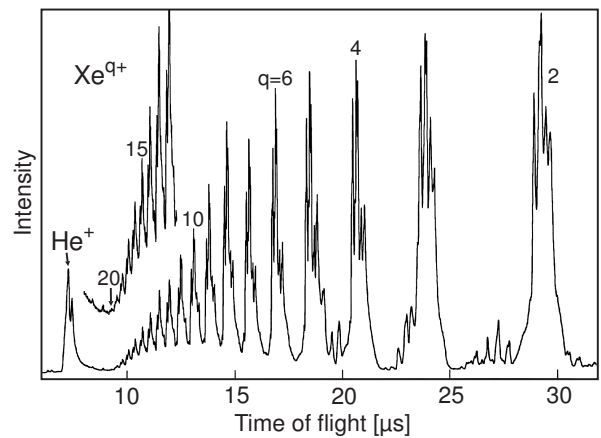}}
  \caption{Mass spectrum of highly charged atomic Xe$^{q+}$ ions resulting from by excitation
    of Xe$_N$ with \mbox{($1\times10^{15}$\,W/cm$^{2}$\,;\,350\,fs\,;\,624\,nm)} showing
    charge states up to $q^{\rm max}=20$. Adapted from~\cite{SnyPRL96}.}
  \label{fig:6A2-SnyPRL96}
\end{figure}

The strong optical absorption leads to high ionization and usually complete disintegration of the clusters. Atomic ions with high ionization stages $q$ are finally detected, see Fig.~\ref{fig:6A2-SnyPRL96} for an early ion spectrum on Xe clusters~\cite{SnyPRL96}.
After excitation with \mbox{($10^{15}$\,W/cm$^{2}$\,;\,350\,fs\,;\,624\,nm)}-pulses a broad charge state distribution emerges, extending up to \mbox{$q^{\rm max}=20$}. Such charge states are much higher than those from atomic Xe under similar conditions. For example, pulse intensities of 10$^{19}$\,W/cm$^{2}$ are required to produce Xe$^{21+}$~\cite{DamPRA01} from atomic gases - in reasonable agreement with the BSI model. From 65\,\AA~Xe clusters, ions with $q^{\rm max}=40$ were reported by~\cite{DitPRL97_2732} after exposure to laser pulses with \mbox{($2\times10^{16}$\,W/cm$^{2}$\,;\,150\,fs\,;\,780\,nm)}. The direct comparison of Ar$_N$ \mbox{(${\rm N}\sim100$)} to Ar gas after irradiation with \mbox{($2\times10^{14}$\,W/cm$^{2}$\,;\,30\,ps\,;\,1064\,nm)}-pulses shows similar trends~\cite{LezJPB97}, i.e., substantially higher maximum charge states with clusters (\mbox{$q^{\rm max}=10$} with clusters over \mbox{$q^{\rm max}=3$} with gas). The ion spectra from metal clusters (Ag$_N$, Au$_N$, Pt$_N$, Pb$_N$) were explored with femtosecond pulses (800\,nm) in several studies~\cite{KoePRL99,RadCPP05,SchEPJD99,LebEPJD02}, leading to values for $q^{\rm max}$ up to 30 for intensities below $10^{16}$\,W/cm$^2$. Further, highly charged ions have been reported for molecular clusters, i.e., atomic iodine up to \mbox{$q=15$} from (CH$_3$I)$_N$~\cite{ForJCP99} with \mbox{($2\times10^{15}$\,W/cm$^{2}$\,;\,130\,fs\,;\,795\,nm)}, and up to ${\rm O}^{6+}$ from to $({\rm H_2O})_N$~\cite{KumPRA03b} with \mbox{($8\times10^{15}$\,W/cm$^{2}$\,;\,100\,fs\,;\,806\,nm)}-pulses.

As a general remark it should be noted that ion distributions like the one in Fig.~\ref{fig:6A2-SnyPRL96} reflect an average over the focal intensity profile in the interaction zone where regions of higher intensity contribute with a smaller effective volume. Only recently, it has been demonstrated that contributions from the different intensities to such spectra can be deconvoluted~\cite{DoeEPJD07,DoePRL09} by using intensity-selective scanning.

Dedicated studies on heterogeneous, doped, and embedded clusters have been performed to investigate the effect of the cluster composition. In~\cite{PurCPL94}, irradiation of HI clusters with \mbox{($1\times10^{15}$\,W/cm$^{2}$\,;\,350\,fs\,;\,624\,nm)} yields I$^{q+}$ with up to \mbox{$q=17$}. In the same work, Ar$_N$ and Ar atoms attached to (HI)$_N$ yield Ar$^{q+}$ up to \mbox{$q=8$} for Ar$_N$(HI)$_M$, whereas no notable contribution from multicharged Ar$^{q+}$ ions is found for bare Ar$_N$. This supports, that the low-IP atoms or molecules act as chromophores and initiate nanoplasma formation. Subsequently also constituents with more strongly bound electrons can be ionized, e.g., via electron impact ionization. Other matrix effects occur in helium nanodroplets: Experiments on embedded clusters have shown evidence for electron transfer processes, where highly charged ions capture electrons from the surrounding helium~\cite{DoePCCP07}. When considerably ionized, the helium shell can produce strong absorption enhancement due to resonant heating of the nanomatrix~\cite{MikPRA08}.

Whereas most of the results have been obtained with optical lasers, first experiments are at hand making use of a VUV Free Electron Laser~\cite{WabN02,LaaPRL04}. Power densities of up to $3\times 10^{13}$\,W/\,cm$^{2}$ at 98\,nm (=12.65\,eV) were used in these experiments on rare gas clusters. Note that IBS heating is less effective at shorter wavelengths because of the low ponderomotive potential so that multiphoton ionization conditions are expected, cf. Sec.~\ref{sec:2C0}. Moreover, resonant collective heating can be ruled out because of the high laser frequency. Still, ionization of clusters is quite effective, leading to ions with charge states up to Xe$^{8+}$ and Ar$^{6+}$. These findings underline that strong cluster excitation is still possible in the domain of large Keldysh parameters. The origin of the high energy absorption required for the observed charging has been investigated by several groups. Various concepts were proposed, ranging from models based on enhanced IBS heating due to strong electron-ion scattering~\cite{SanPRL03}, over efficient IBS heating resulting from a high-density nanoplasma produced by local field enhancement of inner ionization by neighboring ions~\cite{SiePRL04}, to many body heating effects~\cite{BauAPB04,JunJPB05}. It should be noted that the calculations performed by~\cite{SiePRL04} for Xe$_{80}$ show good agreement with the experimental ion spectra when including the experimental focus averaging. Within the approach of~\cite{SanPRL03} there remain clear deviations from the experimental ion spectrum when taking the focus effect into account. For more details see a recent review with emphasize on the VUV-domain~\cite{SaaJPB06}. Further perspectives of VUV and XUV excitations of clusters are subject of Sec.~\ref{sec:7B0}. For now we come back to excitations with optical lasers.

\subsubsection{Ion energy distributions}
\label{sec:6A3}
Another early surprise was the large kinetic energy of atomic species emitted from clusters in intense laser pulses. Due to the strong heating of cluster electrons and high cluster charging on the femtosecond time scale, huge amounts of thermal and Coulomb energy are available to be released within in the explosion of the system. In experiments atomic ions from Xe$_N$ with kinetic energies beyond 1~MeV were observed~\cite{DitN97}. This has opened the route to table-top experiments on cluster-based fusion~\cite{DitN99}. Interestingly, for rare-gas clusters a rather sharp onset of high energy ion emission is observed. For Xe$_N$ a threshold intensity somewhat above $10^{14}$\,W/\,cm$^{2}$ was reported~\cite{TisNIMB03} with 230\,fs pulses at 790\,nm being roughly comparable with the BSI threshold intensity, see Eq.~(\ref{eq:OFIth}).

\begin{figure}[htbp]
\centering \resizebox{0.8\columnwidth}{!}{\includegraphics{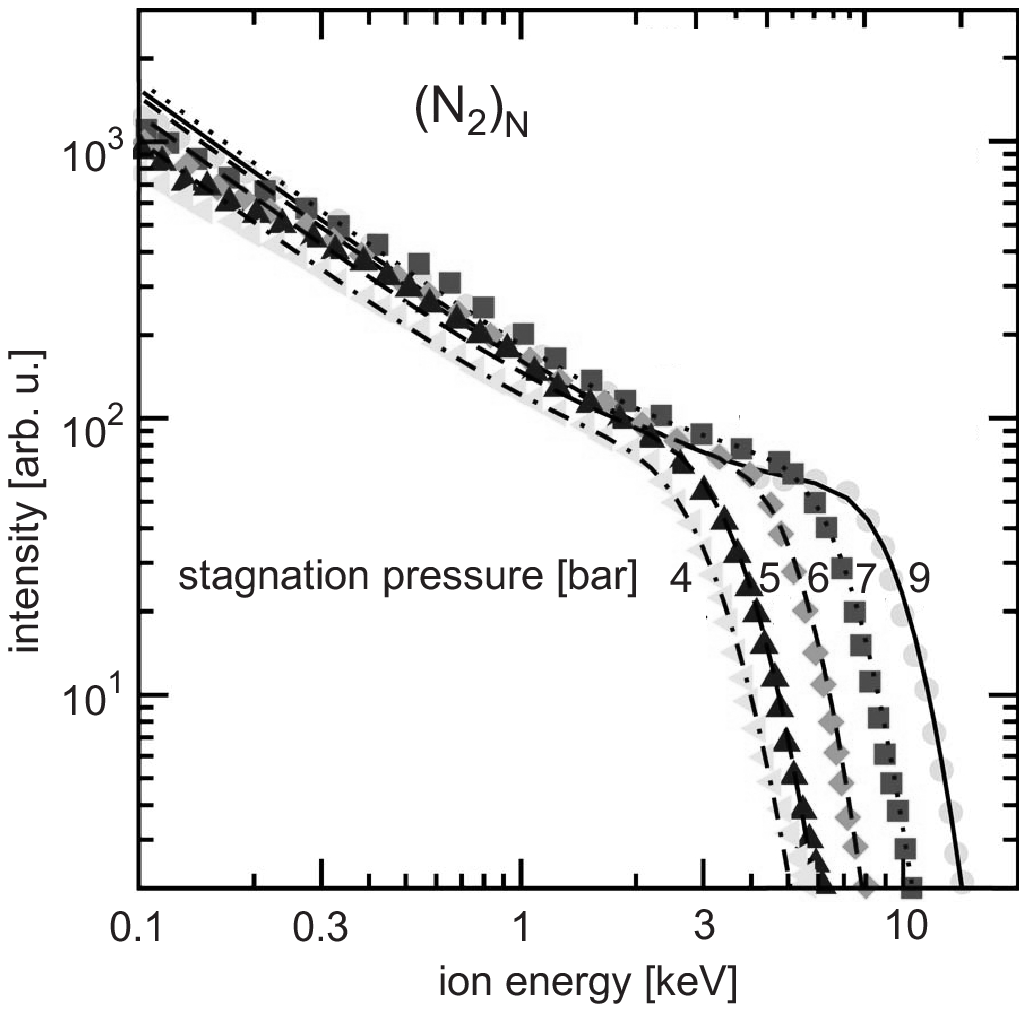}}
\caption{Measured ion energy spectra from (N$_2$)$_N$ (curves) for excitation with \mbox{($1\times10^{16}$\,W/cm$^{2}$\,;\,100\,fs\,;\,800\,nm)}-pulses~\cite{KriPRA04}.
The highest stagnation pressure corresponds to $\langle N \rangle=2300$.
With increasing backing pressure, i.e., for larger clusters, the spectra are shifted to higher energies. Symbols represent fits using a Coulomb explosion model that incorporates averaging due to the laser beam profile and the cluster size distribution. After~\cite{IslPRA06}.
\label{fig_IslPRA06}}
\end{figure}

The charge-state averaged ion energy distributions turn out to be very broad, see Fig.~\ref{fig:exper_fiveways}d for Pb$_N$ and Fig.~\ref{fig_IslPRA06} for (N$_2$)$_N$, and show a smooth decrease with increasing energy, often followed by a cutoff which is frequently termed ''knee''-feature.
Typically the maximum energy, which may be quantified by the energy of the knee-feature, increases with cluster size, see Fig.\,\ref{fig_IslPRA06}. However, in the case of Xe$_N$~\cite{MenPRA01} it was found that the maximum ion energy grows with cluster size until it levels out.
For sufficiently short laser pulses the nanoplasma model predicts that the ion energy decreases beyond a certain size, since the slower expansion of large clusters impedes resonant collective heating. The saturation can be explained by the relatively broad experimental cluster size distribution, i.e., the molecular beam still contains optimally sized particles producing the maximum ion energy.

Several additional effects contribute to the shape of the ion energy spectra, i.e., the spatial laser intensity profile and the degree of cluster ionization. Taking all these effects into account, experimental data on Xe$_N$~\cite{DitN97,SprPRA00}, Ar$_N$~\cite{KumPRL01}, (H$_2$)$_N$~\cite{SakPRA04}, and (N$_2$)$_2$~\cite{KriPRA04} can be reasonably well fitted considering Coulomb explosion~\cite{IslPRA06}.

Neglecting thermal electron excitation and assuming a uniformly charged monoatomic cluster, the final kinetic energy $\epsilon_I$ of an atomic ion is determined by its initial potential energy~\cite{LasJCP97,ZwePRL00a,NisNIMA01}
\begin{equation}\label{eq:6A3-LasPRL01-e1}
    \epsilon_I(r)
    =
    \frac{4\pi}{3} \rho_I \,r^2 q^2\times 14.4{\rm \,eV\,\AA},
\end{equation}
where $r$ is the initial radial ion position, $q$ the charge state, and $\rho_I$ the number density of ions in the cluster. Hence, ions at the cluster surface acquire the highest recoil energies $\epsilon_I^{\rm max}$ and this maximum energy also increases with cluster size. The latter trend was experimentally confirmed by several groups, e.g., on Xe$_N$ and Ar$_N$ by~\cite{DitPRL97_2732,LiChPL03,LezPRL98}. In~\cite{KriPRA04}, a monotonous rise of $\epsilon_I^{\rm max}$ from about 1\,keV to about 8.5\,keV was found with (N$_2$)$_N$ when increasing the size from $N=50$ to $N=2300$, cf. Fig.~\ref{fig_IslPRA06}. With Pb$_N$, $\epsilon_I^{\rm max}$ rises from 70 to 180\,keV when increasing the cluster size from \mbox{$N\approx100$} to \mbox{$N\approx500$}~\cite{TeuEPJD01}, see Fig.~\ref{fig:exper_fiveways}d.

At constant cluster size, a recoil energy enhancement is observed when adding spurious amounts of appropriate dopants~\cite{PurCPL94,JhaAPL06}. For instance, Ar clusters (${\rm N}\sim 2000$) containing about 60 H$_2$O molecules were considered in~\cite{JhaAPL06}. Under exposure to pulses with \mbox{($1\times10^{16}$\,W/cm$^{2}$\,;\,100\,fs\,;\,800\,nm)} the ion yield at 100\,keV scales up by a factor of three and the maximum ion energy is larger when compared to the dopant-free case. This effect was traced back to a longer phase of strong cluster heating and ionization, since dopants with lower ionization thresholds can initiate nanoplasma formation earlier in the laser pulse.

Clusters containing a mixture of low and high atomic number elements can be used to enhance the kinetic energy of the low atomic number ions. This is of particular interest for the acceleration of H$^+$ or D$^+$ for fusion reactions. For pure (D$_2$)$_N$ \mbox{($N\lesssim 10^5$)} ion kinetic energies up to 30\,keV were detected~\cite{ZwePP02} with \mbox{($1\times10^{17}$\,W/cm$^{2}$\,;\,35\,fs\,;\,820\,nm)}-pulses and energies up to 8.1\,keV were found for bare (H$_2$)$_N$ \mbox{($N\sim 10^5$)}~\cite{SakPRA04,SakMTLP06} with \mbox{($6\times10^{16}$\,W/cm$^{2}$\,;\,130\,fs\,;\,850\,nm)}-pulses. The presence of a considerable fraction of highly charged heavy-element ions in the cluster produces a strongly repelling background for the light ions~\cite{GriPRL02,KumPRA03b,MadPP04}. For (D$_2$O)$_N$ an enhancement in $\epsilon_I(D^+)$ of 5.6 over the result from $({\rm D}_2)_N$ of the same radius is predicted in the limit of complete and instantaneous ionization due to higher ionization stages of oxygen~\cite{LasPRA01}. The increase in the kinetic energy of D$^+$ was verified in a study on (D$_2$)$_N$ and (CD$_4$)$_N$~\cite{MadPP04}. From deuterated methane excited with \mbox{($1\times10^{17}$\,W/cm$^{2}$\,;\,35\,fs\,;\,820\,nm)}-pulses, deuterium energies of up to 120\,keV were found~\cite{GriPRL02}. Interestingly, only doubly charged carbon was detected, indicating substantial electron recapture. This is further supported by the fact that the maximum C$^+$ energy (180\,keV) substantially exceeds the value for D$^+$.

\begin{figure}[b] \centering
  \resizebox{0.75\columnwidth}{!}
  {\includegraphics{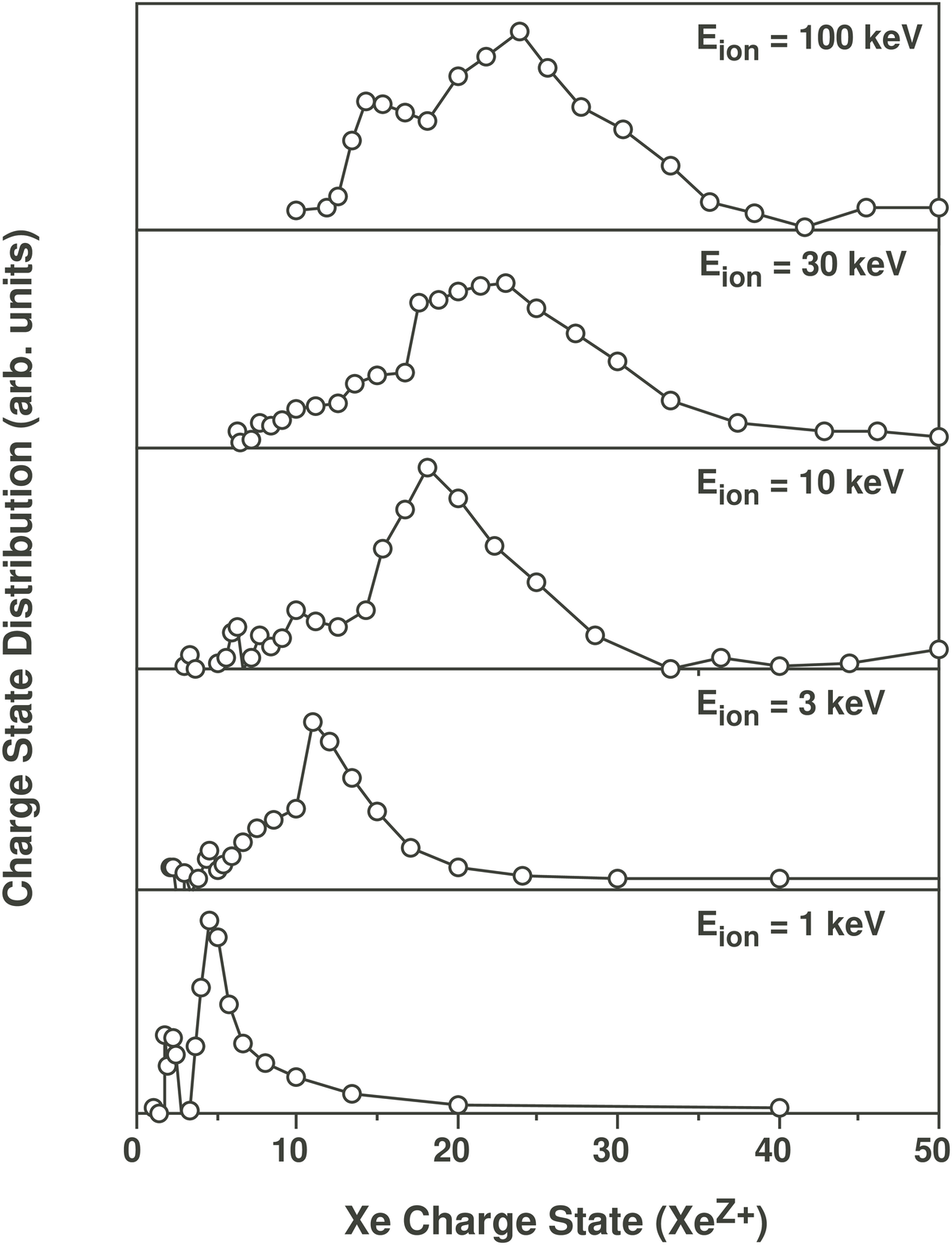}}
  \caption{Charge-resolved ion spectra from Coulomb explosion of Xe$_{2500}$ exposed
    to laser pulses with \mbox{($2\times10^{16}$\,W/cm$^{2}$\,;\,150\,fs\,;\,780\,nm)}
    for different recoil energies $\epsilon_I$
    (as indicated). With increasing ion kinetic energy the charge spectrum shifts to higher ionization stages.
    Adapted from~\cite{DitPRL97_2732}.}
  \label{fig:6A3-DitPRL97_2732}
\end{figure}

Additional insight into the explosion dynamics and the initial ion position within the cluster can be gained by simultaneously measuring charge states and energies. In principle, each emitted ion state has its own characteristic spectrum. To access the charge-resolved spectrum, techniques such as MD-TOF or Thomson spectroscopy can be applied, see Sec.~\ref{sec:4C0}. As an example obtained with another method, i.e., retarding field analysis, Fig.~\ref{fig:6A3-DitPRL97_2732} presents charge-resolved spectra measured at various recoil energies $\epsilon_I$ from irradiation of Xe$_{2500}$ with pulses of \mbox{($2\times10^{16}$\,W/cm$^{2}$\,;\,150\,fs\,;\,800\,nm)}.
With increasing $\epsilon_I$, the charge state distributions shift and broaden. At 100\,keV, ions with \mbox{$q=24$} are the most numerous and charge states up to \mbox{$q\sim40$} are observed. In studies on Ar$_N$ \mbox{($N=1.8\times10^5$)} and Xe$_N$ \mbox{($N=2\times10^6$)}~\cite{LezPRL98} a scaling like $\epsilon_I(q)=180\,{\rm eV}\,q^2$ (Ar) and $\epsilon_I(q)=160\,{\rm eV}\,q^2$ (Xe) was found for $q\le$6, as expected from electrostatic consideration, see Eq.~(\ref{eq:6A3-LasPRL01-e1}). The higher charge states ($q>10$) show a more linear dependence on $q$. A similar behavior was reported by~\cite{LebEPJD02}. This linear dependence has frequently been interpreted as a distinct fingerprint from hydrodynamic cluster expansion (driven by thermal electron energy), as predicted by the nanoplasma model, see Sec.~\ref{sec:3C0}. Such an assignment of parts of the spectrum to a Coulomb explosion or a hydrodynamic expansion has nevertheless to be made with care. From a theoretical point of view~\cite{DitPRA96}, the conclusion that hydrodynamic forces dominate the expansion is based on the assumption of a constant electron temperature in the expanding nanoplasma. This might be too crude to describe the dynamics correctly, as the nanoplasma experiences efficient expansion cooling during cluster explosion, see Sec.~\ref{sec:6B1}. A more elaborate analysis of the expansion process predicts a scaling law for the maximum recoil energy as a function of atomic density, cluster radius and initial temperature~\cite{PeaPRL06,PeaPP07}.  A one-to-one assignment of a quadratic dependence of $\epsilon_i(q)$ with a Coulomb explosion, see Eq.~(\ref{eq:6A3-LasPRL01-e1}), is correct only if the ionization process is quasi-instantaneous, i.e., much shorter than the time scale of the ionic motion, see also~\cite{TeuEPJD01}. For a dynamical cluster charging during the expansion, even pure Coulomb explosion can lead to a linear $\epsilon_I(q)$ scaling.  Note that this interplay between charging time and ionic motion has also been discussed in the context of structure analysis of biomolecules through scattering of XFEL radiation~\cite{NeuN00}.

\subsubsection{Soft x-ray and EUV emission}
\label{sec:6A4}
A decay channel of relevance for diagnostics and possible applications of laser-cluster-interactions is the emission of energetic photons, particularly in the soft x-ray domain. The x-ray spectra contain line emission that reflects recombination of electrons in weakly bound atomic levels with core-level vacancies, as first reported by Rhodes and co-workers~\cite{McPN94}. Note that the charge distribution during the pulse, as partly reflected by the x-ray spectra, may be different from the final ion spectra because of free-bound electron-ion recombinations and charge transfer processes.

\begin{figure}[htbp]  \centering
  \resizebox{0.95\columnwidth}{!}{\includegraphics{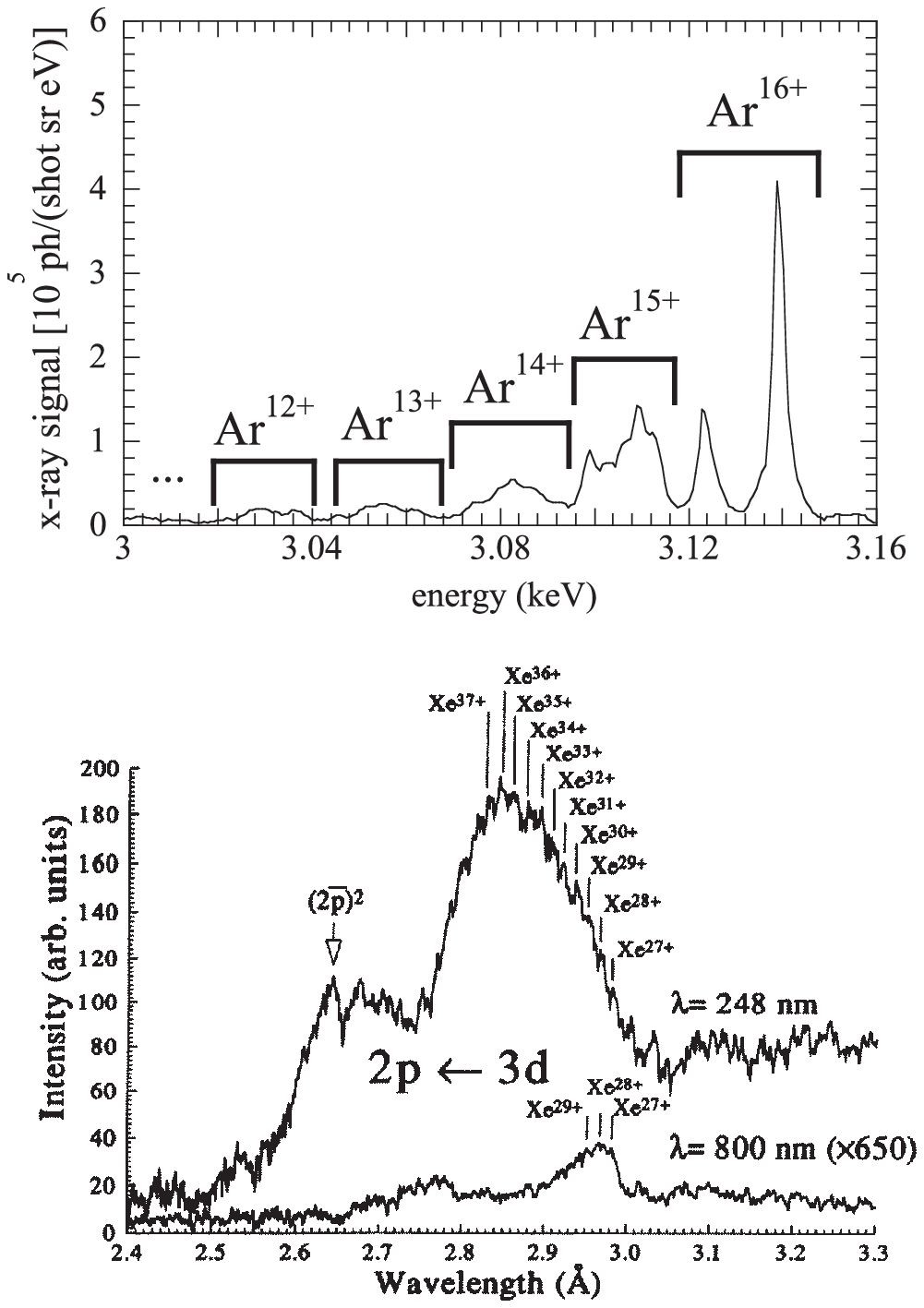}}
  \caption{Top: K-shell line emission from Ar$_N$ \mbox{($\langle N\rangle\sim4\times10^6$)}
  exposed to \mbox{($1.6\times10^{16}$\,W/cm$^{2}$\,;\,500\,fs\,;\,800\,nm)} laser pulses. From~\cite{DorPRE05}.
  Bottom: L-shell line emission from Xe$_N$ \mbox{($\langle N \rangle=12$)} excited with \mbox{($\sim 10^{18}$\,W/cm$^{2}$\,; $\lesssim 1{\rm ps}$\,;\,800\,nm/248\,nm)}. From~\cite{SchJPB98a}, with kind permission from IOP
  Publishing.  \label{fig:6A5-DorPRE05}}
\end{figure}

The two examples in Fig.~\ref{fig:6A5-DorPRE05} display such line emission spectra that can be
attributed to specific ion charge states. The top panel gives a result on K-shell emission from Ar$_N$~\cite{DorPRE05} for laser intensities of $10^{16}\,$W/cm$^2$. The example in the lower panel has been obtained at 10$^{19}$\,W/cm$^{2}$ on small Xe$_N$ and exhibits x-rays down to 2.4~{\AA}(5.2\,keV) and contributions from highly charged Xe$^{q+}$ up to $q\sim40$~\cite{SchJPB98a,SchJPB01}. Note that the excitation at 248\,nm compared to 800\,nm results in lines from much higher charge states. So far, the most energetic x-ray emission from nano-sized targets was observed on Kr$_N$ where strong K$_{\alpha,\beta}$ radiation (12.66\,keV/1.02\,{\AA} and 14.1\,keV/0.88\,{\AA}) was found~\cite{IssPP04}, see also Fig.~\ref{fig:exper_fiveways}e. A comparative study on Ar$_N$, Kr$_N$ and Xe$_N$ with $5\times10^{16}$\,W/cm$^2$ concentrated on M-shell line emission, e.g., Kr$^{9+}$ (3d$\leftarrow$4d), in the wavelength regime of about 10\,nm~\cite{McPAPB93}. Actually this was the first evidence that ionization dynamics in clusters under strong field conditions significantly differs from the behavior of atomic targets. The enhanced x-ray emission disappears when the clusters are destroyed by a weaker pre-pulse~\cite{SkoJETP02}. The major contribution of the x-ray photons is emitted on the ns time scale~\cite{DitPRL95,LarRSI99,KonASS02}, which is comparable to recombination lifetimes.

\begin{figure}[t]  \centering
  \resizebox{0.7\columnwidth}{!}{\includegraphics{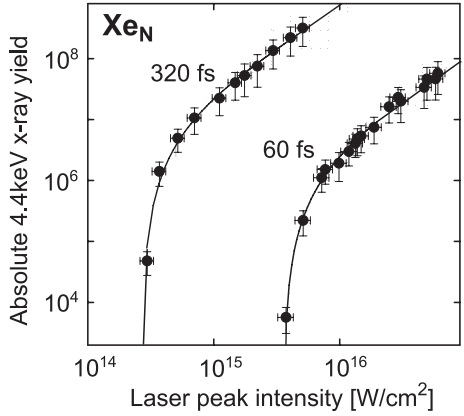}}
  \caption{\label{fig:Lamour07}  Absolute x-ray yield versus the laser intensity Xe$_N$ ($\langle N\rangle \sim 10^5$) for two different pulse durations.
  Solid lines reflect the effective focal volumes with intensity thresholds of $3.5\times 10^{15}$W/cm$^2$ for 60\,fs and $2.5\times 10^{14}$ W/cm$^2$
  for 320\,fs. From~\cite{LamJPCS07}, with permission from IOP Publishing.}
\end{figure}

The yield of x-ray emission with laser intensity scales as $I^{3/2}$ once the saturation regime is attained. This behavior reflects the effective volume in the focus~\cite{RozPS01,DobPRA97} and is similar to atomic targets~\cite{AugJPB92}. On Kr$_N$ (N$\sim10^5$) a threshold for the onset of high energy photon emission was observed slightly below 10$^{16}$\,W/cm$^{2}$, at 790\,nm and 130\,fs pulse width~\cite{DobPRA97}. The onset intensities, however, show a strong dependence on pulse duration and cluster material. Pulse length dependent thresholds for Ar$_N$ and Xe$_N$ were observed in~\cite{LamNIMB05,LamJPCS07,PriPRA08}. An example from Xe$_N$ is displayed in Fig.~\ref{fig:Lamour07},
where threshold intensities of $3.5\times10^{15}$W/cm$^2$ are found with 60\,fs pulses, whereas 320\,fs pulses result in a substantially lower threshold of only $2.5\times10^{14}$W/cm$^2$. The data hint at a transiently resonant heating, see Sec.~\ref{sec:6B1}. Note that the low threshold for the long pulse case is only slightly higher than the BSI threshold for atomic Xe ($9\times10^{13}$\,W/cm$^{2}$).

%
A high number of emitted photons is crucial when aiming at technical applications like EUV lithography (EUVL), see e.g.~\cite{BanJPD04}.
Furthermore, for EUVL applications the emission must occur in a narrow spectral range to avoid aberrations. The spectral range around 13\,nm is of importance since multilayer mirrors of Mo:Be and Mo:Si reach high reflectivities of nearly 70\,\%~\cite{StuJVSTB99}. Conversion efficiencies of about 1\%/\,2$\pi$\,sr have been achieved at only 2\,\% bandwidth, underlining that line emission from laser-driven Xe clusters, droplets, or jets, see e.g.~\cite{HanRSI04}, might by suitable for next generation microprocessor manufacturing~\cite{WuJCST07,Att07}. Within the soft x-ray regime, a recent study has demonstrated the high potential of clusters in intense laser fields as debris-free radiation sources for nanostructure imaging~\cite{FukAPL08}.

\subsubsection{High harmonic generation}
\label{sec:6A5}
High Harmonic Generation (HHG) has been quite extensively studied in atomic and molecular gas jets~\cite{KraPRL92,ChaPRL97,SpiS97,VelPRL01}. The resulting spectra contain odd harmonics because of inversion symmetry and show an initial strong intensity decrease, a plateau-region, and a rapid cutoff near $E_{\rm IP}$+3.17\,U$_P$~\cite{BraRMP00}. This cut-off is governed by the maximum return energy of electrons~\cite{CorPRL93} and reflects the importance of coherent stimulated recombination in atomic and molecular targets~\cite{PukPRL03}. The physics of HHG in clusters and particles has been investigated only in a few experiments, although clusters may act as a unique nonlinear optical medium.

\begin{figure}[b]
  \centering \resizebox{0.9\columnwidth}{!}  {\includegraphics{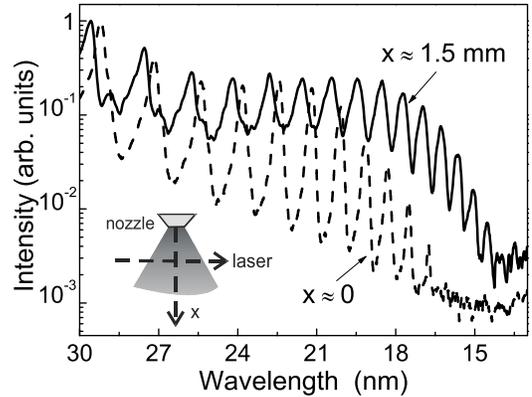}}
  \caption{High harmonic signals from Ar vs. Ar$_N$: spectra resulting mainly from atomic argon (dashed curve) and
  Ar$_N$ ($\langle N \rangle \sim 10^5$, full curve) exposed to pulses with \mbox{($2\times10^{14}$\,W/cm$^{2}$\,;\,25\,fs\,;\,800\,nm)}. The probed target conditions are
  selected by shifting the laser focus  with respect to the nozzle (as indicated). Reprinted with permission~\cite{VozAPL05}. Copyright 2005, American Institute of Physics.}
  \label{fig:6A6-VozAPL05}
\end{figure}

In an early study on HHG in clusters, a substantial harmonic signal of high order, actually 23th harmonic (HH$^{23}$), was reported for Ar$_N$~\cite{DonPRL96}. For laser intensities of up to $1.5\times10^{14}$\,W/cm$^{2}$, the HH$^{23}$ signal scales with $I^{17\pm1}$ and then changes its slope to $I^{4\pm1}$. In atomic gases for comparison, a power dependence of $I^{12}$ is found in the cut-off region~\cite{WahPRA93}. A clear enhancement effect of the HHG yield in clusters was also reported on Xe$_N$~\cite{TisJPB97}, where, for HH$^5$ (78\,nm), an increase of almost one order of magnitude was found with respect to the atomic gas. A comparison of HHG-spectra from atomic argon and Ar$_N$ taken under identical laser conditions is shown in Fig.~\ref{fig:6A6-VozAPL05}.
Close to the nozzle, (i.e. \mbox{$x=0$}, see inset), the cluster formation is not completed and HHG mainly results from atoms. When clusters are present in the beam (at \mbox{$x=1.5$}mm), the cut-off wavelength decreases considerably from 17 nm to 14 nm, thus increasing the highest HH order beyond the ponderomotive limit. A similar result was obtained in~\cite{PaiOL06}. Common with all results is a significant HHG enhancement in clusters when compared to an atomic gas.

In fact, the special linear wave propagation properties of dense cluster media (cluster-cluster separations of the order of the diameter) could be advantageous for efficient HHG. Since a gas of inner-ionized clusters can build up strong depolarization fields, electromagnetic waves can propagate below the plasma cutoff in a particular optical mode--the so-called cluster mode~\cite{TajPP99}. In a homogenous plasma, waves with frequencies below the plasma frequency become evanescent. Thus, a much higher electron density can be established in the cluster media without reflecting the fundamental wave. Moreover, in contrast to atomic gas plasmas, the refractive index of cluster media can be larger than one. A certain mixture of atoms and clusters can be used to tune the refractive index in order to fulfill a major requirement for efficient HHG, i.e., phase-matching. In principle phase-matching can then be attained for any desired harmonic~\cite{TisPRA00}, underlining the promising possibilities of clusters for tailored optical media.

Nonetheless, details on the microscopic mechanisms of HHG in clusters, e.g., concerning the interplay of stimulated recombination and bremsstrahlung, are yet to be explored. For a theoretical study on the contribution from bremsstrahlung see~\cite{PopPRA08}.

\subsection{Analyzing the microscopic cluster response}
\label{sec:6B0}
The above examples highlight the violent and multifaceted nature of laser-cluster interactions in the strong-field regime. The aim of the following sections is to review selected aspects of the ultrafast microscopic dynamics for excitation with strong optical lasers in more detail. This concerns the dynamics of cluster heating, expansion, and ionization. The influence of the pulse structure on the laser energy absorption as well as on the emission of electrons, ions, and x-rays has be studied in great detail in experiments with stretched pulses or by dual-pulse excitation. In most cases the observed signatures support a large impact of resonant plasmon excitations as will be discussed in connection with theoretical concepts and results from numerical simulations. Further, we comment on the persisting difficulties in explaining the origin of highly charged ions, which requires to understand the strong-field ionization on the atomic scale in the presence of a highly excited many-body environment. Further, angular resolved emission of electrons and ions will be addressed, which reveals unique acceleration effects in laser-excited clusters. Such studies can help to identify sensitive parameters for a control of specific decay channels and are important tests for theoretical models.

\subsubsection{The key role of collective excitations}
\label{sec:6B1}
\begin{figure}[t]
\centering \resizebox{0.85\columnwidth}{!}{\includegraphics{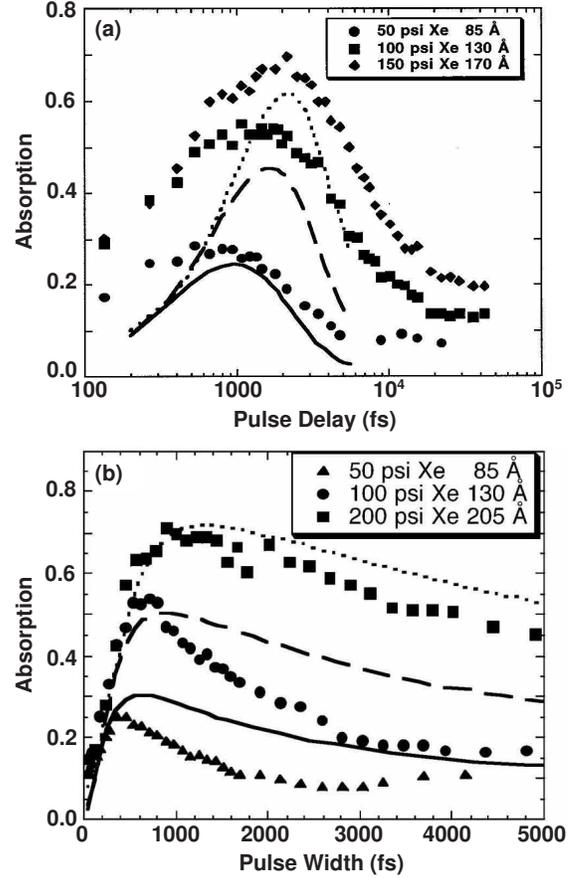}}
  \caption{Laser power absorption (symbols) by Xe clusters of different size
    (as indicated) for dual-pulse excitation (a) and irradiation with stretched pulses (b) at $\lambda$=810nm.
    The dual-pulse results were obtained with 50\,fs pulses of peak intensities $I_{1}=1.6\times 10^{16}{\rm W/cm^2}$
    and $I_{2}=1.6\times 10^{17}{\rm W/cm^2}$.
    The single-pulse experiment was performed with constant pulse energy of
    6.5\,mJ, resulting in a peak intensity of $2.3\times10^{17}{\rm W/cm^2}$
    at 50\,fs pulse duration. The curves show results calculated with the hydrodynamic model from Ditmire. Adapted from~\cite{ZwePRA99}.} \label{fig_Zwe99}
\end{figure}

\paragraph{\bf Evidence for resonant absorption:}
An important concept for explaining the high energy absorption of clusters in intense optical laser pulses involves resonant collective electron excitations. Since an inner ionized cluster is usually overcritical, a certain density lowering from cluster expansion is required to achieve frequency matching of the collective mode with the IR laser pulse, see Sec.~\ref{sec:2C0}. Striking experimental evidence for such transient resonance was found in~\cite{ZwePRA99}, see Fig.~\ref{fig_Zwe99}, where the laser energy absorption in a Xe cluster beam is measured as a function of pulse profile.
The upper panel shows the result of a dual-pulse (pump-and-probe) experiment. In a simplified picture, the leading pulse excites the cluster moderately and initiates its expansion. At a certain time the system reaches resonant conditions leading to a strong peak in the absorption as is clearly seen from the figure. The optimal pulse delay increases with size, showing that larger clusters require more time to reach frequency matching. An alternative way to explore the expansion time is to use one pulse but varying its length, as demonstrated in Fig.~\ref{fig_Zwe99}b. Similarly, it shows an optimal pulse duration to induce maximum absorption which again increases with system size. A word of caution is advised for the interpretation of such stretched pulse measurements, as the intensity decreases with increasing pulse duration. For example, considering a certain threshold intensity for the nanoplasma buildup (e.g. the BSI intensity), the effective interaction volume is strongly dependent on the pulse duration as well. Thus, a long pulse with low peak intensity will probe a smaller number of targets. This problem does not occur with a dual-pulse setup as used in the case of Fig.~\ref{fig_Zwe99}a. Nevertheless, both excitation schemes show a pronounced resonance feature in qualitative accordance with calculations based on the hydrodynamic model from Ditmire discussed in Sec.~\ref{sec:3C0}, see curves in Fig.~\ref{fig_Zwe99}.

\begin{figure}[t]
\centering \resizebox{\columnwidth}{!}{\includegraphics{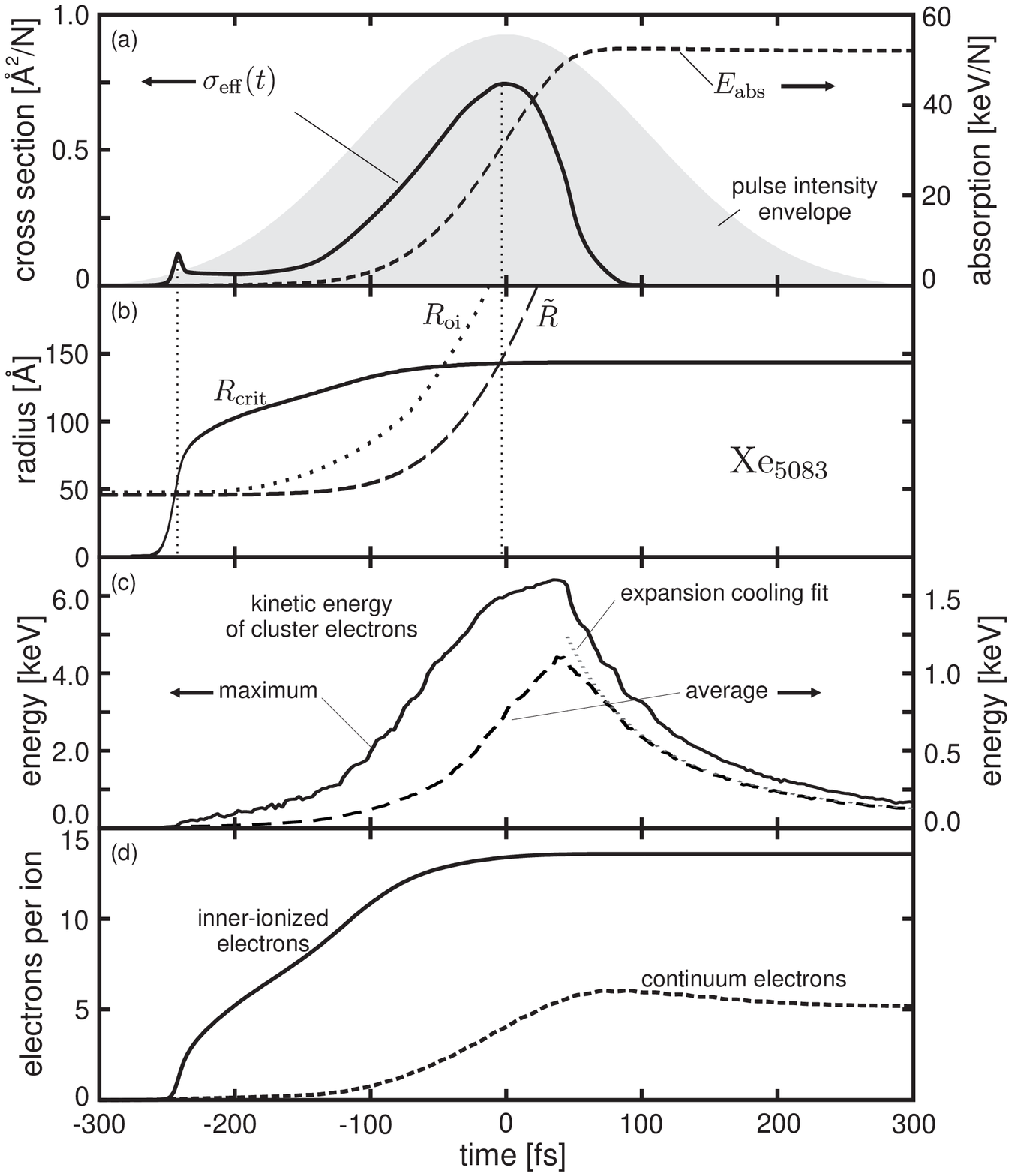}}
\caption{Simulated dynamics of Xe$_{5083}$ exposed to a laser pulse with \mbox{($10^{15}$\,W/cm$^{2}$\,;\,250\,fs\,;\,800\,nm)} using the MD code from~\cite{FenPRL07b}. (a) Pulse intensity profile, cycle-averaged energy capture $E_{\rm abs}$, and associated absorption cross-section $\sigma_{\rm eff}=\dot E_{\rm abs}/I$. (b) Mean cluster radius $\tilde R=\sqrt{5/3}\,R_{\rm rms}$, where $R_{\rm rms}$ is the root-mean-square radius, radial position of outermost ion $R_{\rm oi}$, and critical radius $R_{\rm crit}$ for resonant coupling, Eq.~(\ref{eq:Rcrit}). The vertical dotted lines mark the matching between $\tilde R$ and $R_{\rm crit}$, which coincides with the maxima in the absorption cross-section. The difference between $\tilde R$ and $R_{\rm oi}$ indicates inhomogeneous cluster expansion. (c) Average and maximum electron kinetic within the cluster radius $\tilde R$. The dotted line shows a fit for adiabatic expansion cooling (see text). (d) Number of inner-ionized and continuum electrons (cycle-averaged); the difference of these values reflects the number of quasifree electrons, cf. Fig.~\ref{fig:ioni_mechanisms}. } \label{fig_xe5083abs}
\end{figure}

Modeling the cluster response by a single collective mode is, however, strongly oversimplified. A more comprehensive picture can be drawn from microscopic simulations, such as MD, see Sec.~\ref{sec:3B4}. As an illustrative example, Fig.~\ref{fig_xe5083abs} displays results from a MD simulation of Xe$_{5083}$ exposed to a 250\,fs pulse of $10^{15}{\rm W/cm^2}$, for methodic details see~\cite{FenPRL07b}.

The upper panel shows the effective absorption cross-section (solid line), as derived from the total energy capture (dashed line), together with the laser intensity envelope (centered at time zero). The broad absorption peak between $-80$ and $50$\,fs corresponds to the expansion-induced collective resonance. The dominant energy capture proceeds near the crossing of the mean cluster radius ${\tilde R}$ with the critical radius for resonant collective coupling $R_{\rm crit}$ at $t\approx0\,$fs, see panel (b). The value of $R_{\rm crit}$ is estimated from the Mie formula (Eq.~\ref{eq:miefreq}) by
\begin{equation}
  R_{\rm crit}
  =
  \left(\frac{e^2}
             {16\pi^3\epsilon_0 m_{\rm e}c^2}  \langle q \rangle N \lambda^2\right)^{1/3}
  \;,
\label{eq:Rcrit}
\end{equation}
with $\langle q \rangle$ the average inner ionization of ions. For example, at $\lambda=800\,{\rm nm}$, a xenon cluster ($r_{\rm s}=2.5\,{\rm \AA}$) must expand by a factor of 1.4 for \mbox{$\langle q \rangle=1$} and of 3.0 for \mbox{$\langle q \rangle=10$} to become resonant. The evolution of cluster inner ionization, see panel (d), leads to a time dependent R$_{\rm crit}$. In particular, rapid inner ionization triggered by optical field effects and further enhanced by electron impact induces a sudden rise of R$_{\rm crit}$ in the leading edge of the pulse, see panel (b). That is accompanied by a short period of resonant coupling when $R_\mathrm{crit}$ crosses the actual system radius before the cluster becomes overcritical. This early ionization-driven resonance is reflected in a small feature in the cross-section at \mbox{$t\approx-240\,\mathrm{fs}$} and was also observed in MD calculations by~\cite{SaaPRL03}. Such early resonance would not occur in metal clusters probed with IR pulses, as they are overcritical already in the initial state. However, the major contribution to the energy capture $E_{\rm abs}$ proceeds within the expansion-driven resonance around t$\approx0$\,fs. In total, the ultimately absorbed energy exceeds 50\,keV per atom, similar to findings of~\cite{SaaJOMO06}. Note that the maximum cross-section of $0.75\,{\rm \AA}^2$ per atom is comparable to values typical for collective resonances in the linear regime, cf. Fig.~\ref{fig:exper_fiveways}b. The cross-section in Fig.~\ref{fig_xe5083abs}a drops quickly after the resonance, reflecting the suppressed coupling efficiency at {\it undercritical} density. Collective effects become unimportant and IBS-heating of residual electrons is weak due to rare electron-ion collisions.

Several effects contribute to the observed width of the resonance. To some extent, the broadening can be linked to inhomogeneous cluster expansion, see e.g.~\cite{MilPRE01}, with time-delayed resonant absorption in radial shells of critical density. A comparison of the radial position of the outermost ion $R_{\rm oi}$ with $\tilde R$ in Fig.~\ref{fig_xe5083abs}b shows that outer ions indeed expand more quickly, see also~\cite{IshPRA00}. This is due to a less effective screening of ions near the cluster surface~\cite{PeaPRL06}. Besides the influence of the ionic density profile, the driving of electrons beyond the cluster surface introduces an additional broadening due to nonlinear damping~\cite{MegJPB03,JunPRL04}. For sufficiently high laser intensity, also the occurrence a nonlinear resonance has been discussed~\cite{MulPRL05,KunPRL06}.

Coming back to the example in Fig.~\ref{fig_xe5083abs}, the average kinetic energy of cluster electrons of up to 1\,keV (dashed line in panel (c)) shows that there is strong thermal excitation of the nanoplasma near the resonance. The maximum electron kinetic energy within the cluster of up to 6\,keV (solid line) provides a reasonable measure for the depth of the transient space-charge potential produced from outer ionization and thermal excitation, see also~\cite{SaaEPJD05}. By electron impact excitation, such energetic electrons can directly create deep inner-shell vacancies required for hard x-ray emission. After the resonant heating the expansion of the ionic background leads to an efficient electron cooling. This process can be well described by adiabatic cooling of an ideal gas in an expanding spherical vessel: $\langle \epsilon_{\rm kin}(t)\rangle=a R^{-2}(t)+b$, see dotted line in Fig.~\ref{fig_xe5083abs}c. The offset parameter $b$ accounts for the kinetic energy of electrons that become localized in ionic cells during expansion. As displayed in the lowermost panel, a substantial fraction of the inner-ionized electrons cannot be accelerated to continuum energies and remains bound in the cluster potential. The further evolution of these electrons is of central importance for predicting the final ion charge spectrum and will be reconsidered in Sec.~\ref{sec:6B2}.

The above MD analysis illustrates that the dynamics is clearly dominated by collective energy absorption near the critical density. The cluster expansion to R$_{crit}$ therefore sets a crucial time scale for strong-field laser-cluster interactions in the IR regime. Only at very high intensities, where the laser field exceeds the restoring force from the ion background potential, resonance effects can be disregarded~\cite{HeiJCP07,KraPR02}.

\paragraph{\bf Signatures in emission spectra:}
Having identified the dominant role of collective excitations in the absorption, we now concentrate on corresponding signatures in the emission of x-rays, highly charged ions, and electrons.

\begin{figure}[b]
\centering \resizebox{0.75\columnwidth}{!}{\includegraphics{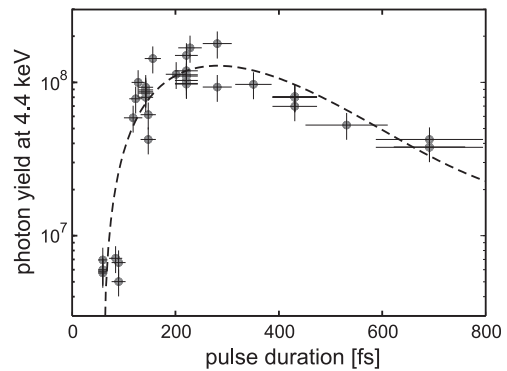}}
\caption{X-ray yield at 4.4\,keV from Xe$_{N}$ with \mbox{$\langle N\rangle\approx 4\times10^{4}$} irradiated with stretched laser pulses of fixed energy (35mJ) at 800\,nm wavelength. The shortest pulse ($\tau$=50\,fs) corresponds to a peak intensity of $3\times10^{16}{\rm W/cm^2}$. The dotted line is guide to the eye. After~\cite{LamNIMB05}, Copyright 2005, with permission from Elsevier.}
\label{fig_Lamour05}
\end{figure}

Time-resolved measurements of the x-ray emission have been performed primarily on rare-gas clusters using stretched pulses~\cite{ZwePRA99,IssPP04,ChePP02,LamNIMB05,ParPRE00,PriPRA08}. An example for Xe clusters is given in Fig.\,\ref{fig_Lamour05} displaying the x-ray yield from \mbox{$\rm 3d\rightarrow 2p$} transitions of Xe$^{\rm q>24+}$~\cite{LamNIMB05}.
The signal can be interpreted as a measure of energetic cluster electrons, as the production of Xe$^{24+}$ plus the 2p-vacancy by electron impact requires considerable energies of 7.3\,keV and $\sim$4.5\,keV, respectively. The energy for the creation of the vacancy should be transferred within one single collision event. Fig.\,\ref{fig_Lamour05} shows a steep increase in the x-ray yield for pulse durations up to 250\,fs, indicating a growing number of multi-keV cluster electrons. Note, that this is compatible with the generation of keV electrons at the instant of resonant heating in Fig.~\ref{fig_xe5083abs}c. The optimal duration thus indicates efficient collective heating, see also~\cite{ZwePRA99,ParPRE00} and similar experiments on EUV emission~\cite{ChePP02}. An alternative electron heating mechanism, namely multiple large-angle electron-ion backscattering in phase with the laser field, was proposed in~\cite{DeiPRL06} in order to explain the x-ray production from pulses that are too short for reaching resonant conditions. Whereas basic aspects of short wavelength emission from clusters can be rationalized, there are still several pending questions.  For example, the role of ionization and excitation over intermediate states or the impact of multielectron collisions on the production of core vacancies have not been resolved so far. Therefore, the physics behind x-ray emission from clusters remains a fascinating subject for further studies.

Measurements of ion kinetic energy spectra as function of pulse duration substantiate the strong impact of the pulse structure.
\begin{figure}[t]
\centering \resizebox{0.7\columnwidth}{!}{\includegraphics{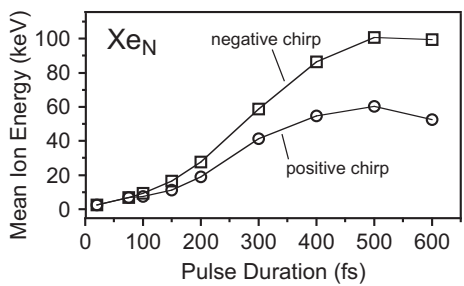}}
\caption{Mean recoil energy of atomic ions emitted from Xe$_N$ clusters \mbox{$(\langle N \rangle \approx 5.5\times10^4)$} for excitation with stretched pulses at 800\,nm (spectral width $\sim 60\,$nm) and constant peak intensity of $2\times 10^{17}{\rm W/cm^2}$. The results have been obtained with positively and negatively chirped pulses (as indicated). Note that a negative chirp corresponds to a decreasing laser frequency within the pulse. For 500\,fs pulses the chirp rate is about 0.13\,nm/fs.  Adapted from~\cite{FukPRA03}.} \label{fig_Fuk03_mod}
\end{figure}
Fig.\,\ref{fig_Fuk03_mod} shows a result of a constant peak intensity experiment on Xe$_N$ performed by~\cite{FukPRA03}. The mean ion energy grows with pulse duration reaching a maximum at about 500\,fs, in accordance with the picture of a delayed resonance. A similar behavior has been reported from a constant fluence measurement~\cite{KumPRA02}. Also a clear effect of the temporal phase of the pulse has been observed~\cite{FukPRA03}. Significant differences in the ion energies were found for positively and negatively chirped pulses. For negative chirp, i.e., a decreasing laser frequency with time, a 60\% enhancement of the mean ion energy was observed, see Fig.\,\ref{fig_Fuk03_mod}. This effect can be explained qualitatively by the joint gradual frequency red-shift of both laser pulse and resonance, which, in turn, extends the time span for resonant collective absorption. Applying the nanoplasma model to these particular experimental parameters, 500\,fs pulses with negative chirp lead to 1.2 times higher total energy capture than the corresponding result with positive chirp.

\begin{figure}[t]
\centering \resizebox{0.6\columnwidth}{!}{\includegraphics{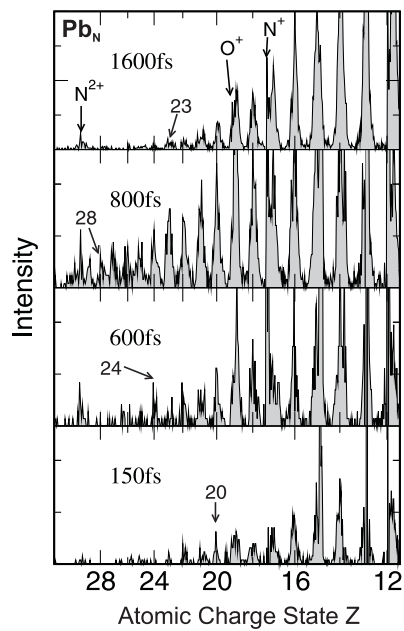}}
\caption{Charge state distribution of atomic ions emitted from lead clusters after exposure to laser pulses of variable duration and constant energy (19\,mJ). The laser peak intensity is $2.6\times10^{16}{\rm W/cm^2}$ for the shortest pulse (150\,fs) and $10^{15}{\rm W/cm^2}$ for the optimal duration (800\,fs), where the latter yields the highest charge states of up to 28. Reprinted from~\cite{DoeAPB00}, with kind permission from Springer Science+Business Media. } \label{fig_DoeAPB00}
\end{figure}
Evidence for an enhanced cluster ionization for certain pulse durations has been reported by several groups, e.g.~\cite{KoePRL99,SchEPJD99,DoeAPB00,LebEPJD02,FukPRA03,DoeEPJD07}. As a typical result, Fig.\,\ref{fig_DoeAPB00} displays spectra of high-q ions as a function of pulse width for an experiment with constant laser fluence~\cite{DoeAPB00}. The shortest and most intense pulses (150\,fs) yield atomic ions up to $q=20$. With increasing pulse duration, the maximum charge state as well as the overall signal intensity grows towards a maximum for an optimal pulse width of 800\,fs, where atomic ions up to \mbox{$q=28$} can be identified.  When further increasing the pulse duration, both the maximum charge state as well as the overall signal degrade.

The efficient charging for a certain pulse duration was in most cases attributed to resonant heating. Another mechanism, i.e., enhanced ionization (ENIO), has been proposed by~\cite{SiePRA03}. It relies on the concept of charge-resonance enhanced ionization (CREI) known from diatomic molecules, see Sec.\,\ref{sec:2C0}, which was also considered for multiple ionization of clusters in~\cite{LasPRA98}. Within ENIO, the increased ionization probability occurs for an optimal interatomic distance, where the tunnelling barrier between neighboring ions and the outer cluster Coulomb barrier of the system are reduced at the same time. The optimal pulse duration is thus related to the instant at which the expanding cluster reaches the optimal interatomic distance. However, because of the large outer Coulomb barriers in highly charged clusters, ENIO is considered to be relevant primarily for small compounds ($N\sim 10$). Enhanced ionization due to resonant heating (plasmon-enhanced ionization), on the other hand, applies to clusters of any size and even to nm-particles~\cite{SurPRL00,ReiAPB01,SaaPRL03,SaaJOMO06,DoePRL05}.

\begin{figure}[b]
\centering \resizebox{0.95\columnwidth}{!}{\includegraphics{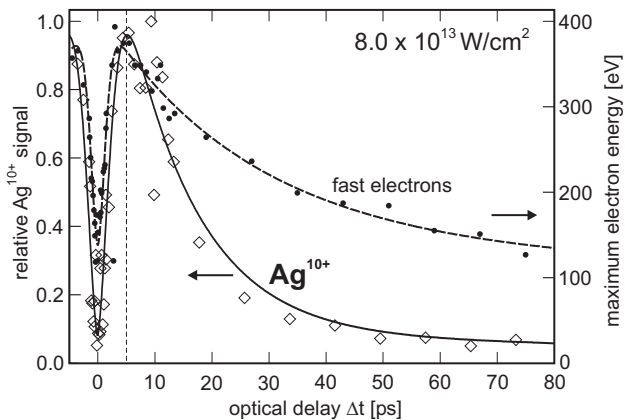}}
\caption{Comparison of the Ag$^{10+}$ yield (diamonds, left axis)
  with the maximum
  kinetic energy of the emitted electrons (dots, right axis)
  following laser excitation of Ag clusters
  \mbox{$(\langle N\rangle \approx2\times10^4)$} with dual 100\,fs laser pulses at intensity
  $8\times10^{13}{\rm W/cm^2}$ and 800\,nm wavelength. The curves are guides to
  the eye.  Adapted from~\cite{DoePRA06}.} \label{fig_DoePRA06}
\end{figure}

The emission spectra discussed so far correspond to experiments with one single pulse of variable duration. Results from a dual-pulse experiment on Ag clusters are given in Fig.\,\ref{fig_DoePRA06} showing the yield of Ag$^{10+}$ and the maximum energy of the emitted electrons as function of pulse separation. The ion signal shows a pronounced maximum for delays of about 5\,ps with an enhancement of more than one order of magnitude. This indicates that cluster activation and enhanced ionization can be disentangled, as is also supported by numerical simulations~\cite{MarPRA05,SiePRA05,DoePRA06,BorLP07,BorCPP07}. Clear indications that a sequence of two pulses may even represent the optimal pulse profile for high-q ion production have been found by~\cite{ZamPRA04} for Xe$_N$ and by \cite{TruPRA09} for Ag$_N$. In both studies genetic feedback algorithms have been used to optimize the temporal pulse structure in order to maximize ion charge states and converged towards a pulse profile containing two sub-pulses.
Dual-pulse excitation further offers a route for targeted control of the cluster dynamics. It has been demonstrated on small silver clusters in helium droplets that the optimal delay can be controlled by the intensity of the leading pulse. As was corroborated by semiclassical Vlasov calculations, a higher intensity of the leading pulse enhances the cluster expansion speed due to higher heating and ionization and thus reduces the time for which resonant coupling conditions are established~\cite{DoePRL05}.

Coming back to Fig.\,\ref{fig_DoePRA06}, also the maximum electron energy is analyzed as function of pulse separation. The coincidence of high ionization yield and maximal electron energy underlines the leading role of collective excitations in both decay channels. A similar correlation between fast electrons and VUV-radiation was reported in~\cite{SprPRA03}.
A common feature is the occurrence of high electron energies. A maximum value of $375\ {\rm eV}\approx60\,U_p$ was observed with Ag$_{N}$ at moderate intensity, see Fig.~\ref{fig_DoePRA06}. For Xe$_N$ energies in the keV range have been reported~\cite{ShaPRL96,KumPRA02,SprPRA03}. Further details on the electron emission, i.e., angular- and time-resolved signatures and underlying acceleration mechanisms are considered in Sec.\ref{sec:6B3}.

\subsubsection{Difficulties of explaining high charge states}
\label{sec:6B2}
Although most of the above trends like higher ionization and energetic particle emission for resonant cluster excitation can qualitatively be explained, the quantitative understanding of the emission spectra remains a challenge. A still largely debated topic is the origin of the very high atomic ionization stages from clusters~\cite{HeiJCP07,FenPRL07b}. In order to calculate realistic ion spectra, inner ionization in the presence of local fields, outer ionization dynamics, as well as recombination effects have to be taken into account consistently. Difficulties arise from at least two facts. First, since inner ionization cannot be treated fully quantum mechanically for practical reasons, simpler approximations like ADK-rates and atomic impact ionization cross-sections have to be used and must be corrected correspondingly. Second, recombination processes, even if treated only classically or with effective rates, proceed at much longer time scales than the interaction with the pulse and are thus numerically extremely time-consuming. However, a few routes towards a more realistic description of high charge states by incorporating these effects have already been explored.

To cope with the first problem, inner ionization has to be corrected for medium contributions like screening or polarization effects in the cluster. This is more or less straightforward for tunnel ionization, as the effective local field resulting from the momentary distribution of charges and the laser is accessible numerically, e.g., from MD simulations. Applying an appropriate temporal or spatial filtering, the effective field can be used for the determination of tunnelling ionization probabilities from the ADK-rates~\cite{AmmJETP86}. More involved is the treatment of electron-impact ionization which is often described by the empirical cross-sections from~\cite{LotZP67}. That requires atomic ionization thresholds which are modified by many-particle effects in the cluster, such as screening and fields from neighboring ions. One way to the determine these shift is the use of statistical approaches like Debye screening or ion sphere models~\cite{GetJPB06,BorLP07}. These, however, assume local thermal equilibrium and neglect the details of ionic correlation. A more direct approach relies on the evaluation of the local field and the resulting shifts directly from a particle based simulation, see~\cite{FenPRL07b}. Irrespective of the particular method, threshold lowering induces substantial enhancement of impact ionization when compared to the bare atomic cross-sections. An example will be discussed below.

\begin{figure}[t]
\centering \resizebox{0.75\columnwidth}{!}{\includegraphics{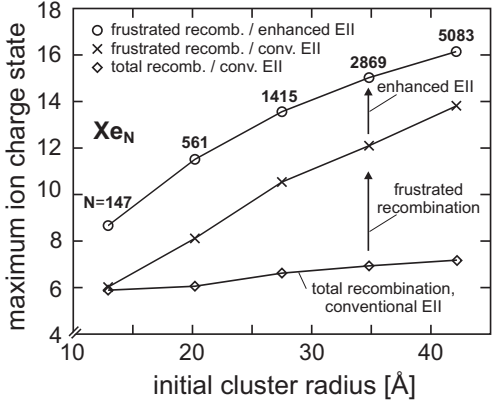}} \caption{
Calculated maximum charge state of atomic ions from Xe$_{N}$
  ($N=147-5083$, as indicated) exposed to 250\,fs laser pulses
 with peak intensity $4\times 10^{14}{\rm W/cm^2}$. The results correspond to
 different treatments of electron-impact ionization (EII) and electron-ion
 recombination. Conventional and enhanced EII correspond to atomic and
 local-field corrected ionization thresholds, respectively.  Total
 recombination assumes that cluster-bound electrons recombine with the
 closest ion after the laser pulse, while the long-term dynamics of quasifree
 electrons in the presence of a weak static background field of ${\rm 3kV/m}$ is taken into account for frustrated recombination. Adapted from~\cite{FenPRL07b}.
 \label{fig_FenPRL07_md}}
\end{figure}

The second problem concerns the handling of electron-ion recombination. Usually it is assumed that only continuum electrons produced during the laser pulse contribute to the final ion spectra and cluster-bound electrons (quasi-free after the laser pulse) fully recombine. With this assumption, however, the high experimental charge states at moderate laser intensities cannot be explained. Under experimental conditions this {\it full recombination} of quasi-free electrons is questionable, as, in particular, weakly bound electrons may not relax to lower ionic levels but can be re-ionized by space-charge field in the interaction zone or by ion extraction fields required for the time-of-flight analysis~\cite{FenPRL07b}.

In the latter work it was found that the combined action of both, enhancement of electron impact ionization trough threshold-lowering and background-field induced frustrated recombination, enhances the maximum ion charge states by up to a factor of two, see Fig.\,\ref{fig_FenPRL07_md}. While enhanced charging of small clusters is dominated by threshold lowering effects, the consideration of the recombination dynamics becomes increasingly important with large clusters. Further contributions such as excitation-autoionization or ionization via intermediate states, the importance of which is known for atomic electron-impact ionization~\cite{GriPRA84,Pindzola08}, have not been studied in detail yet.

\subsubsection{Asymmetric ion and electron emission}
\label{sec:6B3}

\begin{figure}[b]
   \centering
   \resizebox{0.7\columnwidth}{!}{\includegraphics{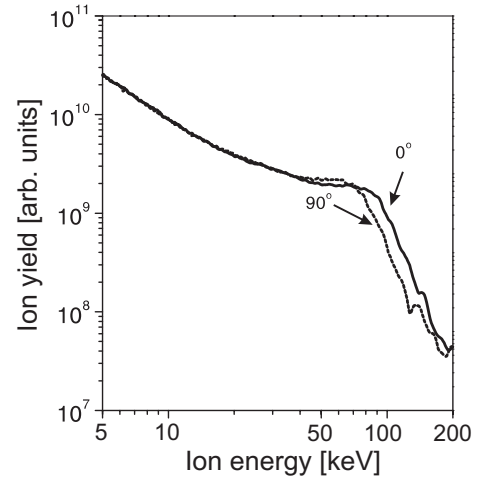}}
   \caption{Angular dependence of the ion energy recoil spectra of Ar
   clusters ($N$=$4\times10^4$) exposed to pulses with
   \mbox{($8\times10^{15}$\,W/cm$^{2}$\,;\,100\,ps\,;\,806\,nm)}.
   At 0$^\circ$, the polarization of the laser is parallel to the time-of-flight
   axis. Adapted from~\cite{KumPRL01}.}
   \label{fig:KumPRL01}
 \end{figure}

An interesting direction for possible applications of clusters is the pulsed generation of energetic ions and electrons. The quest for a detailed understanding of the acceleration mechanisms is therefore not only driven by fundamental interests. The presence of asymmetries in angular resolved ion spectra reveals, that the cluster disintegration notably deviates from an isotropic explosion process. Further, for excitation with appropriate pulses, the electron spectra show strong signatures from field-driven acceleration with high directionality. Corresponding signatures from experimental and theoretical studies as well as the main concepts for their explanation are reviewed below.
\paragraph{\bf Angular resolved ion emission:}
Ion energy spectra exhibit a clear asymmetry, where higher kinetic energies appear for the emission along the laser polarization axis. This was first reported for Xe$_N$~\cite{SprPRA00a,KumPRA02} and later for Ar$_N$~\cite{KumPRL01,HirPRA04}. Fig.~\ref{fig:KumPRL01} displays an example for the directional asymmetry for Ar$_N$ \mbox{($N\sim 40000$)} after excitation with \mbox{($8\times10^{15}$\,W/cm$^{2}$\,;\,100\,ps\,;\,806\,nm)} pulses, where a polarization induced shift of the "knee" in the energy distribution of about 20\% is observed.
Similar shifts between 15\,\% and 40\,\%, depending on pulse duration, were found with Xe$_{N}$~\cite{SprPRA00,LiPST05} and with molecular (N$_2$)$_N$ clusters~\cite{KriPRA04,MatLP06}.

At least three fundamentally different contributions to this asymmetry have been described. In~\cite{IshPRA00}, a mechanism is proposed where the additional acceleration is a direct result of the laser field. Since the net effect of the laser averages out for ions with constant $q$, rapid charge state oscillations of surface atoms were proposed, such that higher effective charge states appear during laser half-cycles with outward electric field component. That accumulates maximum repulsion along the laser polarization axis. However, this mechanism is unlikely to fully explain the experimental asymmetry since the rates for electron-ion recombination are very low at the typically high electron temperatures~\cite{Bet77}.

The second mechanism is an asymmetric Coulomb explosion due to angular dependent charging of ions and was originally discussed for C$_{60}$~\cite{KouJCP00}. Near the cluster poles, i.e., the regions with surface normal parallel to the polarization axis, higher peak electric fields from the laser and the cluster field (polarization and/or space charge) enhances inner ionization. Thus, ions located in this region experience stronger Coulomb repulsion. This view of enhanced ion acceleration along the polarization axis is supported by numerical simulations~\cite{FenEPJD04,JunPRL04} and the observation of asymmetric ion charging~\cite{HirPRA04}.

Finally, forces directly from the cluster polarization field enhance asymmetric ion acceleration~\cite{KumPRA02,FenEPJD04,BrePP05}. In terms of a simple rigid sphere model, cluster ions and electrons can be described by two homogenously charged spheres of opposite charge density and equal radius. The laser-driven oscillation of the electron cloud results in a nonvanishing asymmetric contribution to the radial component of the electric field at the cluster surface, whereby enhanced repulsion follows for surface ions near the cluster poles~\cite{BrePP05}. Even for an isotropic ion charge state distribution, this mechanism supports enhanced repulsion for surface ions near the cluster poles. This repulsion is particularly strong for large-amplitude oscillations of the electron cloud at resonance (see Sec.~\ref{sec:6B1}). Thus, this model can also explain the pulse-length-dependent asymmetry observed in~\cite{KumPRA02}.

\paragraph{\bf Angular resolved electron emission:}
Compared with the ions, the degree of asymmetry is much more pronounced with electrons. The emission is aligned to the laser polarization axis~\cite{ShaPRL96,KumPRA02,SprPRA03}. This preferential ejection is a direct marker for laser-assisted and nonthermal emission and turns out to be strongly dependent also on the pulse duration. On Xe$_N$~\cite{KumPRA03} find a yield ratio \mbox{$Y_{||}/Y_{\perp}\approx 3$} for optimal pulse durations, while almost isotropic emission and less energetic electrons are observed for the shortest and most intense pulse. The authors related this effect to resonant collective enhancement of the polarization field. Enhanced asymmetry for optimal pulse conditions is also supported by simulations~\cite{MarPRA05}.

\begin{figure}[t]
\centering\resizebox{\columnwidth}{!}{\includegraphics{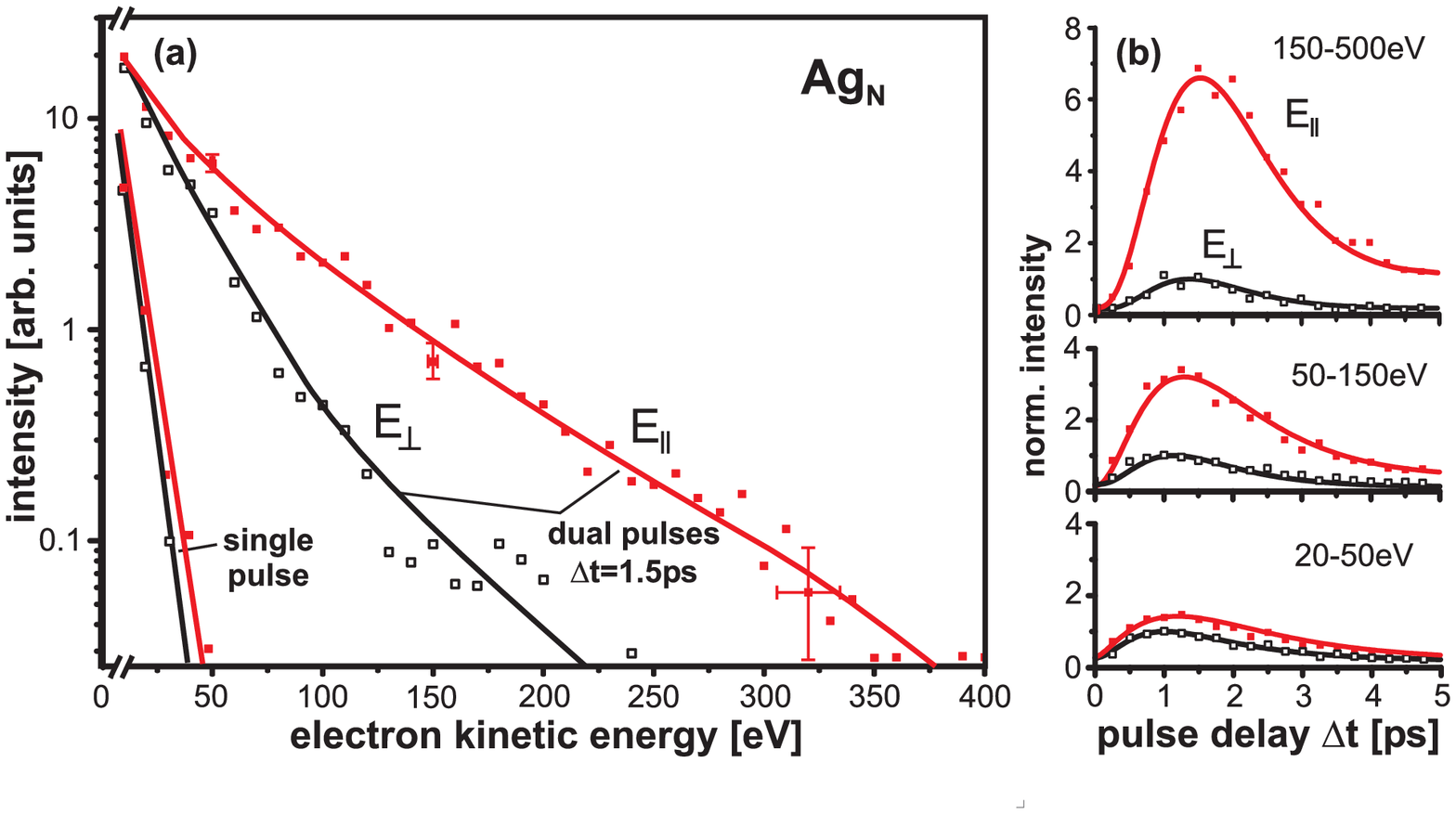}}
 \caption{Photoemission spectra from silver clusters \mbox{($N\approx 10^3$)}
   exposed to 100\,fs laser pulses with peak intensity \mbox{$8\times
     10^{13}{\rm W/cm^2}$} at 800\,nm wavelength: (a) Energy-resolved
   emission parallel (E$_{||}$) and perpendicular (E$_{\perp}$) to the laser
   polarization axis for excitation with a single pulse and dual-pulses with
   optimal temporal delay of $\Delta t=1.5\,{\rm ps}$. (b) Integrated signals
   for three electron energy intervals (as indicated) and normalized to the
   maximum obtained for (E$_{\perp}$) as function of pulse delay.}
 \label{fig_FenPRL07_exp}
\end{figure}

\begin{figure}[t]
\centering\resizebox{0.75\columnwidth}{!}{\includegraphics{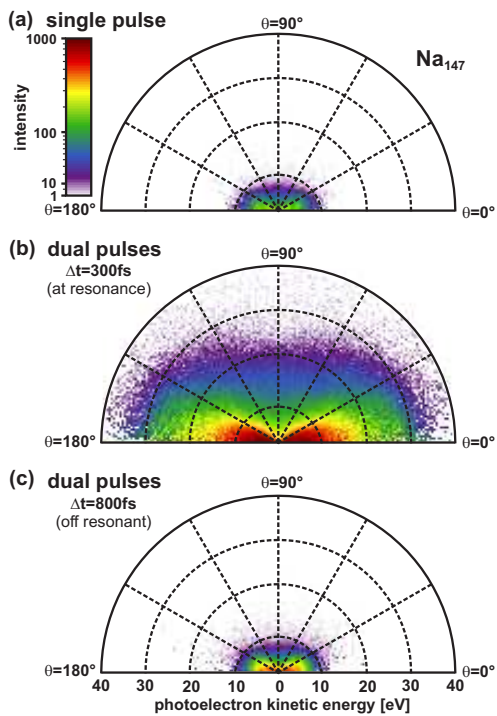}}
\caption{Angular resolved electron emission spectra from Na$_{147}$
  exposed to 25\,fs laser pulses (800\,nm) with peak
  intensity \mbox{$8\times10^{12}\,{\rm W/cm^2}$}, as calculated from
  semiclassical VUU-MD simulations. The data correspond to single-pulse (a)
  and dual-pulse excitations at optimal (b) and a longer nonresonant delay
  (c). The emission angle $\theta$ is given with respect to the laser
  polarization axis. The plot is based on the data from~\cite{FenPRL07a} but shown
  with a more convenient intensity scaling.}
\label{fig_FenPRL07_theo}
\end{figure}

A pronounced resonance effect has further been observed in a dual-pulse experiment on Ag$_N$~\cite{FenPRL07a}. Two pulses with optimal separation yield simultaneously higher electron energies and stronger asymmetry when compared to single-pulse excitation, see Fig\,\ref{fig_FenPRL07_exp}a. Comparison of parallel and perpendicular electron yields for different energy windows as a function of pulse delay  (Fig.\,\ref{fig_FenPRL07_exp}b) shows that the asymmetry increases with electron energy. The strongest anisotropy of about 6.5 is found for the most energetic electrons, uppermost panel in Fig.\,\ref{fig_FenPRL07_exp}b. For all chosen energy windows a maximum yield is observed for similar delays, supporting the presence of plasmon-enhanced electron emission. VUU calculations on a Na$_{147}$, presented in Fig.~\ref{fig_FenPRL07_theo}, show the same qualitative behavior. Off-resonance excitation induces low-energy electron emission and only a small asymmetry (Fig.\,\ref{fig_FenPRL07_theo}a and c), while resonant dual-pulse excitation results in energetic electrons and stronger preference along the polarization axis (Fig.\,\ref{fig_FenPRL07_theo}b).
Note that both experiment and calculation show electron energies beyond 60\,$U_p$ along the polarization axis for resonant excitation.

By trajectory analysis it can be shown that rescattering of electrons by the cluster potential is crucial for the high-energy part of the spectrum. More specifically, corresponding electrons gain the major energy fraction within their final transit through the cluster. The importance of rescattering is well-known from atomic strong-field ionization. A maximum electron energy of $10\,U_{\rm p}$ results from backscattering of tunnel-ionized electrons upon re-encounter with the mother ion at optimal laser phase, see, e.g.,~\cite{WalPRL96}. In contrast to that, a quasi-linear transit along the laser polarization axis with weak deflection turns out to be optimal in clusters. The energies can by far exceed the $10\,U_{\rm p}$ cut-off from atomic backscattering. Two major effects contribute to the energy capture in clusters: (i) acceleration by polarization-fields~\cite{FenPRL07a}, and (ii) laser-field-driven acceleration~\cite{SaaPRL08}.
Within process (i), transit electrons travel in phase with the dynamic cluster polarization field produced from plasmon oscillations. A continuous increase of single-particle energy can be accomplished for fully matched trajectories. This process of surface-plasmon assisted re-scattering in clusters (SPARC) supports preferential ejection of fast electrons along the laser polarization axis. It further provides an explanation for strong acceleration at the instant of resonant plasmon driving due to redistribution of collectively absorbed energy to SPARC-electrons. In the simulation run of Fig.\,\ref{fig_FenPRL07_theo}b, the peak amplitude of the polarization field gradient approaches 35~GeV/\,m. This value corresponds to an effective intensity 25 times higher than that of the laser pulse. Process (ii) results from the laser-driven acceleration of electrons in a static cluster potential. For a passage through a deep global cluster potential, electrons acquire high transit velocities. If the velocity and the polarization axis are parallel and the transit occurs during a beneficial laser half-cycle, electrons can be strongly accelerated by the laser field. Also such type of energy capture from rescattering, which is most effective with a deep cluster potential, produces an alignment of fast electrons. Assuming the formation of a particularly deep cluster space-charge potential for resonant collective electron excitation, this process can result in a plasmon enhancement of the electron kinetic energies as well. A detailed analysis and a corresponding scaling-law for the attainable electron energy are given in~\cite{SaaPRL08}. Besides possible contributions from additional many-body effects, the dynamics will contain a mixture of the processes (i) and (ii). Nevertheless, mechanism (i) dominates for strong collective motion, e.g., in metallic systems at moderate intensity, while (ii) prevails with very deep cluster potentials and high laser intensity.

\section{Perspectives of laser-cluster research}
\label{sec:700}
The previous sections have shown that the field of laser-irradiated clusters is in an actively developing state. Rather than concluding we prefer to discuss a few promising future directions. Among those are prospects of laser pulse shaping or forthcoming novel light sources. Furthermore, complex environments and heterogeneous atomic compositions as well as the use of clusters for relativistic particle acceleration may open new routes for technical applications. Finally we address some prospects and challenges of future theory developments.

\subsection{Laser pulse shaping and control}
\label{sec:7A0}
One of the intriguing perspectives of light-matter coupling pertains to its active manipulation by shaping the pulse in amplitude and phase~\cite{BriAMOP01,BriCPC03}. With molecules, this approach follows the suggestion of~\cite{JudPRL92}, in which a computer-controlled pulse shaper is used in combination with a learning algorithm, see~\cite{BauAPB97,BriN01}, in order to achieve, e.g., a selective molecular reaction. The quantum-mechanical processes can be controlled with the direct feedback from the experiment in an automated fashion, without requiring any model for the system response. This \emph{electron wave-packet engineering} has become a powerful tool to realize the concept of femtochemistry~\cite{ZewPT80}. A more recent technological development further increases the possibilities and prospects of quantum control. With the technique of femtosecond polarization pulse shaping~\cite{BriAPB02,BriPRL04}, it is now possible to vary intensity, instanteneous frequency, and light polarization (i.e., the degree of ellipticity as well as the orientation of the principal axes) as functions of time within a single femtosecond laser pulse. Thus, a full temporal and spatial control is at reach.

For intense laser-cluster interactions, shaping the pulse in amplitude and phase can be a fascinating tool to selectively steer the dynamics of charging, particle-, or photon emission. Basic findings along this line are the control of the Coulomb explosion by varying the laser pulse length as well as the time delay in the dual-pulse experiments as outlined in Sec.~\ref{sec:6B1}. For example, Fig.~\ref{fig_DoePRA06} has shown the dramatic effect of the adjustment of two femtosecond laser pulses on the charging efficiency and the energy of emitted electrons. Adaptive femtosecond control was demonstrated on the Coulomb explosion of Xe$_N$~\cite{ZamPRA04}. Here the signal of highly charged Xe$^{q+}$ could be optimized with the help of a simple genetic algorithm, applied to an initially Fourier transform limited pulse with 100\,fs duration and 230\,\mbox{$\mu$}J energy. The procedure resulted into a sequence of two 120\,fs pulses with similar amplitude and separated in time by about 500\,fs, like in the optimized dual-pulse experiments~\cite{DoePRL05}. It is very interesting to note that this two-pulse optimum has been worked out by the algorithm starting from a 80 parameter unbiased configuration. Corresponding simulations within a semiclassical molecular dynamics approach predicted that for selected combinations of cluster size, laser intensity, and wavelength, the ionization may be optimized by a three-pulse sequence~\cite{MarPRA05}. In another closed-loop optimal control experiment on rare gas clusters, pulse-shaping has shown a significant potential for x-ray yield enhancement~\cite{MooAPB05}.

Whereas the optimal-control studies on clusters were limited to an optimization of the pulse amplitude so far, the simultaneous variation of the pulse phase is still an exciting challenge. First results of such a fully unbiased adaptive fs experiment have demonstrated the controlled adjustment of charge state distributions from the Coulomb explosion of Ag$_N$ embedded in Helium droplets~\cite{TruPRA09}. In this study the optimization of the Ag$^{q+}$ charge spectrum converged to a pulse structure with a weaker pre-pulse and a stronger, negatively chirped main pulse. However, we are far off a full theoretical understanding of the complex dynamics driven by pulses shaped in amplitude and phase. In the future, if sufficient mass-selected cluster intensity can be prepared, single ionization states and narrow-banded high-energy radiation might be realized.

\subsection{Towards VUV-, XUV, and soft x-ray pulses}
\label{sec:7B0}
The nature of the laser cluster coupling fundamentally changes when going from the IR regime towards excitation with VUV-, XUV-, ore even x-ray pulses. This concerns ionization processes as well as the mechanisms of energy absorption. For excitation with IR pulses, field-driven ionization plays a crucial role for the nanoplasma generation, e.g., in rare gas systems. The subsequent energy capture, which eventually removes electrons from the cluster, is of plasma nature and can be strongly enhanced through resonant collective excitations. Because of extensive plasma heating and resulting further ionization high charge states can arise with IR pulses.

When going below about 100\,nm wavelength, a value which was used in the first VUV experiments on rare gas clusters, photoionization becomes the dominant charging mechanism for inner ionization. Concerning the energy absorption, collective effects can be disregarded as the required critical density cannot be reached and pure IBS heating prevails. In fact, the observation of surprisingly high energy capture in the first VUV-experiments on clusters, see Sec.~\ref{sec:6A2}, have sparked a remarkable progress in the understanding of heating- and ionization effects in dense targets~\cite{SanPRL03,SiePRL04,BauAPB04,JunJPB05,RamJPB06,SaaJPB06,ZiaLPB07,GeoPRA07}.

When further increasing the laser frequency, IBS heating becomes more and more suppressed, cf. Eq.~(\ref{eq:Resonance_heating}), so that photoexcitation of tighly bound electrons begins to become even the leading energy capture process. Signatures of this transition have recently been observed on Ar$_N$ in intense femtosecond XUV FEL pulses at $\lambda=32$\,nm ($\hbar\omega_{\rm las}=38$\,eV)~\cite{BosPRL08}. By comparing the experimental photoelectron spectra with complementary Monte-Carlo simulations, the following behavior was found. The cluster ionization first proceeds as a multistep process of direct single-photon absorption events. Electrons are released from the cluster directly without prior inner ionization and the space charge build-up results in an energy downshift for subsequent ionization steps. This shift leads to a highly nonthermal electron energy distribution. At a certain degree of ionization, the cluster potential frustrates further electron release, leading to the formation of a nanoplasma only beyond a certain threshold intensity. Even at higher intensity no strong impact of IBS heating was found. These findings are in agreement with corresponding MD results~\cite{Arb09} and calculations based on kinetic transport equations~\cite{ZiaNJP09}.

Using intense soft x-ray pulses at $\lambda=13$\,nm~\cite{HoeJPB08} found highly efficient charging of Xe$_N$ with ions up Xe$^{9+}$, which can be ascribed to the large absorption cross section of the giant atomic Xe 4-$d$ resonance. By surrounding Xe$_N$ with an additional argon layer it was further shown, that charge recombination dynamics can be studied in the well controllable core-shell system.

Another interesting issue concerns the time-resolved monitoring of the cluster excitation and the subsequent Coulomb explosion by combining different types of pulses. For example, the ionization of rare gas clusters may be driven by VUV radiation, as in the case of~\cite{WabN02}, whereas a subsequent IR pulse probes the collective electron response of the priorily \emph{metallized system}~\cite{SiePRA05}. A combination of VUV and XUV pulses was proposed to monitor the time-dependent ionization stages in small clusters~\cite{GeoPRL07}. Another scheme uses x-ray radiation for Thomson scattering on exploding clusters or droplets, which have been initially excited by strong IR pulses~\cite{HoeHEDP07}. By this, a fundamental understanding can be gained on highly nonstationary strongly coupled plasmas and their transition from degenerate to classical systems. The advent of the X-FEL will open direct access to the temporal development of such complex systems.

\subsection{Clusters in an environment}
\label{sec:7C0}

Embedding clusters into an environment or depositing them at surfaces modifies their optical responses, see~\cite{Krei95}. A major branch of present-days cluster research comprises systems in contact with solid surfaces, for a review see, e.g.,~\cite{Mei00,Mei06,MeiEPJD07}. An extremely rich scenery unfolds when considering the specific effects emerging from the interaction of a cluster with an environment. One finds, e.g., only small shifts for the Mie plasmon resonances of metal clusters embedded in inert matter~\cite{Die02,Feh07a} and larger ones for contact with conducting material~\cite{Pin04a}. Details of the excitation spectrum, however, are rather sensitive to the interface. This affects, e.g., the spectral fragmentation and the plasmon damping, for experimental assessment see, e.g.,~\cite{Hen03a,Zie04a}. Large effects from the environment are to be expected in the reaction dynamics at high excitations.

\begin{figure}
\centering\resizebox{\columnwidth}{!}{\includegraphics{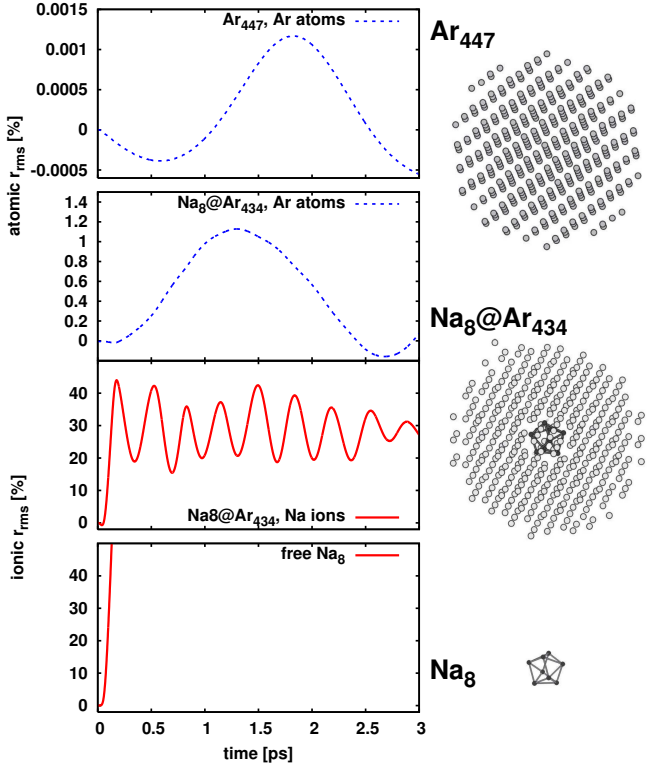}}
\caption{\label{fig:irradvariety}
  Time evolution of the radius mean square (rms) of free Na$_8$ (bottom), Na$_8$ embedded in Ar$_{434}$ (middle), and pure Ar$_{447}$ (top), after irradiation with \mbox{($2.4\times10^{12}\,$W/cm$^2$; 33\,fs; 650nm)}. Calculations have been performed using TDDFT for Na valence electrons and MD for Na$^+$ ions and Ar atoms. From \cite{Feh07b}, with kind permission of The European Physical Journal (EPJ).
}
\end{figure}
A theoretical example is shown in Fig.~\ref{fig:irradvariety}. It compares three test cases, Na$_8$ as a small metal cluster, the Na$_8$ embedded in Ar$_{434}$ (a large rare gas cluster as model for a matrix), and pure Ar$_{447}$, all three exposed to the same laser pulse. The laser pulse leads to a charge state 3+ of Na$_8$. In the free case (bottom panel), this induces a Coulomb explosion. The situation is quite different for Na$_8$ in Ar$_{434}$ matrix. The metal cluster is again highly excited and starts to explode.  But the explosion is stopped by the Ar atoms which efficiently absorb the excitation energy of the system (lower middle panel). The Ar matrix is perturbed and exhibits monopole oscillations, but of much smaller amplitude than Na$_8$ (upper middle panel). The upper panel of Fig.~\ref{fig:irradvariety} finally shows the case of a pure Ar$_{447}$.  Under the same laser conditions, one can see that the Ar$_{447}$ remains essentially unperturbed, showing no electron emission and only extremely weak breathing oscillations. Obviously, the Na$_8$ acts here as a chromophore, absorbing energy from the laser pulse and transferring it to the environment. The example shows that the combination of two materials changes the reaction dynamics of either system dramatically. One can easily imagine that putting clusters in contact with various substrates produces a world of interesting scientific questions and offers technical applications in the field of nanotechnology. We mention in the following briefly a few aspects to give an idea of the enormous possibilities, concentrating on optical properties.

When depositing Au$_N$ on a semiconductor surface the change of optical cluster properties can be exploited to producing enhanced photocurrent~\cite{Sch05a}. There are promising applications, e.g., in medicine where the frequency selective optical coupling of organically coated metal clusters attached to biological tissue may by used for diagnosis~\cite{Bru98,May01,Dub02,Sim07a} or, in the case of stronger laser fields, for localized heating in therapy~\cite{Khl06a}. The field amplification effect is of interest in many other materials and applications, see, e.g., the study of localized melting for the generic combination of Au clusters embedded in ice~\cite{Ric06a}. The strong coupling to light may be used for more than just heating. Ensembles of size and shape selected clusters on a surface are produced by ``shape burning''~\cite{Wen99,Oua05a}. A dedicated modification of the shape for embedded Ag clusters is demonstrated in~\cite{Per00a,Dah06a}. Time scales and mechanisms of energy transport are thus an issue of high interest. Theoretical analysis has yet to cope with the great variety of material combinations. For an example using the generic test system of metal cluster in a rare gas matrix, see~\cite{Feh07b,Feh07c}. A thorough study of surface-deposited cluster subject to strong laser pulses still is a matter of future studies.

\subsection{Relativistic particle acceleration with clusters}
\label{sec:7D0}

Strong laser fields impinging on clusters can drive interesting electron dynamics. For an example from the moderate intensity domain ($\sim 10^{14}\,$W/cm$^2$),~\cite{FenPRL07a} describe a cascade-like acceleration mechanism based on resonant field amplification in individual clusters, see Sec.\ref{sec:6B3}. In the regime of $10^{15-17}\,$W/cm$^2$ electron energies from keV up to some hundreds of keV are reported~\cite{ShaPRL96,SprPRA03,ChePP02}, emitted in transverse direction to the laser propagation axis. Beyond a few tens of keV the emission is most likely due to macroscopic plasma wave-breaking effects in a very dense cluster beam, as is further supported by a pronounced forward peak in the emission~\cite{ChePRE02}. Moreover, there are few examples close to or in the relativistic regime ($10^{19}\,$W/cm$^2$). From studies on bulk and dense atomic gases it is known that charged particles can be accelerated by the plasma wakefield to large kinetic energies, for a detailed theoretical discussion see~\cite{Puk02a} and for a recent experimental example~\cite{Kar07a}. There exist realistic plans to employ the effect to build fairly inexpensive laser-driven table-top accelerators~\cite{Gru07a}.
\begin{figure}[htbp]
\centering\resizebox{0.9\columnwidth}{!}{\includegraphics{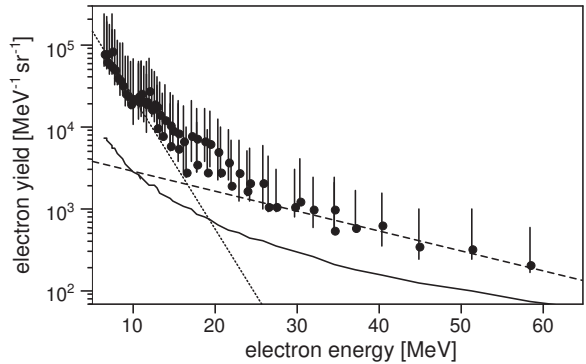}}
\caption{\label{fig:electron-acceleration} Electron kinetic energy distribution resulting from irradiation of large Ar particles (micron sized diameter) in Ar gas environment by an intense laser (3.5$\times 10^{19}$W/cm$^2$). The solid curve shows the detection threshold. Straight lines
indicate fits to thermal distributions, dashed for \mbox{$T=18.8$\,MeV} and dotted for \mbox{$T=2.8$\,MeV}.
Adapted from~\cite{Fuk07a}, with permission from Elsevier. }
\end{figure}
Indications for special relativistic electron acceleration mechanisms with clusters have been reported in a study on large Ar particles in a low-density background gas~\cite{Fuk07a}. The example in Fig.~\ref{fig:electron-acceleration} displays the achieved electron kinetic energies. On the basis of simulation results, the two temperatures have been associated with two different generating mechanisms. The lower-energy electrons stem from acceleration in a distorted wakefield. In contrast to that, the high-energy electrons are removed from the clusters with already relativistic energies and then further accelerated by the laser pulse directly. The kinetic energies observed here are still far below what can emerge from bulk plasma. However, whereas optimum conditions, advantages, and disadvantages have yet to be worked out, the example proves the feasibility of relativistic particle acceleration with clusters. Moreover, the use of clusters as dense electron containers for free-space electron acceleration, e.g., with radially polarized laser beams, might be promising for generating ultrashort electron bunches with durations down to the attosecond domain at up to GeV energy~\cite{VarPRE06,KarLPB07}.

\subsection{Challenges for theory}
\label{sec:6E0}
The theoretical description of laser-cluster dynamics requires to cover very different scales of length, time, and energy. This difficulty usually hampers a fully microscopic treatment of all degrees of freedom. Fortunately, the nature of the response is to great extend determined by the type of excitation or the size and structure of the target, at least at the different scales. We have seen in Sec.~\ref{sec:300} that there exists a bunch of theoretical approaches, ranging from fully microscopic ones to macroscopic ones, which are applicable within certain windows of size and energy. Their limitations result from both, formal constraints, e.g., due to the level on which correlations and quantum effects are resolved, as well as from practical ones like the numerical expense. To promote the development of more elaborate methods and schemes with wider ranges of applicability we see at least two promising directions.

The first and most straightforward path is the formal improvement of particular methods along with the rapid development of high-performance computers. For instance, the impressive growth of numerical power allows the application of fully correlated quantum approaches to systems with several electrons, e.g., with the efficient handling of few-body wavefunctions by MCTDH~\cite{BecPR00} or MCTDHF methods~\cite{Cai05a}. This opens a route to explore truly correlated electron dynamics including continuum and intermediate excited states, for an example on a molecular system see~\cite{SukPRA09}. A more fundamental challenge concerns the inclusion of dynamical correlations in mean-field quantum theories like TDLDA in the sense of a quantum counterpart to the semiclassical description within VUU. On the classical level, efficient numerical schemes and large-scale parallelization promises the feasibility of up to giga-particle simulations.

A second frontier concerns the connection of different treatments in terms of multi-level or multi-scale methods. A well-known example for biological and chemical applications are mixed quantum mechanics/molecular dynamics approaches, see e.g.~\cite{BakJCP96}. Also for clusters the combination of different levels has turned out to be very successful, e.g., within MD and hydrodynamic schemes for strong-field excitations, where the quantum nature of inner ionization is taken into account via effective rates and cross-sections. The connection of different treatments, however, requires interfaces, the validation of which is a great challenge. Firm links between the approaches and reliable interfaces, e.g., within overlapping zones like those indicated in Fig.~\ref{fig:theor_regimes}, are therefore highly desirable and have far-reaching implications. One example could be the connection of an explicit atomic-scale quantum treatment of inner ionization with a more coarse-grained semiclassical or even classical treatment of quasifree and continuum electrons. This would be of high interest for strong-field laser-cluster interactions in a wide range of laser frequencies, i.e., from the IR up to the x-ray domain. Another challenging aspect are strong-field excitations of larger clusters and particles in the IR range, where propagation effects of the light field cannot longer be neglected. Here a combination of molecular dynamics techniques for evaluating the short-range part of the interactions combined with electromagnetic particle-in-cell concepts for describing the long-range component of the Coulomb- and radiation fields might be promising. Last but not least, such neighboring approaches could also be combined in a sequential way, e.g., to resolve the laser excitation microscopically, whereas the long-term behavior is described with a less expensive scheme.

Along these lines the field of laser-cluster dynamics will certainly be inspired by forthcoming developments in other branches like atomic, molecular, and plasma physics and vice versa.

\begin{acknowledgments}
We thank Gustav Gerber for fruitful discussions. T.F., J.T., and K.-H. M.-B. gratefully acknowledge financial support by the Deutsche Forschungsgemeinschaft (DFG) within the SFB 652. Computer time has been provided by the HLRN Computing Center. This work was further supported by the DFG (RE 322/10-1), the French-German exchange program PROCOPE (07523TE), the Institut Universitaire de France, the Agence Nationale de la Recherche (ANR-06-BLAN-0319-02), the Humboldt foundation, a Gay-Lussac price, the French computational facilities CalMip, IDRIS and CINES, and the Computing Center of the University Erlangen.

\end{acknowledgments}

\end{document}